\documentclass[aip,jcp,preprint,floatfix]{revtex4-2}
\usepackage{amsmath, amssymb}
\usepackage{graphicx}
\usepackage{dcolumn}
\usepackage{bm, braket, color}

\begin{document}

\title{Influence of nonlocal excitonic optical response on in-plane exciton polaritons supported by monolayer transition-metal dichalcogenides}
\author{Kuniyuki Miwa}
\email{kuniyukimiwa@ibs.re.kr}
\affiliation{Center for Quantum Conversion Research, Institute for Basic Science, 123 Cheomdangwagi-ro, Buk-gu, Gwangju 61005, Republic of Korea}
\affiliation{Surface and Interface Science Laboratory, RIKEN, 2-1 Hirosawa, Wako, Saitama 351-0198, Japan}

\date{\today}

\begin{abstract}
Achieving nanoscale light confinement and long-range propagation is a central challenge in nanophotonics, with important implications for controlling molecular photophysical processes and energy transfer.
Monolayer transition-metal dichalcogenides (TMDs) are attractive platforms for photonic and optoelectronic applications because of their distinctive optical and electronic properties.
In particular, in-plane exciton-polaritons (IPEPs)---hybrid states of photons and excitons (bound electron--hole pairs) in TMD monolayers---offer a promising route to deeply subwavelength optical modes at visible frequencies.
In this regime, electromagnetic fields vary on short length scales comparable to the intrinsic size of excitons, causing conventional local-response approximations to break down.
Here, we develop a microscopic theory of the nonlocal excitonic optical response and investigate the near-field properties of IPEPs in monolayer WS$_2$ encapsulated by hexagonal boron nitride (hBN).
By combining a two-band massive Dirac fermion description with a generalized Mott--Wannier treatment, we derive a nonlocal electric susceptibility that accounts for both the center-of-mass and internal relative motion of excitons.
Spatial nonlocality in the excitonic optical response shifts the polariton dispersion toward lower in-plane wavevectors compared with the results based on the local-response approximation.
This reduces high-wavevector contributions and produces smoother near-field distributions that extend over larger lateral distances while retaining nanoscale confinement.
Furthermore, at higher excitation energies, multiple exciton-polariton modes contribute at distinct wavevectors, and their relative contributions can be tuned by varying the thickness of the capping hBN layer.
These insights elucidate the fundamental role of spatial nonlocality in the polaritonic properties of van der Waals heterostructures, providing a theoretical foundation for nanophotonic devices such as multimode nanoscale waveguides.
\end{abstract}

\maketitle


\section{Introduction}

Spatial confinement of electromagnetic energy at the nanoscale and its long-range propagation are central to nano-optics~\cite{Novotny2012} and crucial for probing and controlling molecular photophysical processes.~\cite{Nitzan2024, Craig1998}
Because spontaneous emission and nonradiative decay rates in molecular systems are highly sensitive to their surrounding electromagnetic environments,~\cite{Purcell1946, Drexhage1970, Chance1978} manipulating these fields enables the enhancement of optical signals~\cite{Yamaguchi2008, Kumar2023, Moskovits1985, Campion1998, Willets2007, Langer2020, Wang2020, Yi2025, Zhang2013, Lee2019, Jaculbia2020, Gutzler2021, Kumagai2025} and the control of molecular properties.~\cite{Ebbesen2016, Herrera2016, Chen2019, Garcia-Vidal2021, Miwa2023, Sukharev2023, Engelhardt2023, Yuen-Zhou2025}
Furthermore, electromagnetic environments can be engineered to accelerate energy and information transfer between spatially separated molecules.~\cite{Gonzaga-Galeana2013, Li2015, Marocico2016, Hsu2017}
Although metallic planar interfaces and plasmonic nanostructures are widely employed for such purposes, the intrinsic optical losses of metals remain a critical limitation for such applications.~\cite{Drexhage1970, Chance1978, Raether1988, Kelly2003, Maier2007, Shishkov2016, Miwa2021, Reitz2026}

To circumvent these dissipative losses, optically anisotropic dielectrics have emerged as a powerful alternative for manipulating light beyond the diffraction limit.
When one of the principal components of their permittivity tensor has a sign opposite to the other two, the isofrequency surfaces of extraordinary waves form open hyperboloids in reciprocal space, thereby enabling the propagation of modes with large wavevectors.~\cite{Poddubny2013, High2015}
Recently, two-dimensional van der Waals materials have attracted significant interest as natural anisotropic crystals capable of supporting such modes.~\cite{Basov2016, Low2017, Caldwell2019, Zhang2021, Jia2022, Basov2021, Wu2022, Galiffi2023, Wang2024, Basov2025, Zhou2026, Hu2026}
In particular, multilayer hexagonal boron nitride (hBN) has been widely studied for its hyperbolic phonon polariton (HPhP) modes, which are collective modes originating from the coupling of photons with optical phonons within its Reststrahlen bands.~\cite{Dai2014, Caldwell2014, Jacob2014, Caldwell2015}
These HPhPs, characterized by large wavevectors, can confine optical fields to deeply subwavelength scales, offering promising opportunities for sub-diffraction imaging~\cite{Li2015a}, molecular nanospectroscopy~\cite{Autore2017}, interfacial thermal management~\cite{Hutchins2025}, and quantum technologies.~\cite{Narimanov2024}
However, typical HPhPs are restricted to the mid-infrared range due to the characteristic energies of optical phonons.~\cite{Ma2018, Li2020, Hu2020, Chen2020, Martin-Sanchez2021, Ma2021, Hu2023, Guo2025, Kamada2026}
While operation at visible frequencies is highly desirable for practical applications, realizing deeply confined, large-wavevector polaritons in this spectral range remains an important challenge.

In this regard, monolayer transition-metal dichalcogenides (TMDs) represent a promising platform for supporting optical modes with large wavevectors in the visible range.~\cite{Wang2024, DeAbajo2025}
The coupling of photons with bound electron--hole pairs (excitons) in the TMD layer gives rise to in-plane exciton polariton (IPEP) modes near sharp excitonic resonances.~\cite{Gershuni2024}
Although recent experimental results suggest the presence of these modes~\cite{Epstein2020, Meshulam2026}, their strong field confinement and correspondingly large wavevectors pose a fundamental challenge to conventional local-response descriptions.
Because these modes support in-plane wavevectors tens to hundreds of times larger than the free-space wavevector at the same frequency, the characteristic spatial scales of their electromagnetic fields can become comparable to the exciton Bohr radius or the coherence length.~\cite{Wang2018}
Consequently, nonlocal effects in the excitonic optical response become essential for adequately describing their polaritonic properties.
To date, theoretical treatments of the excitonic optical response have typically incorporated such nonlocal effects phenomenologically through modified Lorentz oscillator models that include an exciton kinetic-energy term.~\cite{Eini2022, Gershuni2024}
Although this approach successfully captures nonlocal corrections associated with the exciton's center-of-mass motion, it does not explicitly account for the internal degrees of freedom corresponding to the relative electron--hole motion.
TMD optical spectra, however, are influenced not only by the dominant 1s exciton state but also by transition pathways to higher Rydberg states.~\cite{Chernikov2014, Berkelbach2015, Liu2019}
Furthermore, the steep spatial gradients of near-fields generated by quantum emitters---such as single molecules placed near a TMD interface---provide finite-wavevector coupling pathways to otherwise dipole-forbidden p-type exciton states.~\cite{Basov2021, Shi2022, Basov2025}
These considerations motivate the development of a microscopic nonlocal optical-response theory that explicitly accounts for both the center-of-mass motion and the internal relative electron--hole motion of TMD excitons.

To address this issue, we present a microscopic derivation of the electric susceptibility of monolayer TMDs by combining a band-edge description based on the two-band massive Dirac fermion model~\cite{Xiao2012} with an exciton description based on the Mott--Wannier model.~\cite{Wannier1937, Yu2010, Haug2009}
This formulation yields wavevector-dependent transition moments that incorporate the effects of exciton center-of-mass motion and the Berry phase associated with the band-edge electronic states, thereby enabling a systematic analysis of symmetry-dictated optical selection rules.~\cite{Xiao2012, Ye2014, Srivastava2015}
Crucially, a quantitative description of the exciton envelope function, which characterizes the relative electron--hole motion, is indispensable for realistic polaritonic modeling.~\cite{Elliott1957, Hopfield1958}
To achieve a realistic description, we incorporate material-specific screened electrostatic interactions between electrons and holes into the framework.
By leveraging the Quantum Electrostatic Heterostructure (QEH) model to calculate these screened interactions, we determine the exciton states and their corresponding transition amplitudes.~\cite{Andersen2015, Latini2015, Gjerding2020}
This framework allows us to clarify how the spatial gradients of near-fields and the spatial profile of the exciton envelope function modify the TMD electric susceptibility and the resulting polaritonic properties.

In this study, we apply this theoretical framework to investigate the nonlocal optical response of TMD excitons and the resulting near-field electromagnetic properties of IPEPs.
Specifically, we focus on an hBN-encapsulated WS$_2$ monolayer, a well-established system in which sharp excitonic resonances have been experimentally observed.~\cite{Epstein2020, Meshulam2026}
We first derive the nonlocal electric susceptibility tensor from the microscopic exciton wavefunctions and incorporate it into an electromagnetic solver based on the transfer matrix method (TMM).
We then analyze the spatial distribution of the electromagnetic field generated by a point dipole near the TMD interface.
We show that incorporating the nonlocal excitonic optical response shifts the IPEP dispersion toward lower in-plane wavevectors, leading to smoother field distributions that extend over larger lateral distances while retaining nanoscale confinement.
We further demonstrate that multiple Rydberg exciton states contribute to the polaritonic response at distinct wavevectors and that their relative contributions can be tuned by varying the thickness of the capping hBN layer, providing a route toward multimode nanoscale waveguides with mode-selective control.
These findings suggest potential applications in long-range, spatially modulated F\"{o}rster resonance energy transfer (FRET)~\cite{Katzer2023, Thompson2023, Lin2023} and quantum technologies~\cite{Narimanov2024}.
The remainder of this paper is organized as follows: we present our theoretical formalism for the TMD susceptibility tensor in Section II, analyze the resulting near-field polaritonic properties in Section III, and finally summarize our findings in Section IV.


\section{Theoretical Model}\label{sec:Model}

\begin{figure}[h!]
\centering
\includegraphics[bb=0 30 289 170, clip]{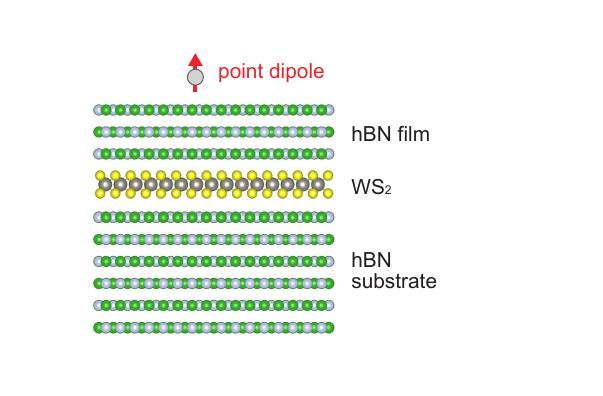}
\caption{
\label{fig:schematic}
Schematic of an hBN-encapsulated monolayer WS$_2$ with a point dipole placed above the heterostructure.
}
\end{figure}

To describe IPEPs in a van der Waals heterostructure, we consider an hBN-encapsulated WS$_2$ monolayer [Fig.~\ref{fig:schematic}].
To capture the microscopic excitonic response under near-field confinement, we construct a nonlocal theoretical framework that combines an effective band-edge description of the TMD with an exciton model incorporating a material-specific screened electron--hole interaction.~\cite{Andersen2015, Latini2015, Gjerding2020}
Although monolayer TMDs are atomically thin, previous studies have shown that the finite extent of the electronic charge distribution in the out-of-plane direction affects the excitonic Rydberg series~\cite{Andersen2015, Latini2015, Gjerding2020}.
We therefore adopt an effective two-dimensional (2D) description of the single-particle electronic states, in which the in-plane electronic structure is represented by two-dimensional Bloch states, while the finite out-of-plane extent is described by a charge-distribution function.

The overall workflow of our theoretical model is structured as follows.
First, we employ the two-band massive Dirac fermion model near the $K_\pm$ valleys to obtain the 2D Bloch states and derive the wavevector-dependent transition momentum matrix elements.
Second, we solve the generalized Mott--Wannier equation for the exciton envelope functions using a material-specific screened Coulomb interaction that accounts for the finite out-of-plane extent of the electronic charge distribution.
Third, by combining the single-particle Bloch states with the exciton envelope functions within linear response theory, we derive the nonlocal electric susceptibility tensor $\chi_{\mu\nu}(\bm{q}_\parallel, \omega)$ of the WS$_2$ monolayer.
Finally, we incorporate this spatially dispersive response into an electromagnetic solver based on the TMM to calculate the dissipation spectrum and near-field propagation in the air/hBN/WS$_2$/hBN system.

For clarity, we use the following wavevector notation throughout this study.
The two-dimensional crystal wavevector in the first Brillouin zone (BZ) is denoted by $\bm{k}_\parallel$.
Near the valley extrema $\bm{K}_\tau$ ($\tau = \pm 1$), the single-particle wavevector measured relative to the valley center is represented by $\bm{\xi} = \bm{k}_\parallel - \bm{K}_\tau$, with components $( \xi_x , \xi_y )$.
The macroscopic in-plane wavevector of the electromagnetic field is denoted by $\bm{q}_\parallel = ( q_x , q_y )$, while its out-of-plane component is denoted by $q_z$.
Through in-plane momentum conservation in the optical transition, $\bm{q}_\parallel$ also corresponds to the center-of-mass wavevector of the exciton.


\subsection{Independent particle description}


To describe the electronic states near the $\bm{K}_\pm$ points of the first Brillouin zone, we consider a minimal band model with the following two states as a basis:
\begin{subequations}
	\begin{align}
		\ket{ u_{\rm c} }
		&=
		\ket{ d_{ z^2 } }
		,
\\
		\ket{ u_{\rm v}^\tau }
		&=
		\frac{1}{ \sqrt{2} }
		\left(
			\ket{ d_{ x^2 - y^2 } }
			+
			i \tau
				\ket{ d_{xy} }
		\right)
		,
	\end{align}
\end{subequations}
where the subscript ${\rm c}$ (${\rm v}$) indicates the conduction (valence) band, and $\tau = \pm 1$ is the valley index.
The conduction band-edge state is made of $\ket{ d_{ z^2 } }$ orbitals of the transition metal, and the valence-band-edge state corresponds to the symmetry-adapted $( \ket{ d_{ x^2 - y^2 } } + i \tau \ket{ d_{xy} } ) / \sqrt{2}$ orbitals.
The conventional two-band massive Dirac Hamiltonian has the form~\cite{Berkelbach2015}
\begin{equation}
	H_\tau ( \bm{k}_\parallel )
	=
	\begin{pmatrix}
		E_{\rm g} / 2
		&
		a t ( \tau \xi_x - i \xi_y )
		\\
		a t ( \tau \xi_x + i \xi_y )
		&
		- E_{\rm g} / 2
	\end{pmatrix}
	,
	\label{eq:Hamiltonian}
\end{equation}
which is linearized in the wavevector difference $\bm{\xi}$ with respect to the nearest $K$ point.
The Hamiltonian is characterized by three parameters: the energy bandgap $E_{\rm g}$, the lattice constant $a$, and the effective hopping integral $t$.
For simplicity, the spin--orbit coupling (SOC) term is not explicitly included in the Hamiltonian.
Although SOC induces a significant spin splitting at the valence-band edges, the spin-up and spin-down components are largely decoupled and the out-of-plane spin remains a good quantum number.~\cite{Glazov2014, Glazov2015}
Therefore, this simplified treatment retains the essential valley-dependent character of the band-edge electronic states and the fundamental optical selection rules.
This Hamiltonian has been used to investigate optical selection rules for monolayer TMDs within the local-response approximation.~\cite{Berkelbach2015}


By diagonalizing Eq.~\eqref{eq:Hamiltonian}, we obtain the symmetric conduction and valence band dispersions
$
	E_{\rm c/v} ( \bm{k}_\parallel )
	=
	\pm \varepsilon( \bm{k}_\parallel )
$
, where the energy scale is defined as
$
	\varepsilon( \bm{k}_\parallel )
	=
	\frac{1}{2} \sqrt{E_{\rm g}^2 + 4 a^2 t^2 \xi^2}
$
 with 
$
	\xi
	=
	|\bm{\xi}|
$
. The corresponding periodic parts of the Bloch eigenstates are given by
\begin{subequations}
	\begin{align}
		\ket{ \psi^\tau_{ {\rm c} , \bm{k}_\parallel } }
		&=
		\frac{ 1 }{ \sqrt{ 2 } }
		\left[
			\sqrt{ \alpha_+ ( \bm{k}_\parallel ) }
				\ket{ u_{\rm c } }
			+
			\sqrt{ \alpha_- ( \bm{k}_\parallel ) }
				e^{ i \tau \phi_{ \bm{k}_\parallel } }
				\ket{ u_{\rm v}^\tau }
		\right]
		,
\\
		\ket{ \psi^\tau_{ {\rm v} , \bm{k}_\parallel } }
		&=
		\frac{ 1 }{ \sqrt{ 2 } }
		\left[
			-
			\sqrt{ \alpha_- ( \bm{k}_\parallel ) }
				\ket{ u_{\rm c } }
			+
			\sqrt{ \alpha_+ ( \bm{k}_\parallel ) }
				e^{ i \tau \phi_{ \bm{k}_\parallel } }
				\ket{ u_{\rm v}^\tau }
		\right]
		,
	\end{align}
\end{subequations}
where the wavevector-dependent expansion coefficients are defined as
$
	\alpha_\pm(\bm{k}_\parallel)
	=
	1 \pm E_{\rm g} / [2\varepsilon(\bm{k}_\parallel)]
$
, and the phase factor
$
	\phi_{\bm{k}_\parallel}
$ 
is determined by the orientation of the wavevector measured from the valley center, 
$
	\tan \phi_{\bm{k}_\parallel}
	=
	\xi_y / \xi_x
$.
Here, we choose the phase of eigenvectors such that the first element is purely real.


The wavefunctions of the full single-particle Bloch states, $\ket{ \Psi^\tau_{ {\rm c} , \bm{k}_\parallel } }$ and $\ket{ \Psi^\tau_{ {\rm v} , \bm{k}_\parallel } }$, are explicitly constructed by combining these periodic parts with the in-plane plane-wave factor and the out-of-plane confinement function as
\begin{subequations}
	\begin{align}
		\Psi^\tau_{ {\rm c} , \bm{k}_\parallel } (\bm{r}_\parallel , z )
		&=
		\frac{ 1 }{ \sqrt{ \Omega_\parallel } } e^{ i \bm{k}_\parallel \cdot \bm{r}_\parallel }
		\psi^\tau_{ {\rm c} , \bm{k}_\parallel } (\bm{r}_\parallel )
		\phi_{ {\rm c} \perp } (z)
		,
\\
		\Psi^\tau_{ {\rm v} , \bm{k}_\parallel } (\bm{r}_\parallel , z )
		&=
		\frac{ 1 }{ \sqrt{ \Omega_\parallel } } e^{ i \bm{k}_\parallel \cdot \bm{r}_\parallel }
		\psi^\tau_{ {\rm v} , \bm{k}_\parallel } (\bm{r}_\parallel )
		\phi_{ {\rm v} \perp } (z)
		,
	\end{align}
\end{subequations}
where $\Omega_\parallel$ is the in-plane normalization area, and $\phi_{ {\rm c/v} \perp }(z)$ represents the confinement function in the out-of-plane direction for the conduction (valence) band.
The influence of the shape of $\phi_{ {\rm c/v} \perp } (z)$ is discussed in the next subsection.


The transition probability and optical selection rules are governed by the in-plane momentum operator $\hat{ \bm{p} }_\parallel$.
To account for the in-plane spatial phase variation of the electromagnetic field, we introduce its in-plane wavevector $\bm{q}_\parallel$.
The inter-band transition momentum matrix elements
$
	\bm{P}_{\rm cv}^\tau
	=
	\braket{
		\Psi^\tau_{ {\rm c} , \bm{k}_{\rm c} }
		\left|
			e^{ i \bm{q}_\parallel \cdot \bm{r}_\parallel } \hat{ \bm{p} }_\parallel
		\right|
		\Psi^\tau_{ {\rm v} , \bm{k}_{\rm v} }
	}
$
are written as
\begin{equation}
	\bm{P}_{\rm cv}^\tau
	=
	\frac{ 1 }{ \Omega_{\rm cell} }
	\frac{ m_0 }{ \hbar }
	\int_{\rm unitcell} d \bm{r}_\parallel
		\psi^{ \tau * }_{ {\rm c} , \bm{k}_{\rm c} } ( \bm{r}_\parallel )
		\left[
			\nabla_{ \bm{k}_\parallel } H ( \bm{k}_\parallel )
		\right]_{ \bm{k}_\parallel = \bm{k}_{\rm v} }
		\psi^\tau_{ {\rm v} , \bm{k}_{\rm v} } ( \bm{r}_\parallel )
	\delta_{ \bm{q}_\parallel , \bm{k}_{\rm c} - \bm{k}_{\rm v} }
	,
\end{equation}
where $\Omega_{\rm cell}$ is the unit cell area and $m_0$ is the free electron mass.
The Kronecker delta ensures the in-plane momentum conservation during the inter-band optical transition, where the difference between the electron wavevectors in the initial and final states must be equal to the electromagnetic-field wavevector $\bm{q}_\parallel$.


Assuming that the magnitude of the electromagnetic-field wavevector $\bm{q}_\parallel$ is much smaller than the characteristic wavevector scale of the Brillouin zone, the periodic part of the conduction band wave function, $\psi^\tau_{ {\rm c} , \bm{k}_{\rm c} } ( \bm{r}_\parallel )$, can be expanded into a Taylor series in $\bm{q}_\parallel$.
Using the momentum conservation relation $\bm{k}_{\rm c} = \bm{k}_{\rm v} + \bm{q}_\parallel$, this expansion up to the first order is given by
\begin{equation}
	\psi^\tau_{ {\rm c} , \bm{k}_{\rm c} } ( \bm{r}_\parallel )
	\simeq
	\psi^\tau_{ {\rm c} , \bm{k}_{\rm v} } ( \bm{r}_\parallel )
	+
	\bm{q}_\parallel \cdot \left[ \nabla_{\bm{k}_\parallel} \psi^\tau_{ {\rm c} , \bm{k}_\parallel } ( \bm{r}_\parallel ) \right]_{ \bm{k}_\parallel = \bm{k}_{\rm v} }
	.
\end{equation}
By substituting this expansion back into the integral, we can separate the inter-band transition momentum matrix element into a zeroth-order term, $\bm{P}_{\rm cv}^{\tau (0)}$, and a first-order term, $\bm{P}_{\rm cv}^{\tau (1)}$, with respect to the electromagnetic-field wavevector $\bm{q}_\parallel$:
\begin{equation}
	\bm{P}_{\rm cv}^\tau
	\simeq
	\bm{P}_{\rm cv}^{\tau (0)} ( \bm{k}_\parallel )
	+
	\bm{P}_{\rm cv}^{\tau (1)} ( \bm{k}_\parallel , \bm{q}_\parallel )
	,
\end{equation}
where we have denoted $\bm{k}_\parallel \equiv \bm{k}_{\rm v}$ for simplicity. 
The zeroth-order term $\bm{P}_{\rm cv}^{\tau (0)} ( \bm{k}_\parallel )$ represents the conventional electric-dipole contribution in the long-wavelength limit, whereas the first-order term $\bm{P}_{\rm cv}^{\tau (1)} ( \bm{k}_\parallel , \bm{q}_\parallel )$ describes the leading finite-wavevector correction arising from spatial gradients in the electromagnetic field.


The $x$ and $y$ components of the zeroth-order term $\bm{P}_{\rm cv}^{\tau (0)} ( \bm{k}_\parallel )$ are explicitly written as
\begin{subequations}\label{eq:transition_momenta0}
	\begin{align}
		P_{ {\rm cv} x }^{\tau (0)} ( \bm{k}_\parallel )
		&=
		\tau
		\frac{ m_0 a t }{ 2 \hbar }
		\left[
			\alpha_+ ( \bm{k}_\parallel )
				e^{ i \tau \phi_{ \bm{k}_\parallel } }
			-
			\alpha_- ( \bm{k}_\parallel )
				e^{ - i \tau \phi_{ \bm{k}_\parallel } }
		\right]
	,
\\
		P_{ {\rm cv} y }^{\tau (0)} ( \bm{k}_\parallel )
		&=
		-i
		\frac{ m_0 a t }{ 2 \hbar }
		\left[
			\alpha_+ ( \bm{k}_\parallel )
				e^{ i \tau \phi_{ \bm{k}_\parallel } }
			+
			\alpha_- ( \bm{k}_\parallel )
				e^{ - i \tau \phi_{ \bm{k}_\parallel } }
		\right]
	.
	\end{align}
\end{subequations}
Near the valley band edges where $\xi \to 0$, Eq.~\eqref{eq:transition_momenta0} reduces to
\begin{subequations}
	\begin{align}
		P_{ {\rm cv} x }^{\tau (0)} ( \bm{k}_\parallel )
		&\simeq
		\tau
		\frac{ m_0 a t }{ \hbar }
			e^{ i \tau \phi_{ \bm{k}_\parallel } }
	,
\\
		P_{ {\rm cv} y }^{\tau (0)} ( \bm{k}_\parallel )
		&\simeq
		-i
		\frac{ m_0 a t }{ \hbar }
			e^{ i \tau \phi_{ \bm{k}_\parallel } }
	.
	\end{align}
\end{subequations}
Furthermore, the $x$ and $y$ components of the first-order term $\bm{P}_{\rm cv}^{\tau (1)} ( \bm{k}_\parallel , \bm{q}_\parallel )$ near the valley band edges are derived as
\begin{subequations}\label{eq:transition_momenta}
	\begin{align}
		P_{ {\rm cv} x}^{\tau (1)} (\bm{k}_\parallel , \bm{q}_\parallel )
		&\simeq
		\tau
		\frac{ m_0 a^3 t^3 }{ \hbar E_{\rm g}^2 }
		\left[
			-
			\left( \bm{q}_\parallel \cdot \bm{\xi} \right)
				e^{ i \tau \phi_{ \bm{k}_\parallel } }
			-
			\left( q_x - i \tau q_y \right)
				\xi
		\right]
		,
\\
		P_{ {\rm cv} y}^{\tau (1)} (\bm{k}_\parallel , \bm{q}_\parallel )
		&\simeq
		-i
		\frac{ m_0 a^3 t^3 }{ \hbar E_{\rm g}^2 }
		\left[
			-
			\left( \bm{q}_\parallel \cdot \bm{\xi} \right)
				e^{ i \tau \phi_{ \bm{k}_\parallel } }
			+
			\left( q_x - i \tau q_y \right)
				\xi
		\right]
		.
	\end{align}
\end{subequations}
As indicated by the conventional Elliott formula, the zeroth-order term $\bm{P}_{\rm cv}^{\tau (0)}$ gives the electric-dipole contribution to the excitonic transition matrix element.~\cite{Elliott1957}
When projected onto the exciton states and integrated over the Brillouin zone, this term couples to the amplitude of the exciton envelope function at the origin, yielding the familiar selection rule favoring s-type excitons.
For strongly confined electromagnetic fields, however, the spatial variation of the field is no longer negligible.
The corresponding in-plane wavevector $\bm{q}_\parallel$ can become large, making the first-order contribution $\bm{P}_{\rm cv}^{\tau (1)}$ important.
Because $\bm{P}_{\rm cv}^{\tau (1)}$ explicitly depends on the relative wavevector $\bm{\xi}$, its contribution to the excitonic transition matrix element is related, upon Fourier transformation, to spatial derivatives of the exciton envelope function.
Consequently, finite-wavevector coupling enables transitions to otherwise dark p-type excitons.
A quantitative evaluation of these transitions requires an explicit description of the bound electron--hole pairs, which we discuss in the following subsection.


\subsection{Exciton states and the Mott--Wannier equation}
\label{subsec:Excitons}

To describe the excitonic optical response, we treat excitons as bound electron-hole pairs.
The wavefunction corresponding to the exciton state $|\Psi_{\lambda, \bm{q}_\parallel}^{ \tau } \rangle$ with center-of-mass wavevector $\bm{q}_\parallel$ and Rydberg state index $\lambda$ (e.g., 1s, 2s, 2p) is expressed as:
\begin{align}
	\Psi^{ \mathrm{ex}, \tau }_{\lambda} ( \bm{r}_{\rm h} , \bm{r}_{\rm e} , \bm{q}_\parallel )
	&=
	\sum_{\bm{k}_\parallel}
		A^\tau_{\lambda} ( \bm{k}_\parallel , \bm{q}_\parallel )
		\Psi^{ \tau * }_{ {\rm v} , \bm{k}_\parallel } ( \bm{r}_{ {\rm h} \parallel } , z_{\rm h} )
		\Psi^{ \tau}_{ {\rm c} , \bm{k}_\parallel + \bm{q}_\parallel } ( \bm{r}_{ {\rm e} \parallel } , z_{\rm e} )
\nonumber
\\
	&=
	\frac{ e^{ i \bm{q}_\parallel \cdot \bm{R}_\parallel } }{ \sqrt{ \Omega_\parallel } }
		\sum_{ \bm{k}_\parallel }
			\frac{ e^{ i \bm{ \kappa }_\parallel \cdot \bm{\rho}_\parallel } }{ \sqrt{ \Omega_\parallel } }
			A^\tau_\lambda ( \bm{k}_\parallel , \bm{q}_\parallel )
		\psi^{ \tau * }_{ {\rm v} , \bm{k}_\parallel } (\bm{r}_{{\rm h} \parallel} )
		\psi^\tau_{ {\rm c} , \bm{k}_\parallel + \bm{q}_\parallel } ( \bm{r}_{{\rm e} \parallel} )
	\phi^*_{ {\rm v} \perp } ( z_{\rm h} )
	\phi_{ {\rm c} \perp } ( z_{\rm e} )
	,
\label{eq:exciton_state}
\end{align}
where $A^\tau_{\lambda} ( \bm{k}_\parallel , \bm{q}_\parallel )$ is the excitonic expansion coefficient in reciprocal space.
The center-of-mass and relative coordinates in real space are defined as
\begin{subequations}
	\begin{align}
		\bm{R}_\parallel
		&=
		\frac{ m_{\rm e} \bm{r}_{ {\rm e} \parallel } + m_{\rm h} \bm{r}_{ {\rm h} \parallel } }{ m_{\rm e} + m_{\rm h} }
		,
		\\
		\bm{\rho}_\parallel
		&=
		\bm{r}_{ {\rm e} \parallel } - \bm{r}_{ {\rm h} \parallel }
		,
	\end{align}
\end{subequations}
where $m_{\rm e}$ ($m_{\rm h}$) denotes the effective mass of the electron (hole).
The mass-weighted relative wavevector $\bm{\kappa}_\parallel$ is given by
\begin{equation}
	\bm{ \kappa }_\parallel
	=
	\bm{k}_\parallel + \frac{ m_{\rm h} }{ m_{\rm e} + m_{\rm h} } \bm{q}_\parallel
	.
\end{equation}
The real-space envelope function $F^\tau_{\lambda} ( \bm{\rho}_\parallel; \bm{q}_\parallel)$, which describes the relative electron--hole motion, is obtained via the 2D Fourier transform:
\begin{equation}
	F^\tau_{\lambda} ( \bm{\rho}_\parallel; \bm{q}_\parallel ) 
	=
	\frac{1}{ \sqrt{ \Omega_\parallel } }
	\sum_{ \bm{k}_\parallel }
	e^{ i \bm{\kappa}_\parallel \cdot \bm{\rho}_\parallel}
	A^\tau_{\lambda} ( \bm{k}_\parallel , \bm{q}_\parallel )
	e^{ - i \tau \phi_{\bm{k}_\parallel} }
	,
\label{eq:Fourier}
\end{equation}
where $e^{ - i \tau \phi_{\bm{k}_\parallel} }$ is the Berry phase factor originating from the single-particle Bloch wavefunctions.~\cite{Berkelbach2015}

Within the effective-mass approximation, the internal envelope function is independent of the center-of-mass wavevector and satisfies the generalized 2D Mott--Wannier equation:
\begin{equation}
	\left[
		-
		\frac{\hbar^2 \nabla^2_{ \bm{\rho}_\parallel } }{ 2 \mu }
		+
		W ( \bm{\rho}_\parallel )
	\right]
	F^\tau_{\lambda} ( \bm{\rho}_\parallel )
	=
	E_\lambda
	F^\tau_{\lambda} ( \bm{\rho}_\parallel )
	,
\label{eq:MottWannier}
\end{equation}
where $\mu = ( m_{\rm e}^{-1} + m_{\rm h}^{-1} )^{-1}$ is the exciton reduced mass and $W(\bm{\rho}_\parallel)$ is the screened Coulomb potential. 
As shown in previous work~\cite{Latini2015}, the screened Coulomb potential in TMDs is isotropic in the monolayer plane; thus, it depends only on the scalar distance, i.e., $W(\rho_\parallel)$ with $\rho_\parallel = |\bm{\rho}_\parallel|$.
Furthermore, in reality, a TMD monolayer has a finite thickness and a continuous out-of-plane electronic charge distribution.
To address this, we explicitly account for the out-of-plane electronic distribution $\rho_{ {\rm c/v} }(z) = |\phi_{ {\rm c/v} \perp} (z)|^2$, which plays a crucial role in softening the unphysical short-range Coulomb divergence as $\rho_\parallel \to 0$.
We construct this material-specific screened potential using the Quantum Electrostatic Heterostructure (QEH) model~\cite{Andersen2015, Latini2015}.
Specifically, the macroscopic dielectric function of the WS$_2$ monolayer is calculated within the QEH model using first-principles dielectric building-block data, from which the 3D screened interaction $W(z, z', q_\parallel)$ is obtained.
Integrating over the out-of-plane wavefunctions gives the effective 2D momentum-space potential $\bar{W} ( q_\parallel )$:
\begin{equation}
	\bar{W} ( q_\parallel )
	=
	\int_{-\infty}^{\infty} dz
	\int_{-\infty}^{\infty} dz'
	\left| \phi_{ {\rm c} \perp } (z) \right|^2
	W( z , z' , q_\parallel )
	\left| \phi_{ {\rm v} \perp } ( z' ) \right|^2
	.
\label{eq:W_q}
\end{equation}
The real-space potential $W ( \rho_\parallel )$ is obtained via the inverse 2D Fourier (Hankel) transform:
\begin{equation}
	W ( \rho_\parallel ) 
	=
	\frac{1}{2\pi}
		\int_0^\infty dq_\parallel \,
		q_\parallel
		J_0( q_\parallel \rho_\parallel )
		\bar{W}( q_\parallel )
	,
\label{eq:Bessel}
\end{equation}
where $J_0(x)$ is the zeroth-order Bessel function of the first kind.
For simplicity, the exciton states are calculated using the screened electron--hole interaction of an isolated WS$_2$ monolayer within the QEH model, while the hBN encapsulation is incorporated only in the subsequent electromagnetic calculations based on the TMM.

To highlight the physical effect of the spatial charge distribution in the out-of-plane direction, we assume a uniform out-of-plane electronic distribution over a monolayer thickness $d_{\rm TMD}$, i.e., $|\phi_{ {\rm c/v} \perp} (z)|^2 = 1 / d_{\rm TMD}$ for $|z| \le d_{\rm TMD} / 2$.
In this case, the integration in Eq.~\eqref{eq:W_q} combined with Eq.~\eqref{eq:Bessel} analytically reduces to the well-known form~\cite{Latini2015}:
\begin{equation}
	W ( \rho_\parallel )
	=
	- \frac{ e^2 }{ 2 \pi \epsilon_0 }
	\int_0^\infty d q_\parallel \,
		\epsilon^{-1}_{\rm M} (q_\parallel )
		\frac{ J_0 ( q_\parallel \rho_\parallel ) }{ q_\parallel d_{\rm TMD} }
		\left[
			1
			-
			\frac{ 1 - e^{ - q_\parallel d_{\rm TMD} } }{ q_\parallel d_{\rm TMD} }
		\right]
	,
\label{eq:W_Q2D}
\end{equation}
where $e$ is the elementary charge, $\epsilon_0$ is the vacuum permittivity, and $\epsilon_{\rm M}(q_\parallel)$ is the 2D macroscopic dielectric function of the TMD monolayer.
This explicitly demonstrates how the out-of-plane spread of charge suppresses large-wavevector (short-range) Coulomb interactions.
It is instructive to note that in the limit of $q_\parallel d_{\rm TMD} \ll 1$, we recover the pure 2D potential energy, whereas for $q_\parallel d_{\rm TMD} \gg 1$, the interaction approaches a three-dimensional-like form~\cite{Latini2015}.


\subsection{Wavevector-dependent nonlocal susceptibility tensor}
\label{subsec:Susceptibility}

Using the single-particle transition momentum matrix elements and the excitonic wavefunctions within linear response theory, we derive the nonlocal electric susceptibility tensor $\chi_{\mu\nu}(\bm{q}_\parallel,\omega)$ of the WS$_2$ monolayer:
\begin{equation}
	\chi_{\mu\nu}(\bm{q}_\parallel, \omega) 
	=
	\chi_{\rm BG}
	\delta_{ \mu \nu }
	+
		\frac{ e^2 \hbar^2 }{ \epsilon_0 d_{\rm TMD} m_0^2 E_{\rm g}^2 }
		f_z ( \bm{q}_\parallel , \omega )
		\sum_{ \tau , \lambda }
		\frac	{ \left[ \mathcal{T}^{ \tau * }_{\lambda} ( \bm{q}_\parallel ) \right]_\mu \left[ \mathcal{T}^{ \tau }_{\lambda} ( \bm{q}_\parallel ) \right]_\nu }
				{ E^{\rm ex}_\lambda ( \bm{q}_\parallel ) - \hbar \omega - i \hbar \eta }
	,
\label{eq:Susceptibility_general}
\end{equation}
where $\delta_{\mu\nu}$ is the Kronecker delta, $\chi_{\rm BG}$ represents the nonresonant background contribution, and $\eta$ is the phenomenological damping parameter.~\cite{Cho1986, Ishihara1990, Inoue2025, Ashida2026} 
We assume equal contributions from the two time-reversal-related valleys, $\tau = \pm 1$, and neglect intervalley coupling.
The total susceptibility is therefore obtained by summing the independent contributions from both valleys.
The exciton resonance energy with center-of-mass wavevector $\bm{q}_\parallel$ is given by
\begin{equation}
	E^{\rm ex}_\lambda ( \bm{q}_\parallel )
	=
	E_{\rm g}
	+
	E_\lambda
	+
	\frac{ \hbar^2 q_\parallel^2 }{ 2 M }
	,
\end{equation}
where $E_\lambda < 0$ denotes the energy of the relative electron--hole motion obtained from the Mott--Wannier equation, and $M = m_{\rm e} + m_{\rm h}$ is the total exciton mass.
The factor $f_z ( \bm{q}_\parallel , \omega )$ is the out-of-plane form factor accounting for the finite TMD thickness and is given by
\begin{equation}
	f_z ( \bm{q}_\parallel , \omega )
	=
	\int dz
	\phi_{ {\rm c} \perp }^* (z)
	e^{i q_z z }
	\phi_{ {\rm v} \perp } (z)
	,
\end{equation}
where
$q_z = \sqrt{ \omega^2 / c^2 - q_\parallel^2 }$, with the branch chosen such that ${\rm Re} \, q_z \ge 0$ and ${\rm Im} \, q_z \ge 0$, and $c$ is the speed of light.
The two-dimensional vector $\bm{\mathcal{T}}^\tau_\lambda ( \bm{q}_\parallel )$ represents the excitonic transition matrix elements and is given by
\begin{align}
	\bm{ \mathcal{T} }^\tau_\lambda ( \bm{q}_\parallel )
	=
	\frac{ m_0 a t }{ \hbar }
	\Bigg\{
	&
		( \tau \bm{e}_x - i \bm{e}_y )
		F_{ \lambda }^{ \tau * } ( \bm{ \rho }_\parallel = 0 )
\nonumber
\\
	& +
		\bm{e}_x
		\frac{ a^2 t^2 }{ E_{\rm g}^2 }
		\left[
			\left( - 2 i \tau q_x - q_y \right) \partial_x F^{ \tau * }_\lambda ( \bm{0} )
			-
			q_x \partial_y F^{ \tau * }_\lambda ( \bm{0} )
			+
			\left( \tau q_x^2 - i q_x q_y \right) F^{ \tau * }_\lambda ( \bm{0} )
		\right]
\nonumber
\\
	& +
		\bm{e}_y
		\frac{ a^2 t^2 }{ E_{\rm g}^2 }
		\left[
			\left( - 2 q_y - i \tau q_x \right) \partial_y F^{ \tau * }_\lambda ( \bm{0} )
			-
			i \tau q_y \partial_x F^{ \tau * }_\lambda ( \bm{0} )
			+
			\left( - i q_y^2 + \tau q_x q_y \right) F^{ \tau * }_\lambda ( \bm{0} )
		\right]
		\Bigg\}
\end{align}
where $\bm{e}_x$ and $\bm{e}_y$ are the in-plane Cartesian unit vectors.
Here, $\partial_{x,y} F_\lambda^{\tau*} ( \bm{0} ) \equiv \left. \partial F_\lambda^{\tau*} ( \bm{\rho}_\parallel ) / \partial \rho_{x,y} \right|_{\bm{\rho}_\parallel = \bm{0}}$ represent the corresponding components of the spatial gradient of the envelope function evaluated at the origin with respect to the relative electron-hole coordinates $\bm{\rho}_\parallel = (\rho_x, \rho_y)$.

We classify the susceptibility tensor into contributions from s-type and p-type exciton states based on their angular-momentum character.
For the conventional s-type exciton states ($\lambda = {\rm 1s}, {\rm 2s}, \dots$), the envelope function has a finite value at the origin, $F^\tau_\lambda ( \bm{0} ) \neq 0$, whereas its spatial gradient vanishes at the origin.
In the long-wavelength limit ($q_\parallel \to 0$), the s-type susceptibility takes the form:
\begin{equation}
	\chi_{ \mu \nu }^{ ( {\rm s} ) } ( \bm{q}_\parallel \to \bm{0}, \omega )
	\simeq
	C_{\rm opt}
	f_z ( \bm{0} , \omega )
	\sum_\tau
	\sum_{\lambda \in \rm s}
	\frac	{ | F^\tau_\lambda ( \bm{0} ) |^2 }
			{ E^{\rm ex}_\lambda ( \bm{0} ) - \hbar \omega - i \hbar \eta } 
	\begin{pmatrix}
		1 & -i\tau \\
		i\tau & 1
	\end{pmatrix}_{\mu\nu}
	,
\label{eq:s_susceptibility}
\end{equation}
where $C_{\rm opt} = e^2 a^2 t^2 / ( \epsilon_0 d_{\rm TMD} E_{\rm g}^2 )$ is the optical coupling constant defined within the two-band model.
For each valley, Eq.~\eqref{eq:s_susceptibility} reproduces the standard dipole-allowed optical selection rules where s-type excitons couple directly to circularly polarized light.
When the two valleys contribute equally, their off-diagonal contributions cancel in the total susceptibility.

In contrast, for the p-type exciton states, the envelope function vanishes at the origin, $F^\tau_\lambda ( \bm{0} ) = 0$, rendering these states optically dark under the local-response approximation ($q_\parallel = 0$).
However, under strong near-field confinement where $q_\parallel$ becomes large, finite-wavevector coupling enables transitions to p-type exciton states.
By expanding the transition matrix elements to first order in $q_\parallel$, we obtain the nonlocal susceptibility contribution for the p-type states:
\begin{equation}
	\chi_{\mu\nu}^{ ( {\rm p} ) } ( \bm{q}_\parallel, \omega ) 
	=
	C_{\rm opt}
		\left( \frac{ a t }{ E_{\rm g} } \right)^4
		f_z ( \bm{q}_\parallel , \omega )
		\sum_\tau
		\sum_{\lambda \in \rm p}
		\frac{ | \nabla_{\bm{\rho}_\parallel} F^\tau_\lambda ( \bm{0} ) |^2 }
				{ E^{\rm ex}_\lambda ( \bm{q}_\parallel ) - \hbar\omega - i \hbar \eta }
		\left[ \mathcal{M}^{\tau *}_\lambda ( \bm{q}_\parallel ) \right]_\mu
		\left[ \mathcal{M}^{\tau }_\lambda ( \bm{q}_\parallel ) \right]_\nu
	,
\label{eq:p_susceptibility}
\end{equation}
where the additional factor $(at/E_{\rm g})^4$ originates from the leading finite-wavevector correction to the $k\cdot p$ interband transition matrix element.
Within the isotropic Mott--Wannier model adopted here, the p-type exciton states are degenerate for a given principal quantum number; we therefore use the equivalent real basis states ${\rm p}_x$ and ${\rm p}_y$.
For the ${\rm p}_x$ state, the corresponding selection vector is
\begin{equation}
	\bm{\mathcal M}^{\tau}_{ \mathrm{p}_x } ( \bm{q}_\parallel )
	=
	-
		( 2 i \tau q_x + q_y )
		\bm{ e }_x
	-
		i \tau q_y \bm{e}_y ,
\end{equation}
while for the ${\rm p}_y$ state,
\begin{equation}
	\bm{\mathcal M}^{\tau}_{ \mathrm{p}_y } ( \bm{q}_\parallel )
	=
	-
		q_x \bm{e}_x
	-
		\left( 2 q_y + i \tau q_x \right)
			\bm{e}_y
	.
\end{equation}
These expressions explicitly demonstrate that the spatial gradient of the exciton envelope function at the origin, $\nabla_{ \bm{ \rho }_\parallel } F^\tau_\lambda ( \bm{0} )$, couples to the finite in-plane wavevector $\bm q_\parallel$.
Consequently, the dipole-forbidden p-type Rydberg series becomes accessible in the finite-wavevector regime, providing a new channel for energy transfer.


\subsection{Transfer matrix method}
\label{subsec:TMM}

To investigate the electromagnetic response of the air/hBN/WS$_2$/hBN heterostructure, we implement the transfer matrix method that self-consistently incorporates the spatial dispersion of the susceptibility tensor $\chi_{\mu\nu}(\bm{q}_\parallel, \omega)$.
The WS$_2$ monolayer located at $z = z_0$ is modeled as an infinitesimally thin sheet conductor characterized by the 2D nonlocal conductivity tensor
$\sigma_{\mu\nu}(\bm{q}_\parallel, \omega) = - i \omega \epsilon_0 d_{\rm TMD} \chi_{\mu\nu}(\bm{q}_\parallel, \omega)$.
The presence of the WS$_2$ monolayer is accounted for by imposing standard electromagnetic boundary conditions at $z = z_0$: the continuity of the tangential electric field, $\bm{e}_z \times \Delta \bm{E} = \bm{0}$, and the jump in the tangential magnetic field induced by the surface current, $\bm{e}_z \times \Delta \bm{H} = \bm{\sigma}(\bm{q}_\parallel, \omega) \bm{E}_\parallel$.
By combining the propagation matrices for the hBN thin film and substrate layers with these boundary conditions, we solve Maxwell's equations to determine the polariton dispersion and the spatial electric field profiles near a quantum emitter.

Furthermore, to analyze the energy transfer from a nearby emitter into TMD exciton states, we evaluate the wavevector-resolved power dissipation spectrum $P ( q_\parallel )$.
For a point dipole with transition dipole moment $\bm{\mu}_{\rm pd}$ placed at position $\bm{r}_{\rm pd}$, the wavevector-resolved power-dissipation density is expressed as
\begin{equation}
	P ( q_\parallel )
	=
	\frac{ \omega q_\parallel }{ 4 \pi }
		{\rm Im}
		\left[
			\bm{ \mu }^*_{\rm pd}
			\cdot
			\bm{E}_{\rm R} ( q_\parallel ; \bm{r}_{\rm pd} )
		\right]
	,
\label{eq:power_dissipation_dipole}
\end{equation}
where $\bm{E}_{\rm R} ( q_\parallel ; \bm{r}_{\rm pd} )$ is the wavevector component of the reflected electric field acting back on the dipole, calculated using the TMM.
Integration of $P ( q_\parallel )$ over $q_\parallel$ gives the power dissipated through the interaction of the emitter with the surrounding structure.
This wavevector-resolved representation allows us to identify the contributions of different TMD exciton states to the energy dissipation of the emitter.
It also distinguishes components inside and outside the free-space light cone, $q_\parallel < \omega / c$ and $q_\parallel > \omega / c$, respectively.


\subsection{Parameters}
\label{subsec:parameters}

To quantitatively evaluate the excitonic optical response, we first solve the generalized 2D Mott--Wannier equation [Eq.~\eqref{eq:MottWannier}] with the screened potential $W(\rho_\parallel)$ using a finite-difference scheme on a logarithmically spaced real-space grid.
For monolayer WS$_2$, we adopt material parameters obtained from first-principles calculations and available in the Computational 2D Materials Database (C2DB), together with the macroscopic dielectric function obtained from the QEH model.~\cite{Andersen2015, Latini2015, Haastrup2018}
The electron and hole effective masses of monolayer WS$_2$ obtained from the C2DB are $m_{\rm e} = 0.328~m_0$ and $m_{\rm h} = 0.336~m_0$, respectively.
Because these values are nearly identical, we assume $m_{\rm e} = m_{\rm h} = 0.332~m_0$, which is consistent with the symmetric conduction- and valence-band dispersions obtained from the two-band massive Dirac Hamiltonian in Eq.~\eqref{eq:Hamiltonian}.
This gives the reduced exciton mass $\mu = 0.166~m_0$ and the total exciton mass $M = 0.664~m_0$.
The effective monolayer thickness is set to $d_{\rm TMD} = 0.616~{\rm nm}$. 
The energy bandgap is set to $E_{\rm g} = 2.532~{\rm eV}$.
The 2D macroscopic dielectric function $\epsilon_{\rm M} ( q_\parallel )$ is calculated within the QEH model using first-principles dielectric building-block data for monolayer WS$_2$.
The calculated exciton binding energies $E_{ \mathrm{b}, \lambda} = - E_\lambda$ for the s-type and p-type Rydberg series are summarized in Table~\ref{tbl:exciton}.
The calculated binding energies are in agreement with reported GW-BSE and experimental values~\cite{Chernikov2014, Ye2014}.

\begin{table}[h!]
  \caption{Calculated exciton binding energies $E_{ \mathrm{b}, \lambda}$ (meV) for a WS$_2$ monolayer using the QEH potential with $d_{\rm TMD} = 0.616~{\rm nm}$ and $\mu = 0.166 \, m_0$.}
  \label{tbl:exciton}
  \begin{tabular}{lrclr}
    \hline
    state~~~&~~~$E_{ \mathrm{b}, \lambda}$	 (meV) &~~~~~~~&
    state~~~&~~~$E_{ \mathrm{b}, \lambda}$	 (meV) 	\\
    \hline
    1s		& 553	&	&2p		& 281	\\
    2s		& 232	&	&3p		& 157	\\
    3s		& 135	&	&4p		& 100	\\
    4s		& 88		&	&5p		& 69		\\
    \hline
  \end{tabular}
\end{table}

For the subsequent electromagnetic calculations based on the TMM, the nonlocal electric susceptibility tensor $\chi_{\mu\nu} ( \bm{q}_\parallel, \omega )$ is constructed using the obtained envelope functions $F_\lambda^\tau ( \rho_\parallel )$ and their gradients at the origin $\nabla_{\bm{\rho}_\parallel} F_\lambda^\tau ( \bm{0} )$.
The phenomenological damping parameter is set to $\hbar\eta = 1.0~{\rm meV}$, corresponding to high-quality hBN-encapsulated samples at cryogenic temperatures.~\cite{Epstein2020, Meshulam2026}
For monolayer WS$_2$, the background contribution to the susceptibility is set to $\chi_{\rm BG} = 10.0$.
The optical coupling constant $C_{\rm opt} = e^2 a^2 t^2 / ( \epsilon_0 d_{\rm TMD} E_{\rm g}^2 )$ is determined to match experimental observations.
Specifically, we set the peak amplitude $A = 2 C_{\rm opt} | F^\tau_{\rm 1s } ( \bm{0} ) |^2 / ( \hbar \eta )$ of the 1s excitonic susceptibility to 500, consistent with the experimentally extracted susceptibility.~\cite{Epstein2020}
The position of the WS$_2$ monolayer is set to $ z_0 = 0$.
For the surrounding dielectric environment, we adopt an isotropic relative permittivity $\varepsilon_{\rm hBN} = 3.87$ for the hBN encapsulation layers.
The representative thickness of the capping hBN layer is set to $d_{\rm hBN} = 2.0~{\rm nm}$ unless stated otherwise.
The main parameters used in the numerical calculations are listed in Table~\ref{tbl:Parameters}.
In the local-response approximation, the susceptibility $\chi_{ \mu \nu }$ is evaluated in the $q_\parallel \to 0 $ limit, while the electromagnetic wavevector dependence is retained in the TMM.

\begin{table}[h!]
  \caption{Parameters utilized in numerical calculations}
  \label{tbl:Parameters}
  \begin{tabular}{lrclr}
    \hline
    parameter	& value	&	~~~~~~~~~~ & parameter	& value\\
    \hline
    $E_{\rm g}$			& 2.532~eV			&& $\varepsilon_{\rm hBN}$	& 3.87		\\
    $\mu$					& 0.166 $m_0$		&& $d_{\rm TMD}$				& 0.616~nm \\
    $M$						& 0.664 $m_0$		&& $d_{\rm hBN}$					& 2.0~nm	\\
    	$\chi_{\rm BG}$	& 10.0					&& $z_0$								& 0~nm		\\
    $\hbar \eta$			& 1.0~meV			&& $A$									& 500		\\
    \hline
  \end{tabular}
\end{table}


\section{Results and discussion}\label{sec:results}


\begin{figure}[h!]
\centering
\includegraphics[bb=0 0 240 170, clip]{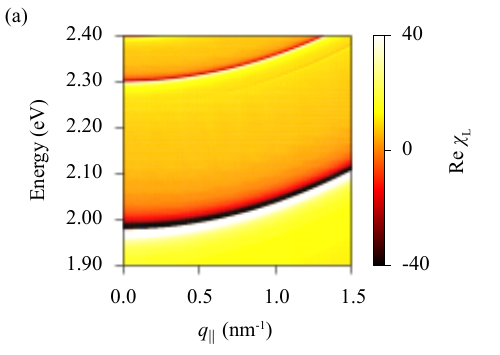}
\includegraphics[bb=0 0 240 170, clip]{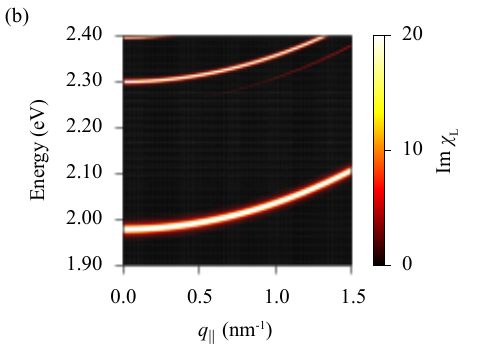}
\caption{
\label{fig:chi}
(a) Real and (b) imaginary parts of the longitudinal component of the nonlocal electric susceptibility, $\chi_{\rm L} ( q_\parallel , \omega )$, of monolayer WS$_2$ as functions of photon energy and the magnitude $q_\parallel$ of the in-plane wavevector.
}
\end{figure}

We first examine the optical response of monolayer WS$_2$.
For the TM-polarized modes considered here, the longitudinal electric susceptibility $\chi_{\rm L}(q_\parallel,\omega)$ is used in the TMM calculations.
For an in-plane wavevector along the $x$ direction, $ \bm{q}_\parallel = q_\parallel \bm{e}_x$, the longitudinal susceptibility is given by $\chi_{\rm L} ( q_\parallel , \omega ) = \chi_{xx} ( q_\parallel , \omega )$.
Figure~\ref{fig:chi} displays the calculated real and imaginary parts of $ \chi_{\rm L} ( q_\parallel , \omega )$ as functions of photon energy $\hbar \omega$ and the magnitude $q_\parallel$ of the in-plane wavevector.
The susceptibility exhibits distinct resonances corresponding to the 1s, 2s, and 2p exciton states.
The s-type states (1s and 2s) exhibit optical responses over a broad range of in-plane wavevectors.
In contrast, the resonance associated with the dipole-forbidden 2p state gradually gains spectral weight as $q_\parallel$ increases, providing a signature of finite-wavevector coupling to the 2p exciton state.


\begin{figure}[h!]
\centering
\includegraphics[bb=0 0 240 170, clip]{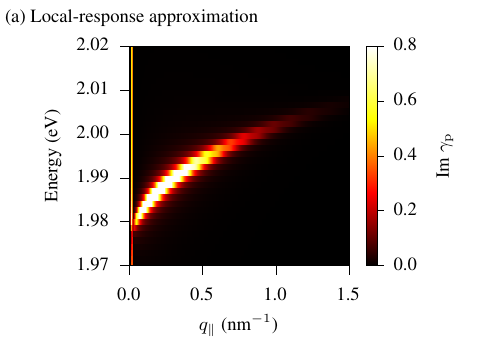}
\includegraphics[bb=0 0 240 170, clip]{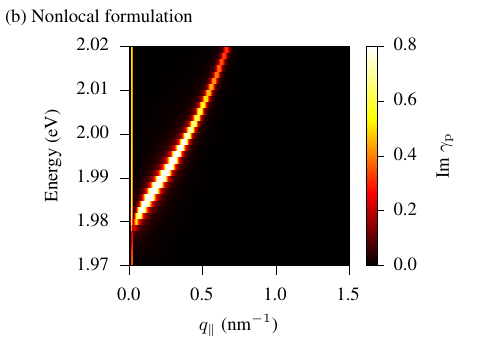}
\caption{
\label{fig:reflection}
Exciton-polariton dispersion maps represented by the imaginary part of the p-polarized reflection coefficient ${\rm Im} \, \gamma_{\rm p} (q_\parallel, \omega)$, obtained within (a) the conventional local-response approximation and (b) the nonlocal formulation.
}
\end{figure}

Next, we investigate how spatial nonlocality in the excitonic optical response modifies the dispersion relation of exciton polaritons.
Figure~\ref{fig:reflection} compares the imaginary part of the p-polarized reflection coefficient ${\rm Im} \, \gamma_{\rm p} (q_\parallel, \omega)$ calculated within the local-response approximation [Fig.~\ref{fig:reflection}(a)] with that obtained using the nonlocal formulation [Fig.~\ref{fig:reflection}(b)].
Under the conventional local-response approximation, the polariton modes extend toward large wavevectors.
In contrast, when nonlocal corrections are incorporated, the polariton dispersion shifts toward lower wavevectors.
This shift primarily originates from the wavevector-dependent kinetic-energy contribution to the exciton energy, which blueshifts the excitonic resonance with increasing $q_\parallel$ and consequently shifts the polariton resonance toward lower $q_\parallel$ at a given photon energy.
The full nonlocal response additionally incorporates the wavevector dependence of the excitonic transition amplitudes, which modifies the spectral weight of the polaritonic modes.


\begin{figure}[h!]
\centering
\includegraphics[bb=0 0 240 170, clip]{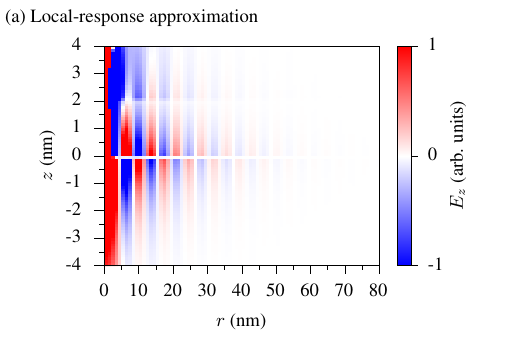}
\includegraphics[bb=0 0 240 170, clip]{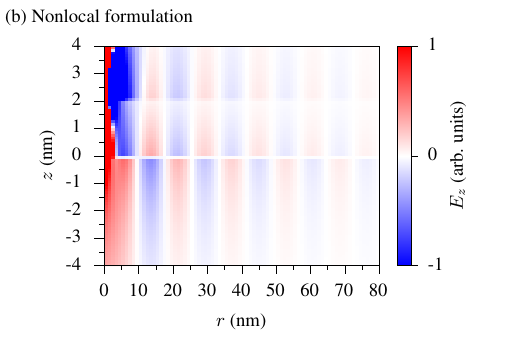}
\caption{
\label{fig:distribution}
Spatial distribution of the $z$ component of the electric field, $E_z ( r , z )$, near the TMD interface generated by a vertical point dipole driven at $2.00~{\rm eV}$, calculated using (a) the local-response approximation and (b) the nonlocal formulation.
The dipole is placed $0.5~{\rm nm}$ above the top hBN layer, the position of the WS$_2$ monolayer is set to $z_0 = 0$, and the thickness of the top hBN layer is set to $d_{\rm hBN} = 2.0~{\rm nm}$.
}
\end{figure}

To clarify how these dispersion modifications impact real-space optical phenomena, we analyze the spatial distribution of the $z$ component of the electric field generated by a vertical point dipole placed $0.5~{\rm nm}$ above the top hBN layer and driven at an energy of $\hbar\omega = 2.00~{\rm eV}$.
Under the local-response approximation [Fig.~\ref{fig:distribution}(a)], the electric field exhibits a steep spatial variation on a length scale of $\sim 10~{\rm nm}$ near the source.
In contrast, when the nonlocal response is included [Fig.~\ref{fig:distribution}(b)], the spatial profile becomes smoother.
The shift of the polariton dispersion toward lower $q_\parallel$ reduces the contribution of large-wavevector components.
This results in a field distribution that extends over larger lateral distances than in the local-response approximation.


\begin{figure}[h!]
\centering
\includegraphics[bb=0 0 240 170, clip]{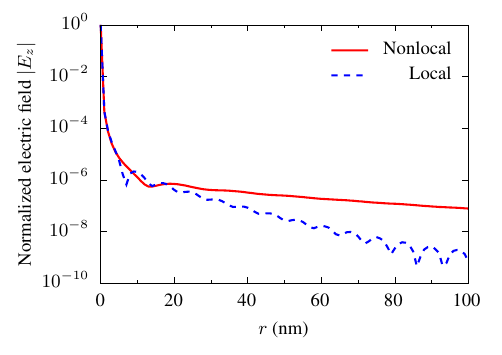}
\caption{
\label{fig:attenuation}
Normalized electric-field amplitude $\left| E_z ( r , z ) \right|$ as a function of the lateral distance $r$ from the dipole source.
Each curve is normalized by its value at $r=0$ and evaluated at a position $0.1~{\rm nm}$ below the dipole source.
The blue dashed and red solid lines correspond to the local-response and nonlocal calculations, respectively.
}
\end{figure}

To quantitatively illustrate the influence of spatial nonlocality on the lateral decay of the polaritonic field, we compare the electric-field amplitude as a function of the lateral distance $r$ from the dipole source in Fig.~\ref{fig:attenuation}.
Compared with the nonlocal calculation (red solid line), the local-response approximation (blue dashed line) predicts stronger attenuation with increasing lateral distance, indicating stronger field confinement near the source.
The nonlocal calculation shows that the polaritonic field persists over larger lateral distances.
This behavior originates from the shift of the polariton dispersion toward lower $q_\parallel$, which reduces the degree of confinement while extending the lateral range of the polaritonic field.
These results demonstrate that IPEPs in TMD heterostructures can reconcile nanoscale field confinement with long-range field propagation.


\begin{figure}[h!]
\centering
\includegraphics[bb=0 0 240 170, clip]{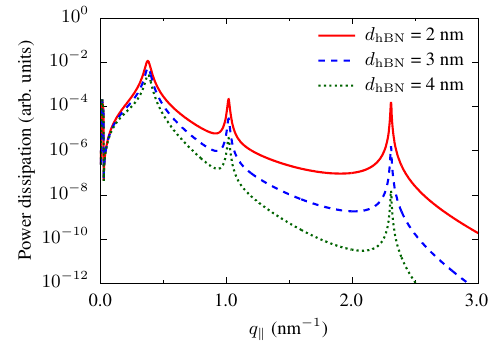}
\caption{
\label{fig:dissipation}
Power dissipation spectrum $P ( q_\parallel )$ at $2.31~{\rm eV}$ showing the distinct contributions from 1s, 2s, and 2p exciton polaritons.
The red solid, blue dashed, and green dotted lines represent the results for capping-layer thickness $d_{\rm hBN}$ = 2, 3, and 4~nm, respectively.
}
\end{figure}

Finally, we explore the contributions of multiple Rydberg exciton states to the near-field power dissipation of a nearby emitter.
Figure~\ref{fig:dissipation} presents the wavevector-resolved power dissipation spectrum $P ( q_\parallel )$ at an excitation energy of $\hbar\omega = 2.31~{\rm eV}$.
The spectrum exhibits multiple peaks originating from the 1s, 2s, and 2p exciton--polariton modes.
The 1s exciton-polariton mode dominates at higher wavevectors, the 2p mode contributes at intermediate wavevectors, and the 2s mode contributes primarily at lower wavevectors.
The coexistence of these modes introduces multiple characteristic spatial scales into the field distribution.
Moreover, changing the thickness $d_{\rm hBN}$ of the capping hBN layer modifies the coupling of the emitter to exciton-polariton modes with different in-plane wavevectors, thereby tuning their relative contributions to power dissipation.
These findings suggest potential applications in multimode nanoscale waveguides with mode-selective control.


\section{Conclusion}
\label{sec:conclusion}

In this work, we developed a microscopic theoretical framework for describing the nonlocal optical response of excitons and the resulting in-plane exciton polaritons in hBN-encapsulated TMD monolayers.
By combining the two-band massive Dirac fermion model with a generalized Mott--Wannier description incorporating material-specific screened electron--hole interactions, we derived a wavevector-dependent electric susceptibility that explicitly accounts for both the center-of-mass motion and the internal relative motion of excitons.
The wavevector-dependent exciton energy shifts the polariton dispersion toward lower $q_\parallel$, thereby reducing the contribution of high-wavevector components.
In real space, this modification smooths the near-field distribution and allows the polaritonic field to extend over larger lateral distances while retaining nanoscale confinement.
Furthermore, at higher excitation energies, multiple exciton-polariton modes associated with the 1s, 2s, and 2p exciton states contribute at distinct wavevectors.
Their relative contributions can be tuned by varying the thickness of the capping hBN layer.
These results demonstrate that nanoscale field confinement, long-range field propagation, and multimode polaritonic responses can coexist in TMD heterostructures, while the relative contributions of the modes can be controlled through the dielectric environment.
The present framework provides a theoretical basis for further studies of exciton-polariton properties in van der Waals heterostructures.
It may also offer useful guidelines for the design of nanoscale waveguides, spatially modulated energy-transfer schemes, and exciton-based nanophotonic devices.

\begin{acknowledgments}
This work was supported by the Institute for Basic Science (IBS) under project code IBS-R037-D1-2026-a00.
The atomic structure was visualized with the use of VESTA.~\cite{Momma2011}
\end{acknowledgments}

\section*{Author declarations}
\subsection*{Conflict of interest}
The author has no conflicts to disclose.



\section*{Data availability statement}
The data that support the findings of this study are available from the corresponding author upon reasonable request.



\begin{thebibliography}{100}%
\makeatletter
\providecommand \@ifxundefined [1]{%
 \@ifx{#1\undefined}
}%
\providecommand \@ifnum [1]{%
 \ifnum #1\expandafter \@firstoftwo
 \else \expandafter \@secondoftwo
 \fi
}%
\providecommand \@ifx [1]{%
 \ifx #1\expandafter \@firstoftwo
 \else \expandafter \@secondoftwo
 \fi
}%
\providecommand \natexlab [1]{#1}%
\providecommand \enquote  [1]{``#1''}%
\providecommand \bibnamefont  [1]{#1}%
\providecommand \bibfnamefont [1]{#1}%
\providecommand \citenamefont [1]{#1}%
\providecommand \href@noop [0]{\@secondoftwo}%
\providecommand \href [0]{\begingroup \@sanitize@url \@href}%
\providecommand \@href[1]{\@@startlink{#1}\@@href}%
\providecommand \@@href[1]{\endgroup#1\@@endlink}%
\providecommand \@sanitize@url [0]{\catcode `\\12\catcode `\$12\catcode
  `\&12\catcode `\#12\catcode `\^12\catcode `\_12\catcode `\%12\relax}%
\providecommand \@@startlink[1]{}%
\providecommand \@@endlink[0]{}%
\providecommand \url  [0]{\begingroup\@sanitize@url \@url }%
\providecommand \@url [1]{\endgroup\@href {#1}{\urlprefix }}%
\providecommand \urlprefix  [0]{URL }%
\providecommand \Eprint [0]{\href }%
\providecommand \doibase [0]{https://doi.org/}%
\providecommand \selectlanguage [0]{\@gobble}%
\providecommand \bibinfo  [0]{\@secondoftwo}%
\providecommand \bibfield  [0]{\@secondoftwo}%
\providecommand \translation [1]{[#1]}%
\providecommand \BibitemOpen [0]{}%
\providecommand \bibitemStop [0]{}%
\providecommand \bibitemNoStop [0]{.\EOS\space}%
\providecommand \EOS [0]{\spacefactor3000\relax}%
\providecommand \BibitemShut  [1]{\csname bibitem#1\endcsname}%
\let\auto@bib@innerbib\@empty
\bibitem [{\citenamefont {Novotny}\ and\ \citenamefont
  {Hecht}(2012)}]{Novotny2012}%
  \BibitemOpen
  \bibfield  {author} {\bibinfo {author} {\bibfnamefont {L.}~\bibnamefont
  {Novotny}}\ and\ \bibinfo {author} {\bibfnamefont {B.}~\bibnamefont
  {Hecht}},\ }\href {https://doi.org/10.1017/CBO9780511794193} {\emph {\bibinfo
  {title} {Principles of {{Nano-Optics}}}}},\ \bibinfo {edition} {2nd}\ ed.\
  (\bibinfo  {publisher} {Cambridge University Press},\ \bibinfo {address}
  {Cambridge},\ \bibinfo {year} {2012})\BibitemShut {NoStop}%
\bibitem [{\citenamefont {Nitzan}(2024)}]{Nitzan2024}%
  \BibitemOpen
  \bibfield  {author} {\bibinfo {author} {\bibfnamefont {A.}~\bibnamefont
  {Nitzan}},\ }\href@noop {} {\emph {\bibinfo {title} {Chemical {{Dynamics}} in
  {{Condensed Phases}}: {{Relaxation}}, {{Transfer}}, and {{Reactions}} in
  {{Condensed Molecular Systems}}}}},\ \bibinfo {edition} {2nd}\ ed.\ (\bibinfo
   {publisher} {Oxford University Press},\ \bibinfo {address} {Oxford},\
  \bibinfo {year} {2024})\BibitemShut {NoStop}%
\bibitem [{\citenamefont {Craig}\ and\ \citenamefont
  {Thirunamachandran}(1998)}]{Craig1998}%
  \BibitemOpen
  \bibfield  {author} {\bibinfo {author} {\bibfnamefont {D.~P.}\ \bibnamefont
  {Craig}}\ and\ \bibinfo {author} {\bibfnamefont {T.}~\bibnamefont
  {Thirunamachandran}},\ }\href@noop {} {\emph {\bibinfo {title} {Molecular
  {{Quantum Electrodynamics}}}}}\ (\bibinfo  {publisher} {Dover Publications},\
  \bibinfo {address} {New York},\ \bibinfo {year} {1998})\BibitemShut {NoStop}%
\bibitem [{\citenamefont {Purcell}(1946)}]{Purcell1946}%
  \BibitemOpen
  \bibfield  {author} {\bibinfo {author} {\bibfnamefont {E.~M.}\ \bibnamefont
  {Purcell}},\ }\bibfield  {title} {\enquote {\bibinfo {title} {Spontaneous
  emission probabilities at radio frequencies},}\ }\href
  {https://doi.org/10.1103/PhysRev.69.674} {\bibfield  {journal} {\bibinfo
  {journal} {Physical Review}\ }\textbf {\bibinfo {volume} {69}},\ \bibinfo
  {pages} {681} (\bibinfo {year} {1946})}\BibitemShut {NoStop}%
\bibitem [{\citenamefont {Drexhage}(1970)}]{Drexhage1970}%
  \BibitemOpen
  \bibfield  {author} {\bibinfo {author} {\bibfnamefont {K.}~\bibnamefont
  {Drexhage}},\ }\bibfield  {title} {\enquote {\bibinfo {title} {Influence of a
  dielectric interface on fluorescence decay time},}\ }\href
  {https://doi.org/10.1016/0022-2313(70)90082-7} {\bibfield  {journal}
  {\bibinfo  {journal} {Journal of Luminescence}\ }\textbf {\bibinfo {volume}
  {1--2}},\ \bibinfo {pages} {693--701} (\bibinfo {year} {1970})}\BibitemShut
  {NoStop}%
\bibitem [{\citenamefont {Chance}, \citenamefont {Prock},\ and\ \citenamefont
  {Silbey}(1978)}]{Chance1978}%
  \BibitemOpen
  \bibfield  {author} {\bibinfo {author} {\bibfnamefont {R.~R.}\ \bibnamefont
  {Chance}}, \bibinfo {author} {\bibfnamefont {A.}~\bibnamefont {Prock}},\ and\
  \bibinfo {author} {\bibfnamefont {R.}~\bibnamefont {Silbey}},\ }\bibfield
  {title} {\enquote {\bibinfo {title} {Molecular {{Fluorescence}} and {{Energy
  Transfer Near Interfaces}}},}\ }in\ \href
  {https://doi.org/10.1002/9780470142561.ch1} {\emph {\bibinfo {booktitle}
  {Advances in {{Chemical Physics}}}}},\ Vol.~\bibinfo {volume} {37},\ \bibinfo
  {editor} {edited by\ \bibinfo {editor} {\bibfnamefont {I.}~\bibnamefont
  {Prigogine}}\ and\ \bibinfo {editor} {\bibfnamefont {S.~A.}\ \bibnamefont
  {Rice}}}\ (\bibinfo  {publisher} {Wiley},\ \bibinfo {year} {1978})\ \bibinfo
  {edition} {1st}\ ed.,\ pp.\ \bibinfo {pages} {1--65}\BibitemShut {NoStop}%
\bibitem [{\citenamefont {Yamaguchi}\ and\ \citenamefont
  {Tahara}(2008)}]{Yamaguchi2008}%
  \BibitemOpen
  \bibfield  {author} {\bibinfo {author} {\bibfnamefont {S.}~\bibnamefont
  {Yamaguchi}}\ and\ \bibinfo {author} {\bibfnamefont {T.}~\bibnamefont
  {Tahara}},\ }\bibfield  {title} {\enquote {\bibinfo {title} {Coherent
  acoustic phonons in a thin gold film probed by femtosecond surface plasmon
  resonance},}\ }\href {https://doi.org/10.1002/jrs.2078} {\bibfield  {journal}
  {\bibinfo  {journal} {Journal of Raman Spectroscopy}\ }\textbf {\bibinfo
  {volume} {39}},\ \bibinfo {pages} {1703--1706} (\bibinfo {year}
  {2008})}\BibitemShut {NoStop}%
\bibitem [{\citenamefont {Kumar}\ \emph {et~al.}(2023)\citenamefont {Kumar},
  \citenamefont {Kuramochi}, \citenamefont {Takeuchi},\ and\ \citenamefont
  {Tahara}}]{Kumar2023}%
  \BibitemOpen
  \bibfield  {author} {\bibinfo {author} {\bibfnamefont {P.}~\bibnamefont
  {Kumar}}, \bibinfo {author} {\bibfnamefont {H.}~\bibnamefont {Kuramochi}},
  \bibinfo {author} {\bibfnamefont {S.}~\bibnamefont {Takeuchi}},\ and\
  \bibinfo {author} {\bibfnamefont {T.}~\bibnamefont {Tahara}},\ }\bibfield
  {title} {\enquote {\bibinfo {title} {Photoexcited {{Plasmon-Driven Ultrafast
  Dynamics}} of the {{Adsorbate Probed}} by {{Femtosecond Time-Resolved
  Surface-Enhanced Time-Domain Raman Spectroscopy}}},}\ }\href
  {https://doi.org/10.1021/acs.jpclett.2c03813} {\bibfield  {journal} {\bibinfo
   {journal} {The Journal of Physical Chemistry Letters}\ }\textbf {\bibinfo
  {volume} {14}},\ \bibinfo {pages} {2845--2853} (\bibinfo {year}
  {2023})}\BibitemShut {NoStop}%
\bibitem [{\citenamefont {Moskovits}(1985)}]{Moskovits1985}%
  \BibitemOpen
  \bibfield  {author} {\bibinfo {author} {\bibfnamefont {M.}~\bibnamefont
  {Moskovits}},\ }\bibfield  {title} {\enquote {\bibinfo {title}
  {Surface-enhanced spectroscopy},}\ }\href
  {https://doi.org/10.1103/RevModPhys.57.783} {\bibfield  {journal} {\bibinfo
  {journal} {Reviews of Modern Physics}\ }\textbf {\bibinfo {volume} {57}},\
  \bibinfo {pages} {783--826} (\bibinfo {year} {1985})}\BibitemShut {NoStop}%
\bibitem [{\citenamefont {Campion}\ and\ \citenamefont
  {Kambhampati}(1998)}]{Campion1998}%
  \BibitemOpen
  \bibfield  {author} {\bibinfo {author} {\bibfnamefont {A.}~\bibnamefont
  {Campion}}\ and\ \bibinfo {author} {\bibfnamefont {P.}~\bibnamefont
  {Kambhampati}},\ }\bibfield  {title} {\enquote {\bibinfo {title}
  {Surface-enhanced {{Raman}} scattering},}\ }\href
  {https://doi.org/10.1039/a827241z} {\bibfield  {journal} {\bibinfo  {journal}
  {Chemical Society Reviews}\ }\textbf {\bibinfo {volume} {27}},\ \bibinfo
  {pages} {241} (\bibinfo {year} {1998})}\BibitemShut {NoStop}%
\bibitem [{\citenamefont {Willets}\ and\ \citenamefont
  {Van~Duyne}(2007)}]{Willets2007}%
  \BibitemOpen
  \bibfield  {author} {\bibinfo {author} {\bibfnamefont {K.~A.}\ \bibnamefont
  {Willets}}\ and\ \bibinfo {author} {\bibfnamefont {R.~P.}\ \bibnamefont
  {Van~Duyne}},\ }\bibfield  {title} {\enquote {\bibinfo {title} {Localized
  {{Surface Plasmon Resonance Spectroscopy}} and {{Sensing}}},}\ }\href
  {https://doi.org/10.1146/annurev.physchem.58.032806.104607} {\bibfield
  {journal} {\bibinfo  {journal} {Annual Review of Physical Chemistry}\
  }\textbf {\bibinfo {volume} {58}},\ \bibinfo {pages} {267--297} (\bibinfo
  {year} {2007})}\BibitemShut {NoStop}%
\bibitem [{\citenamefont {Langer}\ \emph {et~al.}(2020)\citenamefont {Langer},
  \citenamefont {Jimenez De~Aberasturi}, \citenamefont {Aizpurua},
  \citenamefont {{Alvarez-Puebla}}, \citenamefont {Augui{\'e}}, \citenamefont
  {Baumberg}, \citenamefont {Bazan}, \citenamefont {Bell}, \citenamefont
  {Boisen}, \citenamefont {Brolo}, \citenamefont {Choo}, \citenamefont
  {{Cialla-May}}, \citenamefont {Deckert}, \citenamefont {Fabris},
  \citenamefont {Faulds}, \citenamefont {Garc{\'i}a De~Abajo}, \citenamefont
  {Goodacre}, \citenamefont {Graham}, \citenamefont {Haes}, \citenamefont
  {Haynes}, \citenamefont {Huck}, \citenamefont {Itoh}, \citenamefont
  {K{\"a}ll}, \citenamefont {Kneipp}, \citenamefont {Kotov}, \citenamefont
  {Kuang}, \citenamefont {Le~Ru}, \citenamefont {Lee}, \citenamefont {Li},
  \citenamefont {Ling}, \citenamefont {Maier}, \citenamefont {Mayerh{\"o}fer},
  \citenamefont {Moskovits}, \citenamefont {Murakoshi}, \citenamefont {Nam},
  \citenamefont {Nie}, \citenamefont {Ozaki}, \citenamefont
  {{Pastoriza-Santos}}, \citenamefont {{Perez-Juste}}, \citenamefont {Popp},
  \citenamefont {Pucci}, \citenamefont {Reich}, \citenamefont {Ren},
  \citenamefont {Schatz}, \citenamefont {Shegai}, \citenamefont
  {Schl{\"u}cker}, \citenamefont {Tay}, \citenamefont {Thomas}, \citenamefont
  {Tian}, \citenamefont {Van~Duyne}, \citenamefont {{Vo-Dinh}}, \citenamefont
  {Wang}, \citenamefont {Willets}, \citenamefont {Xu}, \citenamefont {Xu},
  \citenamefont {Xu}, \citenamefont {Yamamoto}, \citenamefont {Zhao},\ and\
  \citenamefont {{Liz-Marz{\'a}n}}}]{Langer2020}%
  \BibitemOpen
  \bibfield  {author} {\bibinfo {author} {\bibfnamefont {J.}~\bibnamefont
  {Langer}}, \bibinfo {author} {\bibfnamefont {D.}~\bibnamefont {Jimenez
  De~Aberasturi}}, \bibinfo {author} {\bibfnamefont {J.}~\bibnamefont
  {Aizpurua}}, \bibinfo {author} {\bibfnamefont {R.~A.}\ \bibnamefont
  {{Alvarez-Puebla}}}, \bibinfo {author} {\bibfnamefont {B.}~\bibnamefont
  {Augui{\'e}}}, \bibinfo {author} {\bibfnamefont {J.~J.}\ \bibnamefont
  {Baumberg}}, \bibinfo {author} {\bibfnamefont {G.~C.}\ \bibnamefont {Bazan}},
  \bibinfo {author} {\bibfnamefont {S.~E.~J.}\ \bibnamefont {Bell}}, \bibinfo
  {author} {\bibfnamefont {A.}~\bibnamefont {Boisen}}, \bibinfo {author}
  {\bibfnamefont {A.~G.}\ \bibnamefont {Brolo}}, \bibinfo {author}
  {\bibfnamefont {J.}~\bibnamefont {Choo}}, \bibinfo {author} {\bibfnamefont
  {D.}~\bibnamefont {{Cialla-May}}}, \bibinfo {author} {\bibfnamefont
  {V.}~\bibnamefont {Deckert}}, \bibinfo {author} {\bibfnamefont
  {L.}~\bibnamefont {Fabris}}, \bibinfo {author} {\bibfnamefont
  {K.}~\bibnamefont {Faulds}}, \bibinfo {author} {\bibfnamefont {F.~J.}\
  \bibnamefont {Garc{\'i}a De~Abajo}}, \bibinfo {author} {\bibfnamefont
  {R.}~\bibnamefont {Goodacre}}, \bibinfo {author} {\bibfnamefont
  {D.}~\bibnamefont {Graham}}, \bibinfo {author} {\bibfnamefont {A.~J.}\
  \bibnamefont {Haes}}, \bibinfo {author} {\bibfnamefont {C.~L.}\ \bibnamefont
  {Haynes}}, \bibinfo {author} {\bibfnamefont {C.}~\bibnamefont {Huck}},
  \bibinfo {author} {\bibfnamefont {T.}~\bibnamefont {Itoh}}, \bibinfo {author}
  {\bibfnamefont {M.}~\bibnamefont {K{\"a}ll}}, \bibinfo {author}
  {\bibfnamefont {J.}~\bibnamefont {Kneipp}}, \bibinfo {author} {\bibfnamefont
  {N.~A.}\ \bibnamefont {Kotov}}, \bibinfo {author} {\bibfnamefont
  {H.}~\bibnamefont {Kuang}}, \bibinfo {author} {\bibfnamefont {E.~C.}\
  \bibnamefont {Le~Ru}}, \bibinfo {author} {\bibfnamefont {H.~K.}\ \bibnamefont
  {Lee}}, \bibinfo {author} {\bibfnamefont {J.-F.}\ \bibnamefont {Li}},
  \bibinfo {author} {\bibfnamefont {X.~Y.}\ \bibnamefont {Ling}}, \bibinfo
  {author} {\bibfnamefont {S.~A.}\ \bibnamefont {Maier}}, \bibinfo {author}
  {\bibfnamefont {T.}~\bibnamefont {Mayerh{\"o}fer}}, \bibinfo {author}
  {\bibfnamefont {M.}~\bibnamefont {Moskovits}}, \bibinfo {author}
  {\bibfnamefont {K.}~\bibnamefont {Murakoshi}}, \bibinfo {author}
  {\bibfnamefont {J.-M.}\ \bibnamefont {Nam}}, \bibinfo {author} {\bibfnamefont
  {S.}~\bibnamefont {Nie}}, \bibinfo {author} {\bibfnamefont {Y.}~\bibnamefont
  {Ozaki}}, \bibinfo {author} {\bibfnamefont {I.}~\bibnamefont
  {{Pastoriza-Santos}}}, \bibinfo {author} {\bibfnamefont {J.}~\bibnamefont
  {{Perez-Juste}}}, \bibinfo {author} {\bibfnamefont {J.}~\bibnamefont {Popp}},
  \bibinfo {author} {\bibfnamefont {A.}~\bibnamefont {Pucci}}, \bibinfo
  {author} {\bibfnamefont {S.}~\bibnamefont {Reich}}, \bibinfo {author}
  {\bibfnamefont {B.}~\bibnamefont {Ren}}, \bibinfo {author} {\bibfnamefont
  {G.~C.}\ \bibnamefont {Schatz}}, \bibinfo {author} {\bibfnamefont
  {T.}~\bibnamefont {Shegai}}, \bibinfo {author} {\bibfnamefont
  {S.}~\bibnamefont {Schl{\"u}cker}}, \bibinfo {author} {\bibfnamefont {L.-L.}\
  \bibnamefont {Tay}}, \bibinfo {author} {\bibfnamefont {K.~G.}\ \bibnamefont
  {Thomas}}, \bibinfo {author} {\bibfnamefont {Z.-Q.}\ \bibnamefont {Tian}},
  \bibinfo {author} {\bibfnamefont {R.~P.}\ \bibnamefont {Van~Duyne}}, \bibinfo
  {author} {\bibfnamefont {T.}~\bibnamefont {{Vo-Dinh}}}, \bibinfo {author}
  {\bibfnamefont {Y.}~\bibnamefont {Wang}}, \bibinfo {author} {\bibfnamefont
  {K.~A.}\ \bibnamefont {Willets}}, \bibinfo {author} {\bibfnamefont
  {C.}~\bibnamefont {Xu}}, \bibinfo {author} {\bibfnamefont {H.}~\bibnamefont
  {Xu}}, \bibinfo {author} {\bibfnamefont {Y.}~\bibnamefont {Xu}}, \bibinfo
  {author} {\bibfnamefont {Y.~S.}\ \bibnamefont {Yamamoto}}, \bibinfo {author}
  {\bibfnamefont {B.}~\bibnamefont {Zhao}},\ and\ \bibinfo {author}
  {\bibfnamefont {L.~M.}\ \bibnamefont {{Liz-Marz{\'a}n}}},\ }\bibfield
  {title} {\enquote {\bibinfo {title} {Present and {{Future}} of
  {{Surface-Enhanced Raman Scattering}}},}\ }\href
  {https://doi.org/10.1021/acsnano.9b04224} {\bibfield  {journal} {\bibinfo
  {journal} {ACS Nano}\ }\textbf {\bibinfo {volume} {14}},\ \bibinfo {pages}
  {28--117} (\bibinfo {year} {2020})}\BibitemShut {NoStop}%
\bibitem [{\citenamefont {Wang}\ \emph {et~al.}(2020)\citenamefont {Wang},
  \citenamefont {Huang}, \citenamefont {Hu}, \citenamefont {Yan},\ and\
  \citenamefont {Ren}}]{Wang2020}%
  \BibitemOpen
  \bibfield  {author} {\bibinfo {author} {\bibfnamefont {X.}~\bibnamefont
  {Wang}}, \bibinfo {author} {\bibfnamefont {S.-C.}\ \bibnamefont {Huang}},
  \bibinfo {author} {\bibfnamefont {S.}~\bibnamefont {Hu}}, \bibinfo {author}
  {\bibfnamefont {S.}~\bibnamefont {Yan}},\ and\ \bibinfo {author}
  {\bibfnamefont {B.}~\bibnamefont {Ren}},\ }\bibfield  {title} {\enquote
  {\bibinfo {title} {Fundamental understanding and applications of
  plasmon-enhanced {{Raman}} spectroscopy},}\ }\href
  {https://doi.org/10.1038/s42254-020-0171-y} {\bibfield  {journal} {\bibinfo
  {journal} {Nature Reviews Physics}\ }\textbf {\bibinfo {volume} {2}},\
  \bibinfo {pages} {253--271} (\bibinfo {year} {2020})}\BibitemShut {NoStop}%
\bibitem [{\citenamefont {Yi}\ \emph {et~al.}(2025)\citenamefont {Yi},
  \citenamefont {You}, \citenamefont {Hu}, \citenamefont {Wu}, \citenamefont
  {Liu}, \citenamefont {Yang}, \citenamefont {Zhang}, \citenamefont {Gu},
  \citenamefont {Wang}, \citenamefont {Wang}, \citenamefont {Ma}, \citenamefont
  {Yang}, \citenamefont {Liu}, \citenamefont {Fan}, \citenamefont {Zhan},
  \citenamefont {Tian}, \citenamefont {Qiao}, \citenamefont {Wang},
  \citenamefont {Luo}, \citenamefont {Meng}, \citenamefont {Mao}, \citenamefont
  {Li}, \citenamefont {Ren}, \citenamefont {Aizpurua}, \citenamefont
  {Apkarian}, \citenamefont {Bartlett}, \citenamefont {Baumberg}, \citenamefont
  {Bell}, \citenamefont {Brolo}, \citenamefont {Brus}, \citenamefont {Choo},
  \citenamefont {Cui}, \citenamefont {Deckert}, \citenamefont {Domke},
  \citenamefont {Dong}, \citenamefont {Duan}, \citenamefont {Faulds},
  \citenamefont {Frontiera}, \citenamefont {Halas}, \citenamefont {Haynes},
  \citenamefont {Itoh}, \citenamefont {Kneipp}, \citenamefont {Kneipp},
  \citenamefont {Le~Ru}, \citenamefont {Li}, \citenamefont {Ling},
  \citenamefont {Lipkowski}, \citenamefont {{Liz-Marz{\'a}n}}, \citenamefont
  {Nam}, \citenamefont {Nie}, \citenamefont {Nordlander}, \citenamefont
  {Ozaki}, \citenamefont {Panneerselvam}, \citenamefont {Popp}, \citenamefont
  {Russell}, \citenamefont {Schl{\"u}cker}, \citenamefont {Tian}, \citenamefont
  {Tong}, \citenamefont {Xu}, \citenamefont {Xu}, \citenamefont {Yang},
  \citenamefont {Yao}, \citenamefont {Zhang}, \citenamefont {Zhang},
  \citenamefont {Zhang}, \citenamefont {Zhao}, \citenamefont {Zenobi},
  \citenamefont {Schatz}, \citenamefont {Graham},\ and\ \citenamefont
  {Tian}}]{Yi2025}%
  \BibitemOpen
  \bibfield  {author} {\bibinfo {author} {\bibfnamefont {J.}~\bibnamefont
  {Yi}}, \bibinfo {author} {\bibfnamefont {E.-M.}\ \bibnamefont {You}},
  \bibinfo {author} {\bibfnamefont {R.}~\bibnamefont {Hu}}, \bibinfo {author}
  {\bibfnamefont {D.-Y.}\ \bibnamefont {Wu}}, \bibinfo {author} {\bibfnamefont
  {G.-K.}\ \bibnamefont {Liu}}, \bibinfo {author} {\bibfnamefont {Z.-L.}\
  \bibnamefont {Yang}}, \bibinfo {author} {\bibfnamefont {H.}~\bibnamefont
  {Zhang}}, \bibinfo {author} {\bibfnamefont {Y.}~\bibnamefont {Gu}}, \bibinfo
  {author} {\bibfnamefont {Y.-H.}\ \bibnamefont {Wang}}, \bibinfo {author}
  {\bibfnamefont {X.}~\bibnamefont {Wang}}, \bibinfo {author} {\bibfnamefont
  {H.}~\bibnamefont {Ma}}, \bibinfo {author} {\bibfnamefont {Y.}~\bibnamefont
  {Yang}}, \bibinfo {author} {\bibfnamefont {J.-Y.}\ \bibnamefont {Liu}},
  \bibinfo {author} {\bibfnamefont {F.~R.}\ \bibnamefont {Fan}}, \bibinfo
  {author} {\bibfnamefont {C.}~\bibnamefont {Zhan}}, \bibinfo {author}
  {\bibfnamefont {J.-H.}\ \bibnamefont {Tian}}, \bibinfo {author}
  {\bibfnamefont {Y.}~\bibnamefont {Qiao}}, \bibinfo {author} {\bibfnamefont
  {H.}~\bibnamefont {Wang}}, \bibinfo {author} {\bibfnamefont {S.-H.}\
  \bibnamefont {Luo}}, \bibinfo {author} {\bibfnamefont {Z.-D.}\ \bibnamefont
  {Meng}}, \bibinfo {author} {\bibfnamefont {B.-W.}\ \bibnamefont {Mao}},
  \bibinfo {author} {\bibfnamefont {J.-F.}\ \bibnamefont {Li}}, \bibinfo
  {author} {\bibfnamefont {B.}~\bibnamefont {Ren}}, \bibinfo {author}
  {\bibfnamefont {J.}~\bibnamefont {Aizpurua}}, \bibinfo {author}
  {\bibfnamefont {V.~A.}\ \bibnamefont {Apkarian}}, \bibinfo {author}
  {\bibfnamefont {P.~N.}\ \bibnamefont {Bartlett}}, \bibinfo {author}
  {\bibfnamefont {J.}~\bibnamefont {Baumberg}}, \bibinfo {author}
  {\bibfnamefont {S.~E.~J.}\ \bibnamefont {Bell}}, \bibinfo {author}
  {\bibfnamefont {A.~G.}\ \bibnamefont {Brolo}}, \bibinfo {author}
  {\bibfnamefont {L.~E.}\ \bibnamefont {Brus}}, \bibinfo {author}
  {\bibfnamefont {J.}~\bibnamefont {Choo}}, \bibinfo {author} {\bibfnamefont
  {L.}~\bibnamefont {Cui}}, \bibinfo {author} {\bibfnamefont {V.}~\bibnamefont
  {Deckert}}, \bibinfo {author} {\bibfnamefont {K.~F.}\ \bibnamefont {Domke}},
  \bibinfo {author} {\bibfnamefont {Z.-C.}\ \bibnamefont {Dong}}, \bibinfo
  {author} {\bibfnamefont {S.}~\bibnamefont {Duan}}, \bibinfo {author}
  {\bibfnamefont {K.}~\bibnamefont {Faulds}}, \bibinfo {author} {\bibfnamefont
  {R.}~\bibnamefont {Frontiera}}, \bibinfo {author} {\bibfnamefont
  {N.}~\bibnamefont {Halas}}, \bibinfo {author} {\bibfnamefont
  {C.}~\bibnamefont {Haynes}}, \bibinfo {author} {\bibfnamefont
  {T.}~\bibnamefont {Itoh}}, \bibinfo {author} {\bibfnamefont {J.}~\bibnamefont
  {Kneipp}}, \bibinfo {author} {\bibfnamefont {K.}~\bibnamefont {Kneipp}},
  \bibinfo {author} {\bibfnamefont {E.~C.}\ \bibnamefont {Le~Ru}}, \bibinfo
  {author} {\bibfnamefont {Z.-P.}\ \bibnamefont {Li}}, \bibinfo {author}
  {\bibfnamefont {X.~Y.}\ \bibnamefont {Ling}}, \bibinfo {author}
  {\bibfnamefont {J.}~\bibnamefont {Lipkowski}}, \bibinfo {author}
  {\bibfnamefont {L.~M.}\ \bibnamefont {{Liz-Marz{\'a}n}}}, \bibinfo {author}
  {\bibfnamefont {J.-M.}\ \bibnamefont {Nam}}, \bibinfo {author} {\bibfnamefont
  {S.}~\bibnamefont {Nie}}, \bibinfo {author} {\bibfnamefont {P.}~\bibnamefont
  {Nordlander}}, \bibinfo {author} {\bibfnamefont {Y.}~\bibnamefont {Ozaki}},
  \bibinfo {author} {\bibfnamefont {R.}~\bibnamefont {Panneerselvam}}, \bibinfo
  {author} {\bibfnamefont {J.}~\bibnamefont {Popp}}, \bibinfo {author}
  {\bibfnamefont {A.~E.}\ \bibnamefont {Russell}}, \bibinfo {author}
  {\bibfnamefont {S.}~\bibnamefont {Schl{\"u}cker}}, \bibinfo {author}
  {\bibfnamefont {Y.}~\bibnamefont {Tian}}, \bibinfo {author} {\bibfnamefont
  {L.}~\bibnamefont {Tong}}, \bibinfo {author} {\bibfnamefont {H.}~\bibnamefont
  {Xu}}, \bibinfo {author} {\bibfnamefont {Y.}~\bibnamefont {Xu}}, \bibinfo
  {author} {\bibfnamefont {L.}~\bibnamefont {Yang}}, \bibinfo {author}
  {\bibfnamefont {J.}~\bibnamefont {Yao}}, \bibinfo {author} {\bibfnamefont
  {J.}~\bibnamefont {Zhang}}, \bibinfo {author} {\bibfnamefont
  {Y.}~\bibnamefont {Zhang}}, \bibinfo {author} {\bibfnamefont
  {Y.}~\bibnamefont {Zhang}}, \bibinfo {author} {\bibfnamefont
  {B.}~\bibnamefont {Zhao}}, \bibinfo {author} {\bibfnamefont {R.}~\bibnamefont
  {Zenobi}}, \bibinfo {author} {\bibfnamefont {G.~C.}\ \bibnamefont {Schatz}},
  \bibinfo {author} {\bibfnamefont {D.}~\bibnamefont {Graham}},\ and\ \bibinfo
  {author} {\bibfnamefont {Z.-Q.}\ \bibnamefont {Tian}},\ }\bibfield  {title}
  {\enquote {\bibinfo {title} {Surface-enhanced {{Raman}} spectroscopy: A
  half-century historical perspective},}\ }\href
  {https://doi.org/10.1039/D4CS00883A} {\bibfield  {journal} {\bibinfo
  {journal} {Chemical Society Reviews}\ }\textbf {\bibinfo {volume} {54}},\
  \bibinfo {pages} {1453--1551} (\bibinfo {year} {2025})}\BibitemShut {NoStop}%
\bibitem [{\citenamefont {Zhang}\ \emph {et~al.}(2013)\citenamefont {Zhang},
  \citenamefont {Zhang}, \citenamefont {Dong}, \citenamefont {Jiang},
  \citenamefont {Zhang}, \citenamefont {Chen}, \citenamefont {Zhang},
  \citenamefont {Liao}, \citenamefont {Aizpurua}, \citenamefont {Luo},
  \citenamefont {Yang},\ and\ \citenamefont {Hou}}]{Zhang2013}%
  \BibitemOpen
  \bibfield  {author} {\bibinfo {author} {\bibfnamefont {R.}~\bibnamefont
  {Zhang}}, \bibinfo {author} {\bibfnamefont {Y.}~\bibnamefont {Zhang}},
  \bibinfo {author} {\bibfnamefont {Z.~C.}\ \bibnamefont {Dong}}, \bibinfo
  {author} {\bibfnamefont {S.}~\bibnamefont {Jiang}}, \bibinfo {author}
  {\bibfnamefont {C.}~\bibnamefont {Zhang}}, \bibinfo {author} {\bibfnamefont
  {L.~G.}\ \bibnamefont {Chen}}, \bibinfo {author} {\bibfnamefont
  {L.}~\bibnamefont {Zhang}}, \bibinfo {author} {\bibfnamefont
  {Y.}~\bibnamefont {Liao}}, \bibinfo {author} {\bibfnamefont {J.}~\bibnamefont
  {Aizpurua}}, \bibinfo {author} {\bibfnamefont {Y.}~\bibnamefont {Luo}},
  \bibinfo {author} {\bibfnamefont {J.~L.}\ \bibnamefont {Yang}},\ and\
  \bibinfo {author} {\bibfnamefont {J.~G.}\ \bibnamefont {Hou}},\ }\bibfield
  {title} {\enquote {\bibinfo {title} {Chemical mapping of a single molecule by
  plasmon-enhanced {{Raman}} scattering},}\ }\href
  {https://doi.org/10.1038/nature12151} {\bibfield  {journal} {\bibinfo
  {journal} {Nature}\ }\textbf {\bibinfo {volume} {498}},\ \bibinfo {pages}
  {82--86} (\bibinfo {year} {2013})}\BibitemShut {NoStop}%
\bibitem [{\citenamefont {Lee}\ \emph {et~al.}(2019)\citenamefont {Lee},
  \citenamefont {Crampton}, \citenamefont {Tallarida},\ and\ \citenamefont
  {Apkarian}}]{Lee2019}%
  \BibitemOpen
  \bibfield  {author} {\bibinfo {author} {\bibfnamefont {J.}~\bibnamefont
  {Lee}}, \bibinfo {author} {\bibfnamefont {K.~T.}\ \bibnamefont {Crampton}},
  \bibinfo {author} {\bibfnamefont {N.}~\bibnamefont {Tallarida}},\ and\
  \bibinfo {author} {\bibfnamefont {V.~A.}\ \bibnamefont {Apkarian}},\
  }\bibfield  {title} {\enquote {\bibinfo {title} {Visualizing vibrational
  normal modes of a single molecule with atomically confined light},}\ }\href
  {https://doi.org/10.1038/s41586-019-1059-9} {\bibfield  {journal} {\bibinfo
  {journal} {Nature}\ }\textbf {\bibinfo {volume} {568}},\ \bibinfo {pages}
  {78--82} (\bibinfo {year} {2019})}\BibitemShut {NoStop}%
\bibitem [{\citenamefont {Jaculbia}\ \emph {et~al.}(2020)\citenamefont
  {Jaculbia}, \citenamefont {Imada}, \citenamefont {Miwa}, \citenamefont
  {Iwasa}, \citenamefont {Takenaka}, \citenamefont {Yang}, \citenamefont
  {Kazuma}, \citenamefont {Hayazawa}, \citenamefont {Taketsugu},\ and\
  \citenamefont {Kim}}]{Jaculbia2020}%
  \BibitemOpen
  \bibfield  {author} {\bibinfo {author} {\bibfnamefont {R.~B.}\ \bibnamefont
  {Jaculbia}}, \bibinfo {author} {\bibfnamefont {H.}~\bibnamefont {Imada}},
  \bibinfo {author} {\bibfnamefont {K.}~\bibnamefont {Miwa}}, \bibinfo {author}
  {\bibfnamefont {T.}~\bibnamefont {Iwasa}}, \bibinfo {author} {\bibfnamefont
  {M.}~\bibnamefont {Takenaka}}, \bibinfo {author} {\bibfnamefont
  {B.}~\bibnamefont {Yang}}, \bibinfo {author} {\bibfnamefont {E.}~\bibnamefont
  {Kazuma}}, \bibinfo {author} {\bibfnamefont {N.}~\bibnamefont {Hayazawa}},
  \bibinfo {author} {\bibfnamefont {T.}~\bibnamefont {Taketsugu}},\ and\
  \bibinfo {author} {\bibfnamefont {Y.}~\bibnamefont {Kim}},\ }\bibfield
  {title} {\enquote {\bibinfo {title} {Single-molecule resonance {{Raman}}
  effect in a plasmonic nanocavity},}\ }\href
  {https://doi.org/10.1038/s41565-019-0614-8} {\bibfield  {journal} {\bibinfo
  {journal} {Nature Nanotechnology}\ }\textbf {\bibinfo {volume} {15}},\
  \bibinfo {pages} {105--110} (\bibinfo {year} {2020})}\BibitemShut {NoStop}%
\bibitem [{\citenamefont {Gutzler}\ \emph {et~al.}(2021)\citenamefont
  {Gutzler}, \citenamefont {Garg}, \citenamefont {Ast}, \citenamefont
  {Kuhnke},\ and\ \citenamefont {Kern}}]{Gutzler2021}%
  \BibitemOpen
  \bibfield  {author} {\bibinfo {author} {\bibfnamefont {R.}~\bibnamefont
  {Gutzler}}, \bibinfo {author} {\bibfnamefont {M.}~\bibnamefont {Garg}},
  \bibinfo {author} {\bibfnamefont {C.~R.}\ \bibnamefont {Ast}}, \bibinfo
  {author} {\bibfnamefont {K.}~\bibnamefont {Kuhnke}},\ and\ \bibinfo {author}
  {\bibfnamefont {K.}~\bibnamefont {Kern}},\ }\bibfield  {title} {\enquote
  {\bibinfo {title} {Light--matter interaction at atomic scales},}\ }\href
  {https://doi.org/10.1038/s42254-021-00306-5} {\bibfield  {journal} {\bibinfo
  {journal} {Nature Reviews Physics}\ }\textbf {\bibinfo {volume} {3}},\
  \bibinfo {pages} {441--453} (\bibinfo {year} {2021})}\BibitemShut {NoStop}%
\bibitem [{\citenamefont {Kumagai}, \citenamefont {Miwa},\ and\ \citenamefont
  {Cirera}(2025)}]{Kumagai2025}%
  \BibitemOpen
  \bibfield  {author} {\bibinfo {author} {\bibfnamefont {T.}~\bibnamefont
  {Kumagai}}, \bibinfo {author} {\bibfnamefont {K.}~\bibnamefont {Miwa}},\ and\
  \bibinfo {author} {\bibfnamefont {B.}~\bibnamefont {Cirera}},\ }\bibfield
  {title} {\enquote {\bibinfo {title} {Point-{{Contact Tip-Enhanced Raman
  Spectroscopy}}: {{Picoscale Light}}--{{Matter Interactions}} within
  {{Plasmonic Cavities}}},}\ }\href
  {https://doi.org/10.1021/acs.nanolett.5c04045} {\bibfield  {journal}
  {\bibinfo  {journal} {Nano Letters}\ }\textbf {\bibinfo {volume} {25}},\
  \bibinfo {pages} {15449--15459} (\bibinfo {year} {2025})}\BibitemShut
  {NoStop}%
\bibitem [{\citenamefont {Ebbesen}(2016)}]{Ebbesen2016}%
  \BibitemOpen
  \bibfield  {author} {\bibinfo {author} {\bibfnamefont {T.~W.}\ \bibnamefont
  {Ebbesen}},\ }\bibfield  {title} {\enquote {\bibinfo {title} {Hybrid
  {{Light}}--{{Matter States}} in a {{Molecular}} and {{Material Science
  Perspective}}},}\ }\href {https://doi.org/10.1021/acs.accounts.6b00295}
  {\bibfield  {journal} {\bibinfo  {journal} {Accounts of Chemical Research}\
  }\textbf {\bibinfo {volume} {49}},\ \bibinfo {pages} {2403--2412} (\bibinfo
  {year} {2016})}\BibitemShut {NoStop}%
\bibitem [{\citenamefont {Herrera}\ and\ \citenamefont
  {Spano}(2016)}]{Herrera2016}%
  \BibitemOpen
  \bibfield  {author} {\bibinfo {author} {\bibfnamefont {F.}~\bibnamefont
  {Herrera}}\ and\ \bibinfo {author} {\bibfnamefont {F.~C.}\ \bibnamefont
  {Spano}},\ }\bibfield  {title} {\enquote {\bibinfo {title}
  {Cavity-{{Controlled Chemistry}} in {{Molecular Ensembles}}},}\ }\href
  {https://doi.org/10.1103/PhysRevLett.116.238301} {\bibfield  {journal}
  {\bibinfo  {journal} {Physical Review Letters}\ }\textbf {\bibinfo {volume}
  {116}},\ \bibinfo {pages} {238301} (\bibinfo {year} {2016})}\BibitemShut
  {NoStop}%
\bibitem [{\citenamefont {Chen}, \citenamefont {Miwa},\ and\ \citenamefont
  {Galperin}(2019)}]{Chen2019}%
  \BibitemOpen
  \bibfield  {author} {\bibinfo {author} {\bibfnamefont {F.}~\bibnamefont
  {Chen}}, \bibinfo {author} {\bibfnamefont {K.}~\bibnamefont {Miwa}},\ and\
  \bibinfo {author} {\bibfnamefont {M.}~\bibnamefont {Galperin}},\ }\bibfield
  {title} {\enquote {\bibinfo {title} {Electronic friction in interacting
  systems},}\ }\href {https://doi.org/10.1063/1.5095425} {\bibfield  {journal}
  {\bibinfo  {journal} {The Journal of Chemical Physics}\ }\textbf {\bibinfo
  {volume} {150}},\ \bibinfo {pages} {174101} (\bibinfo {year}
  {2019})}\BibitemShut {NoStop}%
\bibitem [{\citenamefont {{Garcia-Vidal}}, \citenamefont {Ciuti},\ and\
  \citenamefont {Ebbesen}(2021)}]{Garcia-Vidal2021}%
  \BibitemOpen
  \bibfield  {author} {\bibinfo {author} {\bibfnamefont {F.~J.}\ \bibnamefont
  {{Garcia-Vidal}}}, \bibinfo {author} {\bibfnamefont {C.}~\bibnamefont
  {Ciuti}},\ and\ \bibinfo {author} {\bibfnamefont {T.~W.}\ \bibnamefont
  {Ebbesen}},\ }\bibfield  {title} {\enquote {\bibinfo {title} {Manipulating
  matter by strong coupling to vacuum fields},}\ }\href
  {https://doi.org/10.1126/science.abd0336} {\bibfield  {journal} {\bibinfo
  {journal} {Science}\ }\textbf {\bibinfo {volume} {373}},\ \bibinfo {pages}
  {eabd0336} (\bibinfo {year} {2021})}\BibitemShut {NoStop}%
\bibitem [{\citenamefont {Miwa}, \citenamefont {Sakamoto},\ and\ \citenamefont
  {Ishizaki}(2023)}]{Miwa2023}%
  \BibitemOpen
  \bibfield  {author} {\bibinfo {author} {\bibfnamefont {K.}~\bibnamefont
  {Miwa}}, \bibinfo {author} {\bibfnamefont {S.}~\bibnamefont {Sakamoto}},\
  and\ \bibinfo {author} {\bibfnamefont {A.}~\bibnamefont {Ishizaki}},\
  }\bibfield  {title} {\enquote {\bibinfo {title} {Control and {{Enhancement}}
  of {{Single-Molecule Electroluminescence}} through {{Strong Light}}--{{Matter
  Coupling}}},}\ }\href {https://doi.org/10.1021/acs.nanolett.2c05089}
  {\bibfield  {journal} {\bibinfo  {journal} {Nano Letters}\ }\textbf {\bibinfo
  {volume} {23}},\ \bibinfo {pages} {3231--3238} (\bibinfo {year}
  {2023})}\BibitemShut {NoStop}%
\bibitem [{\citenamefont {Sukharev}, \citenamefont {Subotnik},\ and\
  \citenamefont {Nitzan}(2023)}]{Sukharev2023}%
  \BibitemOpen
  \bibfield  {author} {\bibinfo {author} {\bibfnamefont {M.}~\bibnamefont
  {Sukharev}}, \bibinfo {author} {\bibfnamefont {J.}~\bibnamefont {Subotnik}},\
  and\ \bibinfo {author} {\bibfnamefont {A.}~\bibnamefont {Nitzan}},\
  }\bibfield  {title} {\enquote {\bibinfo {title} {Dissociation slowdown by
  collective optical response under strong coupling conditions},}\ }\href
  {https://doi.org/10.1063/5.0133972} {\bibfield  {journal} {\bibinfo
  {journal} {The Journal of Chemical Physics}\ }\textbf {\bibinfo {volume}
  {158}},\ \bibinfo {pages} {084104} (\bibinfo {year} {2023})}\BibitemShut
  {NoStop}%
\bibitem [{\citenamefont {Engelhardt}\ and\ \citenamefont
  {Cao}(2023)}]{Engelhardt2023}%
  \BibitemOpen
  \bibfield  {author} {\bibinfo {author} {\bibfnamefont {G.}~\bibnamefont
  {Engelhardt}}\ and\ \bibinfo {author} {\bibfnamefont {J.}~\bibnamefont
  {Cao}},\ }\bibfield  {title} {\enquote {\bibinfo {title} {Polariton
  {{Localization}} and {{Dispersion Properties}} of {{Disordered Quantum
  Emitters}} in {{Multimode Microcavities}}},}\ }\href
  {https://doi.org/10.1103/PhysRevLett.130.213602} {\bibfield  {journal}
  {\bibinfo  {journal} {Physical Review Letters}\ }\textbf {\bibinfo {volume}
  {130}},\ \bibinfo {pages} {213602} (\bibinfo {year} {2023})}\BibitemShut
  {NoStop}%
\bibitem [{\citenamefont {Yuen-Zhou}, \citenamefont {Giebink},\ and\
  \citenamefont {Ribeiro}(2025)}]{Yuen-Zhou2025}%
  \BibitemOpen
  \bibinfo {editor} {\bibfnamefont {J.}~\bibnamefont {Yuen-Zhou}}, \bibinfo
  {editor} {\bibfnamefont {N.~C.}\ \bibnamefont {Giebink}},\ and\ \bibinfo
  {editor} {\bibfnamefont {R.~F.}\ \bibnamefont {Ribeiro}},\ eds.,\ \href
  {https://doi.org/10.1002/9781119783329} {\emph {\bibinfo {title} {Polariton
  {{Chemistry}}: {{Molecules}} in {{Cavities}}}}},\ \bibinfo {edition} {1st}\
  ed.\ (\bibinfo  {publisher} {Wiley},\ \bibinfo {year} {2025})\BibitemShut
  {NoStop}%
\bibitem [{\citenamefont {{Gonzaga-Galeana}}\ and\ \citenamefont
  {{Zurita-S{\'a}nchez}}(2013)}]{Gonzaga-Galeana2013}%
  \BibitemOpen
  \bibfield  {author} {\bibinfo {author} {\bibfnamefont {J.~A.}\ \bibnamefont
  {{Gonzaga-Galeana}}}\ and\ \bibinfo {author} {\bibfnamefont {J.~R.}\
  \bibnamefont {{Zurita-S{\'a}nchez}}},\ }\bibfield  {title} {\enquote
  {\bibinfo {title} {A revisitation of the {{F\"orster}} energy transfer near a
  metallic spherical nanoparticle: (1) {{Efficiency}} enhancement or reduction?
  (2) {{The}} control of the {{F\"orster}} radius of the unbounded medium. (3)
  {{The}} impact of the local density of states},}\ }\href
  {https://doi.org/10.1063/1.4847875} {\bibfield  {journal} {\bibinfo
  {journal} {The Journal of Chemical Physics}\ }\textbf {\bibinfo {volume}
  {139}},\ \bibinfo {pages} {244302} (\bibinfo {year} {2013})}\BibitemShut
  {NoStop}%
\bibitem [{\citenamefont {Li}\ \emph {et~al.}(2015{\natexlab{a}})\citenamefont
  {Li}, \citenamefont {Cushing}, \citenamefont {Meng}, \citenamefont {Senty},
  \citenamefont {Bristow},\ and\ \citenamefont {Wu}}]{Li2015}%
  \BibitemOpen
  \bibfield  {author} {\bibinfo {author} {\bibfnamefont {J.}~\bibnamefont
  {Li}}, \bibinfo {author} {\bibfnamefont {S.~K.}\ \bibnamefont {Cushing}},
  \bibinfo {author} {\bibfnamefont {F.}~\bibnamefont {Meng}}, \bibinfo {author}
  {\bibfnamefont {T.~R.}\ \bibnamefont {Senty}}, \bibinfo {author}
  {\bibfnamefont {A.~D.}\ \bibnamefont {Bristow}},\ and\ \bibinfo {author}
  {\bibfnamefont {N.}~\bibnamefont {Wu}},\ }\bibfield  {title} {\enquote
  {\bibinfo {title} {Plasmon-induced resonance energy transfer for solar energy
  conversion},}\ }\href {https://doi.org/10.1038/nphoton.2015.142} {\bibfield
  {journal} {\bibinfo  {journal} {Nature Photonics}\ }\textbf {\bibinfo
  {volume} {9}},\ \bibinfo {pages} {601--607} (\bibinfo {year}
  {2015}{\natexlab{a}})}\BibitemShut {NoStop}%
\bibitem [{\citenamefont {Marocico}, \citenamefont {Zhang},\ and\ \citenamefont
  {Bradley}(2016)}]{Marocico2016}%
  \BibitemOpen
  \bibfield  {author} {\bibinfo {author} {\bibfnamefont {C.~A.}\ \bibnamefont
  {Marocico}}, \bibinfo {author} {\bibfnamefont {X.}~\bibnamefont {Zhang}},\
  and\ \bibinfo {author} {\bibfnamefont {A.~L.}\ \bibnamefont {Bradley}},\
  }\bibfield  {title} {\enquote {\bibinfo {title} {A theoretical investigation
  of the influence of gold nanosphere size on the decay and energy transfer
  rates and efficiencies of quantum emitters},}\ }\href
  {https://doi.org/10.1063/1.4939206} {\bibfield  {journal} {\bibinfo
  {journal} {The Journal of Chemical Physics}\ }\textbf {\bibinfo {volume}
  {144}},\ \bibinfo {pages} {024108} (\bibinfo {year} {2016})}\BibitemShut
  {NoStop}%
\bibitem [{\citenamefont {Hsu}, \citenamefont {Ding},\ and\ \citenamefont
  {Schatz}(2017)}]{Hsu2017}%
  \BibitemOpen
  \bibfield  {author} {\bibinfo {author} {\bibfnamefont {L.-Y.}\ \bibnamefont
  {Hsu}}, \bibinfo {author} {\bibfnamefont {W.}~\bibnamefont {Ding}},\ and\
  \bibinfo {author} {\bibfnamefont {G.~C.}\ \bibnamefont {Schatz}},\ }\bibfield
   {title} {\enquote {\bibinfo {title} {Plasmon-{{Coupled Resonance Energy
  Transfer}}},}\ }\href {https://doi.org/10.1021/acs.jpclett.7b00526}
  {\bibfield  {journal} {\bibinfo  {journal} {The Journal of Physical Chemistry
  Letters}\ }\textbf {\bibinfo {volume} {8}},\ \bibinfo {pages} {2357--2367}
  (\bibinfo {year} {2017})}\BibitemShut {NoStop}%
\bibitem [{\citenamefont {Raether}(1988)}]{Raether1988}%
  \BibitemOpen
  \bibfield  {author} {\bibinfo {author} {\bibfnamefont {H.}~\bibnamefont
  {Raether}},\ }\href {https://doi.org/10.1007/BFb0048317} {\emph {\bibinfo
  {title} {Surface {{Plasmons}} on {{Smooth}} and {{Rough Surfaces}} and on
  {{Gratings}}}}},\ \bibinfo {series} {Springer {{Tracts}} in {{Modern
  Physics}}}, Vol.\ \bibinfo {volume} {111}\ (\bibinfo  {publisher} {Springer
  Berlin Heidelberg},\ \bibinfo {address} {Berlin, Heidelberg},\ \bibinfo
  {year} {1988})\BibitemShut {NoStop}%
\bibitem [{\citenamefont {Kelly}\ \emph {et~al.}(2003)\citenamefont {Kelly},
  \citenamefont {Coronado}, \citenamefont {Zhao},\ and\ \citenamefont
  {Schatz}}]{Kelly2003}%
  \BibitemOpen
  \bibfield  {author} {\bibinfo {author} {\bibfnamefont {K.~L.}\ \bibnamefont
  {Kelly}}, \bibinfo {author} {\bibfnamefont {E.}~\bibnamefont {Coronado}},
  \bibinfo {author} {\bibfnamefont {L.~L.}\ \bibnamefont {Zhao}},\ and\
  \bibinfo {author} {\bibfnamefont {G.~C.}\ \bibnamefont {Schatz}},\ }\bibfield
   {title} {\enquote {\bibinfo {title} {The {{Optical Properties}} of {{Metal
  Nanoparticles}}:\, {{The Influence}} of {{Size}}, {{Shape}}, and {{Dielectric
  Environment}}},}\ }\href {https://doi.org/10.1021/jp026731y} {\bibfield
  {journal} {\bibinfo  {journal} {The Journal of Physical Chemistry B}\
  }\textbf {\bibinfo {volume} {107}},\ \bibinfo {pages} {668--677} (\bibinfo
  {year} {2003})}\BibitemShut {NoStop}%
\bibitem [{\citenamefont {Maier}(2007)}]{Maier2007}%
  \BibitemOpen
  \bibfield  {author} {\bibinfo {author} {\bibfnamefont {S.~A.}\ \bibnamefont
  {Maier}},\ }\href {https://doi.org/10.1007/0-387-37825-1} {\emph {\bibinfo
  {title} {Plasmonics: {{Fundamentals}} and {{Applications}}}}}\ (\bibinfo
  {publisher} {Springer US},\ \bibinfo {address} {New York, NY},\ \bibinfo
  {year} {2007})\BibitemShut {NoStop}%
\bibitem [{\citenamefont {Shishkov}\ \emph {et~al.}(2016)\citenamefont
  {Shishkov}, \citenamefont {Andrianov}, \citenamefont {Pukhov},\ and\
  \citenamefont {Vinogradov}}]{Shishkov2016}%
  \BibitemOpen
  \bibfield  {author} {\bibinfo {author} {\bibfnamefont {V.~Y.}\ \bibnamefont
  {Shishkov}}, \bibinfo {author} {\bibfnamefont {E.~S.}\ \bibnamefont
  {Andrianov}}, \bibinfo {author} {\bibfnamefont {A.~A.}\ \bibnamefont
  {Pukhov}},\ and\ \bibinfo {author} {\bibfnamefont {A.~P.}\ \bibnamefont
  {Vinogradov}},\ }\bibfield  {title} {\enquote {\bibinfo {title} {Hermitian
  description of localized plasmons in dispersive dissipative subwavelength
  spherical nanostructures},}\ }\href
  {https://doi.org/10.1103/PhysRevB.94.235443} {\bibfield  {journal} {\bibinfo
  {journal} {Physical Review B}\ }\textbf {\bibinfo {volume} {94}},\ \bibinfo
  {pages} {235443} (\bibinfo {year} {2016})}\BibitemShut {NoStop}%
\bibitem [{\citenamefont {Miwa}\ and\ \citenamefont {Schatz}(2021)}]{Miwa2021}%
  \BibitemOpen
  \bibfield  {author} {\bibinfo {author} {\bibfnamefont {K.}~\bibnamefont
  {Miwa}}\ and\ \bibinfo {author} {\bibfnamefont {G.~C.}\ \bibnamefont
  {Schatz}},\ }\bibfield  {title} {\enquote {\bibinfo {title} {Quantum
  electrodynamics description of localized surface plasmons at a metal
  nanosphere},}\ }\href {https://doi.org/10.1103/PhysRevA.103.L041501}
  {\bibfield  {journal} {\bibinfo  {journal} {Physical Review A}\ }\textbf
  {\bibinfo {volume} {103}},\ \bibinfo {pages} {L041501} (\bibinfo {year}
  {2021})}\BibitemShut {NoStop}%
\bibitem [{\citenamefont {Reitz}\ \emph {et~al.}(2026)\citenamefont {Reitz},
  \citenamefont {Van Den~Wildenberg}, \citenamefont {Koner}, \citenamefont
  {Schatz},\ and\ \citenamefont {Yuen-Zhou}}]{Reitz2026}%
  \BibitemOpen
  \bibfield  {author} {\bibinfo {author} {\bibfnamefont {M.}~\bibnamefont
  {Reitz}}, \bibinfo {author} {\bibfnamefont {S.}~\bibnamefont {Van
  Den~Wildenberg}}, \bibinfo {author} {\bibfnamefont {A.}~\bibnamefont
  {Koner}}, \bibinfo {author} {\bibfnamefont {G.~C.}\ \bibnamefont {Schatz}},\
  and\ \bibinfo {author} {\bibfnamefont {J.}~\bibnamefont {Yuen-Zhou}},\
  }\bibfield  {title} {\enquote {\bibinfo {title} {Quantum {{Theory}} of
  {{Surface Lattice Resonances}}},}\ }\href
  {https://doi.org/10.1002/nap2.70008} {\bibfield  {journal} {\bibinfo
  {journal} {Nanophotonics}\ }\textbf {\bibinfo {volume} {15}},\ \bibinfo
  {pages} {e70008} (\bibinfo {year} {2026})}\BibitemShut {NoStop}%
\bibitem [{\citenamefont {Poddubny}\ \emph {et~al.}(2013)\citenamefont
  {Poddubny}, \citenamefont {Iorsh}, \citenamefont {Belov},\ and\ \citenamefont
  {Kivshar}}]{Poddubny2013}%
  \BibitemOpen
  \bibfield  {author} {\bibinfo {author} {\bibfnamefont {A.}~\bibnamefont
  {Poddubny}}, \bibinfo {author} {\bibfnamefont {I.}~\bibnamefont {Iorsh}},
  \bibinfo {author} {\bibfnamefont {P.}~\bibnamefont {Belov}},\ and\ \bibinfo
  {author} {\bibfnamefont {Y.}~\bibnamefont {Kivshar}},\ }\bibfield  {title}
  {\enquote {\bibinfo {title} {Hyperbolic metamaterials},}\ }\href
  {https://doi.org/10.1038/nphoton.2013.243} {\bibfield  {journal} {\bibinfo
  {journal} {Nature Photonics}\ }\textbf {\bibinfo {volume} {7}},\ \bibinfo
  {pages} {948--957} (\bibinfo {year} {2013})}\BibitemShut {NoStop}%
\bibitem [{\citenamefont {High}\ \emph {et~al.}(2015)\citenamefont {High},
  \citenamefont {Devlin}, \citenamefont {Dibos}, \citenamefont {Polking},
  \citenamefont {Wild}, \citenamefont {Perczel}, \citenamefont {De~Leon},
  \citenamefont {Lukin},\ and\ \citenamefont {Park}}]{High2015}%
  \BibitemOpen
  \bibfield  {author} {\bibinfo {author} {\bibfnamefont {A.~A.}\ \bibnamefont
  {High}}, \bibinfo {author} {\bibfnamefont {R.~C.}\ \bibnamefont {Devlin}},
  \bibinfo {author} {\bibfnamefont {A.}~\bibnamefont {Dibos}}, \bibinfo
  {author} {\bibfnamefont {M.}~\bibnamefont {Polking}}, \bibinfo {author}
  {\bibfnamefont {D.~S.}\ \bibnamefont {Wild}}, \bibinfo {author}
  {\bibfnamefont {J.}~\bibnamefont {Perczel}}, \bibinfo {author} {\bibfnamefont
  {N.~P.}\ \bibnamefont {De~Leon}}, \bibinfo {author} {\bibfnamefont {M.~D.}\
  \bibnamefont {Lukin}},\ and\ \bibinfo {author} {\bibfnamefont
  {H.}~\bibnamefont {Park}},\ }\bibfield  {title} {\enquote {\bibinfo {title}
  {Visible-frequency hyperbolic metasurface},}\ }\href
  {https://doi.org/10.1038/nature14477} {\bibfield  {journal} {\bibinfo
  {journal} {Nature}\ }\textbf {\bibinfo {volume} {522}},\ \bibinfo {pages}
  {192--196} (\bibinfo {year} {2015})}\BibitemShut {NoStop}%
\bibitem [{\citenamefont {Basov}, \citenamefont {Fogler},\ and\ \citenamefont
  {Garc{\'i}a De~Abajo}(2016)}]{Basov2016}%
  \BibitemOpen
  \bibfield  {author} {\bibinfo {author} {\bibfnamefont {D.~N.}\ \bibnamefont
  {Basov}}, \bibinfo {author} {\bibfnamefont {M.~M.}\ \bibnamefont {Fogler}},\
  and\ \bibinfo {author} {\bibfnamefont {F.~J.}\ \bibnamefont {Garc{\'i}a
  De~Abajo}},\ }\bibfield  {title} {\enquote {\bibinfo {title} {Polaritons in
  van der {{Waals}} materials},}\ }\href
  {https://doi.org/10.1126/science.aag1992} {\bibfield  {journal} {\bibinfo
  {journal} {Science}\ }\textbf {\bibinfo {volume} {354}},\ \bibinfo {pages}
  {aag1992} (\bibinfo {year} {2016})}\BibitemShut {NoStop}%
\bibitem [{\citenamefont {Low}\ \emph {et~al.}(2017)\citenamefont {Low},
  \citenamefont {Chaves}, \citenamefont {Caldwell}, \citenamefont {Kumar},
  \citenamefont {Fang}, \citenamefont {Avouris}, \citenamefont {Heinz},
  \citenamefont {Guinea}, \citenamefont {{Martin-Moreno}},\ and\ \citenamefont
  {Koppens}}]{Low2017}%
  \BibitemOpen
  \bibfield  {author} {\bibinfo {author} {\bibfnamefont {T.}~\bibnamefont
  {Low}}, \bibinfo {author} {\bibfnamefont {A.}~\bibnamefont {Chaves}},
  \bibinfo {author} {\bibfnamefont {J.~D.}\ \bibnamefont {Caldwell}}, \bibinfo
  {author} {\bibfnamefont {A.}~\bibnamefont {Kumar}}, \bibinfo {author}
  {\bibfnamefont {N.~X.}\ \bibnamefont {Fang}}, \bibinfo {author}
  {\bibfnamefont {P.}~\bibnamefont {Avouris}}, \bibinfo {author} {\bibfnamefont
  {T.~F.}\ \bibnamefont {Heinz}}, \bibinfo {author} {\bibfnamefont
  {F.}~\bibnamefont {Guinea}}, \bibinfo {author} {\bibfnamefont
  {L.}~\bibnamefont {{Martin-Moreno}}},\ and\ \bibinfo {author} {\bibfnamefont
  {F.}~\bibnamefont {Koppens}},\ }\bibfield  {title} {\enquote {\bibinfo
  {title} {Polaritons in layered two-dimensional materials},}\ }\href
  {https://doi.org/10.1038/nmat4792} {\bibfield  {journal} {\bibinfo  {journal}
  {Nature Materials}\ }\textbf {\bibinfo {volume} {16}},\ \bibinfo {pages}
  {182--194} (\bibinfo {year} {2017})}\BibitemShut {NoStop}%
\bibitem [{\citenamefont {Caldwell}\ \emph {et~al.}(2019)\citenamefont
  {Caldwell}, \citenamefont {Aharonovich}, \citenamefont {Cassabois},
  \citenamefont {Edgar}, \citenamefont {Gil},\ and\ \citenamefont
  {Basov}}]{Caldwell2019}%
  \BibitemOpen
  \bibfield  {author} {\bibinfo {author} {\bibfnamefont {J.~D.}\ \bibnamefont
  {Caldwell}}, \bibinfo {author} {\bibfnamefont {I.}~\bibnamefont
  {Aharonovich}}, \bibinfo {author} {\bibfnamefont {G.}~\bibnamefont
  {Cassabois}}, \bibinfo {author} {\bibfnamefont {J.~H.}\ \bibnamefont
  {Edgar}}, \bibinfo {author} {\bibfnamefont {B.}~\bibnamefont {Gil}},\ and\
  \bibinfo {author} {\bibfnamefont {D.~N.}\ \bibnamefont {Basov}},\ }\bibfield
  {title} {\enquote {\bibinfo {title} {Photonics with hexagonal boron
  nitride},}\ }\href {https://doi.org/10.1038/s41578-019-0124-1} {\bibfield
  {journal} {\bibinfo  {journal} {Nature Reviews Materials}\ }\textbf {\bibinfo
  {volume} {4}},\ \bibinfo {pages} {552--567} (\bibinfo {year}
  {2019})}\BibitemShut {NoStop}%
\bibitem [{\citenamefont {Zhang}\ \emph {et~al.}(2021)\citenamefont {Zhang},
  \citenamefont {Hu}, \citenamefont {Ma}, \citenamefont {Li}, \citenamefont
  {Krasnok}, \citenamefont {Hillenbrand}, \citenamefont {Al{\`u}},\ and\
  \citenamefont {Qiu}}]{Zhang2021}%
  \BibitemOpen
  \bibfield  {author} {\bibinfo {author} {\bibfnamefont {Q.}~\bibnamefont
  {Zhang}}, \bibinfo {author} {\bibfnamefont {G.}~\bibnamefont {Hu}}, \bibinfo
  {author} {\bibfnamefont {W.}~\bibnamefont {Ma}}, \bibinfo {author}
  {\bibfnamefont {P.}~\bibnamefont {Li}}, \bibinfo {author} {\bibfnamefont
  {A.}~\bibnamefont {Krasnok}}, \bibinfo {author} {\bibfnamefont
  {R.}~\bibnamefont {Hillenbrand}}, \bibinfo {author} {\bibfnamefont
  {A.}~\bibnamefont {Al{\`u}}},\ and\ \bibinfo {author} {\bibfnamefont {C.-W.}\
  \bibnamefont {Qiu}},\ }\bibfield  {title} {\enquote {\bibinfo {title}
  {Interface nano-optics with van der {{Waals}} polaritons},}\ }\href
  {https://doi.org/10.1038/s41586-021-03581-5} {\bibfield  {journal} {\bibinfo
  {journal} {Nature}\ }\textbf {\bibinfo {volume} {597}},\ \bibinfo {pages}
  {187--195} (\bibinfo {year} {2021})}\BibitemShut {NoStop}%
\bibitem [{\citenamefont {Jia}\ \emph {et~al.}(2022)\citenamefont {Jia},
  \citenamefont {Luo}, \citenamefont {Wang}, \citenamefont {Ma}, \citenamefont
  {Liu}, \citenamefont {Dai},\ and\ \citenamefont {Asgari}}]{Jia2022}%
  \BibitemOpen
  \bibfield  {author} {\bibinfo {author} {\bibfnamefont {G.}~\bibnamefont
  {Jia}}, \bibinfo {author} {\bibfnamefont {J.}~\bibnamefont {Luo}}, \bibinfo
  {author} {\bibfnamefont {H.}~\bibnamefont {Wang}}, \bibinfo {author}
  {\bibfnamefont {Q.}~\bibnamefont {Ma}}, \bibinfo {author} {\bibfnamefont
  {Q.}~\bibnamefont {Liu}}, \bibinfo {author} {\bibfnamefont {H.}~\bibnamefont
  {Dai}},\ and\ \bibinfo {author} {\bibfnamefont {R.}~\bibnamefont {Asgari}},\
  }\bibfield  {title} {\enquote {\bibinfo {title} {Two-dimensional natural
  hyperbolic materials: From polaritons modulation to applications},}\ }\href
  {https://doi.org/10.1039/D2NR04181B} {\bibfield  {journal} {\bibinfo
  {journal} {Nanoscale}\ }\textbf {\bibinfo {volume} {14}},\ \bibinfo {pages}
  {17096--17118} (\bibinfo {year} {2022})}\BibitemShut {NoStop}%
\bibitem [{\citenamefont {Basov}\ \emph {et~al.}(2021)\citenamefont {Basov},
  \citenamefont {Asenjo-Garcia}, \citenamefont {Schuck}, \citenamefont {Zhu},\
  and\ \citenamefont {Rubio}}]{Basov2021}%
  \BibitemOpen
  \bibfield  {author} {\bibinfo {author} {\bibfnamefont {D.~N.}\ \bibnamefont
  {Basov}}, \bibinfo {author} {\bibfnamefont {A.}~\bibnamefont
  {Asenjo-Garcia}}, \bibinfo {author} {\bibfnamefont {P.~J.}\ \bibnamefont
  {Schuck}}, \bibinfo {author} {\bibfnamefont {X.}~\bibnamefont {Zhu}},\ and\
  \bibinfo {author} {\bibfnamefont {A.}~\bibnamefont {Rubio}},\ }\bibfield
  {title} {\enquote {\bibinfo {title} {Polariton panorama},}\ }\href
  {https://doi.org/10.1515/nanoph-2020-0449} {\bibfield  {journal} {\bibinfo
  {journal} {Nanophotonics}\ }\textbf {\bibinfo {volume} {10}},\ \bibinfo
  {pages} {549--577} (\bibinfo {year} {2021})}\BibitemShut {NoStop}%
\bibitem [{\citenamefont {Wu}\ \emph {et~al.}(2022)\citenamefont {Wu},
  \citenamefont {Duan}, \citenamefont {Ma}, \citenamefont {Ou}, \citenamefont
  {Li}, \citenamefont {{Alonso-Gonz{\'a}lez}}, \citenamefont {Caldwell},\ and\
  \citenamefont {Bao}}]{Wu2022}%
  \BibitemOpen
  \bibfield  {author} {\bibinfo {author} {\bibfnamefont {Y.}~\bibnamefont
  {Wu}}, \bibinfo {author} {\bibfnamefont {J.}~\bibnamefont {Duan}}, \bibinfo
  {author} {\bibfnamefont {W.}~\bibnamefont {Ma}}, \bibinfo {author}
  {\bibfnamefont {Q.}~\bibnamefont {Ou}}, \bibinfo {author} {\bibfnamefont
  {P.}~\bibnamefont {Li}}, \bibinfo {author} {\bibfnamefont {P.}~\bibnamefont
  {{Alonso-Gonz{\'a}lez}}}, \bibinfo {author} {\bibfnamefont {J.~D.}\
  \bibnamefont {Caldwell}},\ and\ \bibinfo {author} {\bibfnamefont
  {Q.}~\bibnamefont {Bao}},\ }\bibfield  {title} {\enquote {\bibinfo {title}
  {Manipulating polaritons at the extreme scale in van der {{Waals}}
  materials},}\ }\href {https://doi.org/10.1038/s42254-022-00472-0} {\bibfield
  {journal} {\bibinfo  {journal} {Nature Reviews Physics}\ }\textbf {\bibinfo
  {volume} {4}},\ \bibinfo {pages} {578--594} (\bibinfo {year}
  {2022})}\BibitemShut {NoStop}%
\bibitem [{\citenamefont {Galiffi}\ \emph {et~al.}(2023)\citenamefont
  {Galiffi}, \citenamefont {Carini}, \citenamefont {Ni}, \citenamefont
  {{\'A}lvarez-P{\'e}rez}, \citenamefont {Yves}, \citenamefont {Renzi},
  \citenamefont {Nolen}, \citenamefont {Wasserroth}, \citenamefont {Wolf},
  \citenamefont {{Alonso-Gonzalez}}, \citenamefont {Paarmann},\ and\
  \citenamefont {Al{\`u}}}]{Galiffi2023}%
  \BibitemOpen
  \bibfield  {author} {\bibinfo {author} {\bibfnamefont {E.}~\bibnamefont
  {Galiffi}}, \bibinfo {author} {\bibfnamefont {G.}~\bibnamefont {Carini}},
  \bibinfo {author} {\bibfnamefont {X.}~\bibnamefont {Ni}}, \bibinfo {author}
  {\bibfnamefont {G.}~\bibnamefont {{\'A}lvarez-P{\'e}rez}}, \bibinfo {author}
  {\bibfnamefont {S.}~\bibnamefont {Yves}}, \bibinfo {author} {\bibfnamefont
  {E.~M.}\ \bibnamefont {Renzi}}, \bibinfo {author} {\bibfnamefont
  {R.}~\bibnamefont {Nolen}}, \bibinfo {author} {\bibfnamefont
  {S.}~\bibnamefont {Wasserroth}}, \bibinfo {author} {\bibfnamefont
  {M.}~\bibnamefont {Wolf}}, \bibinfo {author} {\bibfnamefont {P.}~\bibnamefont
  {{Alonso-Gonzalez}}}, \bibinfo {author} {\bibfnamefont {A.}~\bibnamefont
  {Paarmann}},\ and\ \bibinfo {author} {\bibfnamefont {A.}~\bibnamefont
  {Al{\`u}}},\ }\bibfield  {title} {\enquote {\bibinfo {title} {Extreme light
  confinement and control in low-symmetry phonon-polaritonic crystals},}\
  }\href {https://doi.org/10.1038/s41578-023-00620-7} {\bibfield  {journal}
  {\bibinfo  {journal} {Nature Reviews Materials}\ }\textbf {\bibinfo {volume}
  {9}},\ \bibinfo {pages} {9--28} (\bibinfo {year} {2023})}\BibitemShut
  {NoStop}%
\bibitem [{\citenamefont {Wang}\ \emph {et~al.}(2024)\citenamefont {Wang},
  \citenamefont {Kumar}, \citenamefont {Dai}, \citenamefont {Lin},
  \citenamefont {Jacob}, \citenamefont {Oh}, \citenamefont {Menon},
  \citenamefont {Narimanov}, \citenamefont {Kim}, \citenamefont {Wang},
  \citenamefont {Avouris}, \citenamefont {Martin~Moreno}, \citenamefont
  {Caldwell},\ and\ \citenamefont {Low}}]{Wang2024}%
  \BibitemOpen
  \bibfield  {author} {\bibinfo {author} {\bibfnamefont {H.}~\bibnamefont
  {Wang}}, \bibinfo {author} {\bibfnamefont {A.}~\bibnamefont {Kumar}},
  \bibinfo {author} {\bibfnamefont {S.}~\bibnamefont {Dai}}, \bibinfo {author}
  {\bibfnamefont {X.}~\bibnamefont {Lin}}, \bibinfo {author} {\bibfnamefont
  {Z.}~\bibnamefont {Jacob}}, \bibinfo {author} {\bibfnamefont {S.-H.}\
  \bibnamefont {Oh}}, \bibinfo {author} {\bibfnamefont {V.}~\bibnamefont
  {Menon}}, \bibinfo {author} {\bibfnamefont {E.}~\bibnamefont {Narimanov}},
  \bibinfo {author} {\bibfnamefont {Y.~D.}\ \bibnamefont {Kim}}, \bibinfo
  {author} {\bibfnamefont {J.-P.}\ \bibnamefont {Wang}}, \bibinfo {author}
  {\bibfnamefont {P.}~\bibnamefont {Avouris}}, \bibinfo {author} {\bibfnamefont
  {L.}~\bibnamefont {Martin~Moreno}}, \bibinfo {author} {\bibfnamefont
  {J.}~\bibnamefont {Caldwell}},\ and\ \bibinfo {author} {\bibfnamefont
  {T.}~\bibnamefont {Low}},\ }\bibfield  {title} {\enquote {\bibinfo {title}
  {Planar hyperbolic polaritons in {{2D}} van der {{Waals}} materials},}\
  }\href {https://doi.org/10.1038/s41467-023-43992-8} {\bibfield  {journal}
  {\bibinfo  {journal} {Nature Communications}\ }\textbf {\bibinfo {volume}
  {15}},\ \bibinfo {pages} {69} (\bibinfo {year} {2024})}\BibitemShut {NoStop}%
\bibitem [{\citenamefont {Basov}\ \emph {et~al.}(2025)\citenamefont {Basov},
  \citenamefont {Asenjo-Garcia}, \citenamefont {Schuck}, \citenamefont {Zhu},
  \citenamefont {Rubio}, \citenamefont {Cavalleri}, \citenamefont {Delor},
  \citenamefont {Fogler},\ and\ \citenamefont {Liu}}]{Basov2025}%
  \BibitemOpen
  \bibfield  {author} {\bibinfo {author} {\bibfnamefont {D.}~\bibnamefont
  {Basov}}, \bibinfo {author} {\bibfnamefont {A.}~\bibnamefont
  {Asenjo-Garcia}}, \bibinfo {author} {\bibfnamefont {P.~J.}\ \bibnamefont
  {Schuck}}, \bibinfo {author} {\bibfnamefont {X.}~\bibnamefont {Zhu}},
  \bibinfo {author} {\bibfnamefont {A.}~\bibnamefont {Rubio}}, \bibinfo
  {author} {\bibfnamefont {A.}~\bibnamefont {Cavalleri}}, \bibinfo {author}
  {\bibfnamefont {M.}~\bibnamefont {Delor}}, \bibinfo {author} {\bibfnamefont
  {M.~M.}\ \bibnamefont {Fogler}},\ and\ \bibinfo {author} {\bibfnamefont
  {M.}~\bibnamefont {Liu}},\ }\bibfield  {title} {\enquote {\bibinfo {title}
  {Polaritonic quantum matter},}\ }\href
  {https://doi.org/10.1515/nanoph-2025-0001} {\bibfield  {journal} {\bibinfo
  {journal} {Nanophotonics}\ }\textbf {\bibinfo {volume} {14}},\ \bibinfo
  {pages} {3723--3760} (\bibinfo {year} {2025})}\BibitemShut {NoStop}%
\bibitem [{\citenamefont {Zhou}\ \emph {et~al.}(2026)\citenamefont {Zhou},
  \citenamefont {Guo}, \citenamefont {{Tarazaga Mart{\'i}n-Luengo}},
  \citenamefont {Lanza}, \citenamefont {{\'A}lvarez-P{\'e}rez}, \citenamefont
  {Yu}, \citenamefont {Li}, \citenamefont {Xia}, \citenamefont
  {{\'A}lvarez~Cuervo}, \citenamefont {Duan}, \citenamefont {Wang},
  \citenamefont {{Mart{\'i}n-S{\'a}nchez}}, \citenamefont {Nikitin},
  \citenamefont {Yao}, \citenamefont {Li}, \citenamefont {{Alonso-Gonzalez}},\
  and\ \citenamefont {Duan}}]{Zhou2026}%
  \BibitemOpen
  \bibfield  {author} {\bibinfo {author} {\bibfnamefont {Y.}~\bibnamefont
  {Zhou}}, \bibinfo {author} {\bibfnamefont {Z.}~\bibnamefont {Guo}}, \bibinfo
  {author} {\bibfnamefont {A.}~\bibnamefont {{Tarazaga Mart{\'i}n-Luengo}}},
  \bibinfo {author} {\bibfnamefont {C.}~\bibnamefont {Lanza}}, \bibinfo
  {author} {\bibfnamefont {G.}~\bibnamefont {{\'A}lvarez-P{\'e}rez}}, \bibinfo
  {author} {\bibfnamefont {C.}~\bibnamefont {Yu}}, \bibinfo {author}
  {\bibfnamefont {C.}~\bibnamefont {Li}}, \bibinfo {author} {\bibfnamefont
  {W.}~\bibnamefont {Xia}}, \bibinfo {author} {\bibfnamefont {J.}~\bibnamefont
  {{\'A}lvarez~Cuervo}}, \bibinfo {author} {\bibfnamefont {X.}~\bibnamefont
  {Duan}}, \bibinfo {author} {\bibfnamefont {Y.}~\bibnamefont {Wang}}, \bibinfo
  {author} {\bibfnamefont {J.}~\bibnamefont {{Mart{\'i}n-S{\'a}nchez}}},
  \bibinfo {author} {\bibfnamefont {A.~Y.}\ \bibnamefont {Nikitin}}, \bibinfo
  {author} {\bibfnamefont {Y.}~\bibnamefont {Yao}}, \bibinfo {author}
  {\bibfnamefont {J.}~\bibnamefont {Li}}, \bibinfo {author} {\bibfnamefont
  {P.}~\bibnamefont {{Alonso-Gonzalez}}},\ and\ \bibinfo {author}
  {\bibfnamefont {J.}~\bibnamefont {Duan}},\ }\bibfield  {title} {\enquote
  {\bibinfo {title} {Fundamental optical phenomena of strongly anisotropic
  polaritons at the nanoscale},}\ }\href
  {https://doi.org/10.1038/s41565-025-02039-3} {\bibfield  {journal} {\bibinfo
  {journal} {Nature Nanotechnology}\ }\textbf {\bibinfo {volume} {21}},\
  \bibinfo {pages} {23--38} (\bibinfo {year} {2026})}\BibitemShut {NoStop}%
\bibitem [{\citenamefont {Hu}\ \emph {et~al.}(2026)\citenamefont {Hu},
  \citenamefont {Teng}, \citenamefont {Chen}, \citenamefont {Xue},
  \citenamefont {Sun}, \citenamefont {Garc{\'i}a De~Abajo},\ and\ \citenamefont
  {Dai}}]{Hu2026}%
  \BibitemOpen
  \bibfield  {author} {\bibinfo {author} {\bibfnamefont {H.}~\bibnamefont
  {Hu}}, \bibinfo {author} {\bibfnamefont {H.}~\bibnamefont {Teng}}, \bibinfo
  {author} {\bibfnamefont {N.}~\bibnamefont {Chen}}, \bibinfo {author}
  {\bibfnamefont {Z.}~\bibnamefont {Xue}}, \bibinfo {author} {\bibfnamefont
  {Z.}~\bibnamefont {Sun}}, \bibinfo {author} {\bibfnamefont {F.~J.}\
  \bibnamefont {Garc{\'i}a De~Abajo}},\ and\ \bibinfo {author} {\bibfnamefont
  {Q.}~\bibnamefont {Dai}},\ }\bibfield  {title} {\enquote {\bibinfo {title}
  {Transport of polaritons in hyperbolic media},}\ }\href
  {https://doi.org/10.1038/s41578-026-00907-5} {\bibfield  {journal} {\bibinfo
  {journal} {Nature Reviews Materials}\ }\textbf {\bibinfo {volume} {11}},\
  \bibinfo {pages} {554--574} (\bibinfo {year} {2026})}\BibitemShut {NoStop}%
\bibitem [{\citenamefont {Dai}\ \emph {et~al.}(2014)\citenamefont {Dai},
  \citenamefont {Fei}, \citenamefont {Ma}, \citenamefont {Rodin}, \citenamefont
  {Wagner}, \citenamefont {McLeod}, \citenamefont {Liu}, \citenamefont
  {Gannett}, \citenamefont {Regan}, \citenamefont {Watanabe}, \citenamefont
  {Taniguchi}, \citenamefont {Thiemens}, \citenamefont {Dominguez},
  \citenamefont {Neto}, \citenamefont {Zettl}, \citenamefont {Keilmann},
  \citenamefont {{Jarillo-Herrero}}, \citenamefont {Fogler},\ and\
  \citenamefont {Basov}}]{Dai2014}%
  \BibitemOpen
  \bibfield  {author} {\bibinfo {author} {\bibfnamefont {S.}~\bibnamefont
  {Dai}}, \bibinfo {author} {\bibfnamefont {Z.}~\bibnamefont {Fei}}, \bibinfo
  {author} {\bibfnamefont {Q.}~\bibnamefont {Ma}}, \bibinfo {author}
  {\bibfnamefont {A.~S.}\ \bibnamefont {Rodin}}, \bibinfo {author}
  {\bibfnamefont {M.}~\bibnamefont {Wagner}}, \bibinfo {author} {\bibfnamefont
  {A.~S.}\ \bibnamefont {McLeod}}, \bibinfo {author} {\bibfnamefont {M.~K.}\
  \bibnamefont {Liu}}, \bibinfo {author} {\bibfnamefont {W.}~\bibnamefont
  {Gannett}}, \bibinfo {author} {\bibfnamefont {W.}~\bibnamefont {Regan}},
  \bibinfo {author} {\bibfnamefont {K.}~\bibnamefont {Watanabe}}, \bibinfo
  {author} {\bibfnamefont {T.}~\bibnamefont {Taniguchi}}, \bibinfo {author}
  {\bibfnamefont {M.}~\bibnamefont {Thiemens}}, \bibinfo {author}
  {\bibfnamefont {G.}~\bibnamefont {Dominguez}}, \bibinfo {author}
  {\bibfnamefont {A.~H.~C.}\ \bibnamefont {Neto}}, \bibinfo {author}
  {\bibfnamefont {A.}~\bibnamefont {Zettl}}, \bibinfo {author} {\bibfnamefont
  {F.}~\bibnamefont {Keilmann}}, \bibinfo {author} {\bibfnamefont
  {P.}~\bibnamefont {{Jarillo-Herrero}}}, \bibinfo {author} {\bibfnamefont
  {M.~M.}\ \bibnamefont {Fogler}},\ and\ \bibinfo {author} {\bibfnamefont
  {D.~N.}\ \bibnamefont {Basov}},\ }\bibfield  {title} {\enquote {\bibinfo
  {title} {Tunable {{Phonon Polaritons}} in {{Atomically Thin}} van der {{Waals
  Crystals}} of {{Boron Nitride}}},}\ }\href
  {https://doi.org/10.1126/science.1246833} {\bibfield  {journal} {\bibinfo
  {journal} {Science}\ }\textbf {\bibinfo {volume} {343}},\ \bibinfo {pages}
  {1125--1129} (\bibinfo {year} {2014})}\BibitemShut {NoStop}%
\bibitem [{\citenamefont {Caldwell}\ \emph {et~al.}(2014)\citenamefont
  {Caldwell}, \citenamefont {Kretinin}, \citenamefont {Chen}, \citenamefont
  {Giannini}, \citenamefont {Fogler}, \citenamefont {Francescato},
  \citenamefont {Ellis}, \citenamefont {Tischler}, \citenamefont {Woods},
  \citenamefont {Giles}, \citenamefont {Hong}, \citenamefont {Watanabe},
  \citenamefont {Taniguchi}, \citenamefont {Maier},\ and\ \citenamefont
  {Novoselov}}]{Caldwell2014}%
  \BibitemOpen
  \bibfield  {author} {\bibinfo {author} {\bibfnamefont {J.~D.}\ \bibnamefont
  {Caldwell}}, \bibinfo {author} {\bibfnamefont {A.~V.}\ \bibnamefont
  {Kretinin}}, \bibinfo {author} {\bibfnamefont {Y.}~\bibnamefont {Chen}},
  \bibinfo {author} {\bibfnamefont {V.}~\bibnamefont {Giannini}}, \bibinfo
  {author} {\bibfnamefont {M.~M.}\ \bibnamefont {Fogler}}, \bibinfo {author}
  {\bibfnamefont {Y.}~\bibnamefont {Francescato}}, \bibinfo {author}
  {\bibfnamefont {C.~T.}\ \bibnamefont {Ellis}}, \bibinfo {author}
  {\bibfnamefont {J.~G.}\ \bibnamefont {Tischler}}, \bibinfo {author}
  {\bibfnamefont {C.~R.}\ \bibnamefont {Woods}}, \bibinfo {author}
  {\bibfnamefont {A.~J.}\ \bibnamefont {Giles}}, \bibinfo {author}
  {\bibfnamefont {M.}~\bibnamefont {Hong}}, \bibinfo {author} {\bibfnamefont
  {K.}~\bibnamefont {Watanabe}}, \bibinfo {author} {\bibfnamefont
  {T.}~\bibnamefont {Taniguchi}}, \bibinfo {author} {\bibfnamefont {S.~A.}\
  \bibnamefont {Maier}},\ and\ \bibinfo {author} {\bibfnamefont {K.~S.}\
  \bibnamefont {Novoselov}},\ }\bibfield  {title} {\enquote {\bibinfo {title}
  {Sub-diffractional volume-confined polaritons in the natural hyperbolic
  material hexagonal boron nitride},}\ }\href
  {https://doi.org/10.1038/ncomms6221} {\bibfield  {journal} {\bibinfo
  {journal} {Nature Communications}\ }\textbf {\bibinfo {volume} {5}},\
  \bibinfo {pages} {5221} (\bibinfo {year} {2014})}\BibitemShut {NoStop}%
\bibitem [{\citenamefont {Jacob}(2014)}]{Jacob2014}%
  \BibitemOpen
  \bibfield  {author} {\bibinfo {author} {\bibfnamefont {Z.}~\bibnamefont
  {Jacob}},\ }\bibfield  {title} {\enquote {\bibinfo {title} {Hyperbolic
  phonon--polaritons},}\ }\href {https://doi.org/10.1038/nmat4149} {\bibfield
  {journal} {\bibinfo  {journal} {Nature Materials}\ }\textbf {\bibinfo
  {volume} {13}},\ \bibinfo {pages} {1081--1083} (\bibinfo {year}
  {2014})}\BibitemShut {NoStop}%
\bibitem [{\citenamefont {Caldwell}\ \emph {et~al.}(2015)\citenamefont
  {Caldwell}, \citenamefont {Lindsay}, \citenamefont {Giannini}, \citenamefont
  {Vurgaftman}, \citenamefont {Reinecke}, \citenamefont {Maier},\ and\
  \citenamefont {Glembocki}}]{Caldwell2015}%
  \BibitemOpen
  \bibfield  {author} {\bibinfo {author} {\bibfnamefont {J.~D.}\ \bibnamefont
  {Caldwell}}, \bibinfo {author} {\bibfnamefont {L.}~\bibnamefont {Lindsay}},
  \bibinfo {author} {\bibfnamefont {V.}~\bibnamefont {Giannini}}, \bibinfo
  {author} {\bibfnamefont {I.}~\bibnamefont {Vurgaftman}}, \bibinfo {author}
  {\bibfnamefont {T.~L.}\ \bibnamefont {Reinecke}}, \bibinfo {author}
  {\bibfnamefont {S.~A.}\ \bibnamefont {Maier}},\ and\ \bibinfo {author}
  {\bibfnamefont {O.~J.}\ \bibnamefont {Glembocki}},\ }\bibfield  {title}
  {\enquote {\bibinfo {title} {Low-loss, infrared and terahertz nanophotonics
  using surface phonon polaritons},}\ }\href
  {https://doi.org/10.1515/nanoph-2014-0003} {\bibfield  {journal} {\bibinfo
  {journal} {Nanophotonics}\ }\textbf {\bibinfo {volume} {4}},\ \bibinfo
  {pages} {44--68} (\bibinfo {year} {2015})}\BibitemShut {NoStop}%
\bibitem [{\citenamefont {Li}\ \emph {et~al.}(2015{\natexlab{b}})\citenamefont
  {Li}, \citenamefont {Lewin}, \citenamefont {Kretinin}, \citenamefont
  {Caldwell}, \citenamefont {Novoselov}, \citenamefont {Taniguchi},
  \citenamefont {Watanabe}, \citenamefont {Gaussmann},\ and\ \citenamefont
  {Taubner}}]{Li2015a}%
  \BibitemOpen
  \bibfield  {author} {\bibinfo {author} {\bibfnamefont {P.}~\bibnamefont
  {Li}}, \bibinfo {author} {\bibfnamefont {M.}~\bibnamefont {Lewin}}, \bibinfo
  {author} {\bibfnamefont {A.~V.}\ \bibnamefont {Kretinin}}, \bibinfo {author}
  {\bibfnamefont {J.~D.}\ \bibnamefont {Caldwell}}, \bibinfo {author}
  {\bibfnamefont {K.~S.}\ \bibnamefont {Novoselov}}, \bibinfo {author}
  {\bibfnamefont {T.}~\bibnamefont {Taniguchi}}, \bibinfo {author}
  {\bibfnamefont {K.}~\bibnamefont {Watanabe}}, \bibinfo {author}
  {\bibfnamefont {F.}~\bibnamefont {Gaussmann}},\ and\ \bibinfo {author}
  {\bibfnamefont {T.}~\bibnamefont {Taubner}},\ }\bibfield  {title} {\enquote
  {\bibinfo {title} {Hyperbolic phonon-polaritons in boron nitride for
  near-field optical imaging and focusing},}\ }\href
  {https://doi.org/10.1038/ncomms8507} {\bibfield  {journal} {\bibinfo
  {journal} {Nature Communications}\ }\textbf {\bibinfo {volume} {6}},\
  \bibinfo {pages} {7507} (\bibinfo {year} {2015}{\natexlab{b}})}\BibitemShut
  {NoStop}%
\bibitem [{\citenamefont {Autore}\ \emph {et~al.}(2017)\citenamefont {Autore},
  \citenamefont {Li}, \citenamefont {Dolado}, \citenamefont {{Alfaro-Mozaz}},
  \citenamefont {Esteban}, \citenamefont {Atxabal}, \citenamefont {Casanova},
  \citenamefont {Hueso}, \citenamefont {{Alonso-Gonz{\'a}lez}}, \citenamefont
  {Aizpurua}, \citenamefont {Nikitin}, \citenamefont {V{\'e}lez},\ and\
  \citenamefont {Hillenbrand}}]{Autore2017}%
  \BibitemOpen
  \bibfield  {author} {\bibinfo {author} {\bibfnamefont {M.}~\bibnamefont
  {Autore}}, \bibinfo {author} {\bibfnamefont {P.}~\bibnamefont {Li}}, \bibinfo
  {author} {\bibfnamefont {I.}~\bibnamefont {Dolado}}, \bibinfo {author}
  {\bibfnamefont {F.~J.}\ \bibnamefont {{Alfaro-Mozaz}}}, \bibinfo {author}
  {\bibfnamefont {R.}~\bibnamefont {Esteban}}, \bibinfo {author} {\bibfnamefont
  {A.}~\bibnamefont {Atxabal}}, \bibinfo {author} {\bibfnamefont
  {F.}~\bibnamefont {Casanova}}, \bibinfo {author} {\bibfnamefont {L.~E.}\
  \bibnamefont {Hueso}}, \bibinfo {author} {\bibfnamefont {P.}~\bibnamefont
  {{Alonso-Gonz{\'a}lez}}}, \bibinfo {author} {\bibfnamefont {J.}~\bibnamefont
  {Aizpurua}}, \bibinfo {author} {\bibfnamefont {A.~Y.}\ \bibnamefont
  {Nikitin}}, \bibinfo {author} {\bibfnamefont {S.}~\bibnamefont {V{\'e}lez}},\
  and\ \bibinfo {author} {\bibfnamefont {R.}~\bibnamefont {Hillenbrand}},\
  }\bibfield  {title} {\enquote {\bibinfo {title} {Boron nitride nanoresonators
  for phonon-enhanced molecular vibrational spectroscopy at the strong coupling
  limit},}\ }\href {https://doi.org/10.1038/lsa.2017.172} {\bibfield  {journal}
  {\bibinfo  {journal} {Light: Science \& Applications}\ }\textbf {\bibinfo
  {volume} {7}},\ \bibinfo {pages} {17172--17172} (\bibinfo {year}
  {2017})}\BibitemShut {NoStop}%
\bibitem [{\citenamefont {Hutchins}\ \emph {et~al.}(2025)\citenamefont
  {Hutchins}, \citenamefont {Zare}, \citenamefont {Hirt}, \citenamefont
  {Tomko}, \citenamefont {Matson}, \citenamefont {{Diaz-Granados}},
  \citenamefont {Long}, \citenamefont {He}, \citenamefont {Pfeifer},
  \citenamefont {Li}, \citenamefont {Edgar}, \citenamefont {Maria},
  \citenamefont {Caldwell},\ and\ \citenamefont {Hopkins}}]{Hutchins2025}%
  \BibitemOpen
  \bibfield  {author} {\bibinfo {author} {\bibfnamefont {W.}~\bibnamefont
  {Hutchins}}, \bibinfo {author} {\bibfnamefont {S.}~\bibnamefont {Zare}},
  \bibinfo {author} {\bibfnamefont {D.~M.}\ \bibnamefont {Hirt}}, \bibinfo
  {author} {\bibfnamefont {J.~A.}\ \bibnamefont {Tomko}}, \bibinfo {author}
  {\bibfnamefont {J.~R.}\ \bibnamefont {Matson}}, \bibinfo {author}
  {\bibfnamefont {K.}~\bibnamefont {{Diaz-Granados}}}, \bibinfo {author}
  {\bibfnamefont {M.}~\bibnamefont {Long}}, \bibinfo {author} {\bibfnamefont
  {M.}~\bibnamefont {He}}, \bibinfo {author} {\bibfnamefont {T.}~\bibnamefont
  {Pfeifer}}, \bibinfo {author} {\bibfnamefont {J.}~\bibnamefont {Li}},
  \bibinfo {author} {\bibfnamefont {J.~H.}\ \bibnamefont {Edgar}}, \bibinfo
  {author} {\bibfnamefont {J.-P.}\ \bibnamefont {Maria}}, \bibinfo {author}
  {\bibfnamefont {J.~D.}\ \bibnamefont {Caldwell}},\ and\ \bibinfo {author}
  {\bibfnamefont {P.~E.}\ \bibnamefont {Hopkins}},\ }\bibfield  {title}
  {\enquote {\bibinfo {title} {Ultrafast evanescent heat transfer across solid
  interfaces via hyperbolic phonon--polariton modes in hexagonal boron
  nitride},}\ }\href {https://doi.org/10.1038/s41563-025-02154-5} {\bibfield
  {journal} {\bibinfo  {journal} {Nature Materials}\ }\textbf {\bibinfo
  {volume} {24}},\ \bibinfo {pages} {698--706} (\bibinfo {year}
  {2025})}\BibitemShut {NoStop}%
\bibitem [{\citenamefont {Narimanov}\ and\ \citenamefont
  {Demler}(2024)}]{Narimanov2024}%
  \BibitemOpen
  \bibfield  {author} {\bibinfo {author} {\bibfnamefont {E.~E.}\ \bibnamefont
  {Narimanov}}\ and\ \bibinfo {author} {\bibfnamefont {E.~A.}\ \bibnamefont
  {Demler}},\ }\href {https://doi.org/10.48550/arXiv.2412.14098} {\enquote
  {\bibinfo {title} {Hyperbolic {{Quantum Processor}}},}\ } (\bibinfo {year}
  {2024}),\ \Eprint {https://arxiv.org/abs/2412.14098} {arXiv:2412.14098
  [quant-ph]} \BibitemShut {NoStop}%
\bibitem [{\citenamefont {Ma}\ \emph {et~al.}(2018)\citenamefont {Ma},
  \citenamefont {{Alonso-Gonz{\'a}lez}}, \citenamefont {Li}, \citenamefont
  {Nikitin}, \citenamefont {Yuan}, \citenamefont {{Mart{\'i}n-S{\'a}nchez}},
  \citenamefont {{Taboada-Guti{\'e}rrez}}, \citenamefont {Amenabar},
  \citenamefont {Li}, \citenamefont {V{\'e}lez}, \citenamefont {Tollan},
  \citenamefont {Dai}, \citenamefont {Zhang}, \citenamefont {Sriram},
  \citenamefont {{Kalantar-Zadeh}}, \citenamefont {Lee}, \citenamefont
  {Hillenbrand},\ and\ \citenamefont {Bao}}]{Ma2018}%
  \BibitemOpen
  \bibfield  {author} {\bibinfo {author} {\bibfnamefont {W.}~\bibnamefont
  {Ma}}, \bibinfo {author} {\bibfnamefont {P.}~\bibnamefont
  {{Alonso-Gonz{\'a}lez}}}, \bibinfo {author} {\bibfnamefont {S.}~\bibnamefont
  {Li}}, \bibinfo {author} {\bibfnamefont {A.~Y.}\ \bibnamefont {Nikitin}},
  \bibinfo {author} {\bibfnamefont {J.}~\bibnamefont {Yuan}}, \bibinfo {author}
  {\bibfnamefont {J.}~\bibnamefont {{Mart{\'i}n-S{\'a}nchez}}}, \bibinfo
  {author} {\bibfnamefont {J.}~\bibnamefont {{Taboada-Guti{\'e}rrez}}},
  \bibinfo {author} {\bibfnamefont {I.}~\bibnamefont {Amenabar}}, \bibinfo
  {author} {\bibfnamefont {P.}~\bibnamefont {Li}}, \bibinfo {author}
  {\bibfnamefont {S.}~\bibnamefont {V{\'e}lez}}, \bibinfo {author}
  {\bibfnamefont {C.}~\bibnamefont {Tollan}}, \bibinfo {author} {\bibfnamefont
  {Z.}~\bibnamefont {Dai}}, \bibinfo {author} {\bibfnamefont {Y.}~\bibnamefont
  {Zhang}}, \bibinfo {author} {\bibfnamefont {S.}~\bibnamefont {Sriram}},
  \bibinfo {author} {\bibfnamefont {K.}~\bibnamefont {{Kalantar-Zadeh}}},
  \bibinfo {author} {\bibfnamefont {S.-T.}\ \bibnamefont {Lee}}, \bibinfo
  {author} {\bibfnamefont {R.}~\bibnamefont {Hillenbrand}},\ and\ \bibinfo
  {author} {\bibfnamefont {Q.}~\bibnamefont {Bao}},\ }\bibfield  {title}
  {\enquote {\bibinfo {title} {In-plane anisotropic and ultra-low-loss
  polaritons in a natural van der waals crystal},}\ }\href
  {https://doi.org/10.1038/s41586-018-0618-9} {\bibfield  {journal} {\bibinfo
  {journal} {Nature}\ }\textbf {\bibinfo {volume} {562}},\ \bibinfo {pages}
  {557--562} (\bibinfo {year} {2018})}\BibitemShut {NoStop}%
\bibitem [{\citenamefont {Li}\ \emph {et~al.}(2020)\citenamefont {Li},
  \citenamefont {Hu}, \citenamefont {Dolado}, \citenamefont {Tymchenko},
  \citenamefont {Qiu}, \citenamefont {{Alfaro-Mozaz}}, \citenamefont
  {Casanova}, \citenamefont {Hueso}, \citenamefont {Liu}, \citenamefont
  {Edgar}, \citenamefont {V{\'e}lez}, \citenamefont {Alu},\ and\ \citenamefont
  {Hillenbrand}}]{Li2020}%
  \BibitemOpen
  \bibfield  {author} {\bibinfo {author} {\bibfnamefont {P.}~\bibnamefont
  {Li}}, \bibinfo {author} {\bibfnamefont {G.}~\bibnamefont {Hu}}, \bibinfo
  {author} {\bibfnamefont {I.}~\bibnamefont {Dolado}}, \bibinfo {author}
  {\bibfnamefont {M.}~\bibnamefont {Tymchenko}}, \bibinfo {author}
  {\bibfnamefont {C.-W.}\ \bibnamefont {Qiu}}, \bibinfo {author} {\bibfnamefont
  {F.~J.}\ \bibnamefont {{Alfaro-Mozaz}}}, \bibinfo {author} {\bibfnamefont
  {F.}~\bibnamefont {Casanova}}, \bibinfo {author} {\bibfnamefont {L.~E.}\
  \bibnamefont {Hueso}}, \bibinfo {author} {\bibfnamefont {S.}~\bibnamefont
  {Liu}}, \bibinfo {author} {\bibfnamefont {J.~H.}\ \bibnamefont {Edgar}},
  \bibinfo {author} {\bibfnamefont {S.}~\bibnamefont {V{\'e}lez}}, \bibinfo
  {author} {\bibfnamefont {A.}~\bibnamefont {Alu}},\ and\ \bibinfo {author}
  {\bibfnamefont {R.}~\bibnamefont {Hillenbrand}},\ }\bibfield  {title}
  {\enquote {\bibinfo {title} {Collective near-field coupling and nonlocal
  phenomena in infrared-phononic metasurfaces for nano-light canalization},}\
  }\href {https://doi.org/10.1038/s41467-020-17425-9} {\bibfield  {journal}
  {\bibinfo  {journal} {Nature Communications}\ }\textbf {\bibinfo {volume}
  {11}},\ \bibinfo {pages} {3663} (\bibinfo {year} {2020})}\BibitemShut
  {NoStop}%
\bibitem [{\citenamefont {Hu}\ \emph {et~al.}(2020)\citenamefont {Hu},
  \citenamefont {Ou}, \citenamefont {Si}, \citenamefont {Wu}, \citenamefont
  {Wu}, \citenamefont {Dai}, \citenamefont {Krasnok}, \citenamefont {Mazor},
  \citenamefont {Zhang}, \citenamefont {Bao}, \citenamefont {Qiu},\ and\
  \citenamefont {Al{\`u}}}]{Hu2020}%
  \BibitemOpen
  \bibfield  {author} {\bibinfo {author} {\bibfnamefont {G.}~\bibnamefont
  {Hu}}, \bibinfo {author} {\bibfnamefont {Q.}~\bibnamefont {Ou}}, \bibinfo
  {author} {\bibfnamefont {G.}~\bibnamefont {Si}}, \bibinfo {author}
  {\bibfnamefont {Y.}~\bibnamefont {Wu}}, \bibinfo {author} {\bibfnamefont
  {J.}~\bibnamefont {Wu}}, \bibinfo {author} {\bibfnamefont {Z.}~\bibnamefont
  {Dai}}, \bibinfo {author} {\bibfnamefont {A.}~\bibnamefont {Krasnok}},
  \bibinfo {author} {\bibfnamefont {Y.}~\bibnamefont {Mazor}}, \bibinfo
  {author} {\bibfnamefont {Q.}~\bibnamefont {Zhang}}, \bibinfo {author}
  {\bibfnamefont {Q.}~\bibnamefont {Bao}}, \bibinfo {author} {\bibfnamefont
  {C.-W.}\ \bibnamefont {Qiu}},\ and\ \bibinfo {author} {\bibfnamefont
  {A.}~\bibnamefont {Al{\`u}}},\ }\bibfield  {title} {\enquote {\bibinfo
  {title} {Topological polaritons and photonic magic angles in twisted
  {$\alpha$}-{{MoO3}} bilayers},}\ }\href
  {https://doi.org/10.1038/s41586-020-2359-9} {\bibfield  {journal} {\bibinfo
  {journal} {Nature}\ }\textbf {\bibinfo {volume} {582}},\ \bibinfo {pages}
  {209--213} (\bibinfo {year} {2020})}\BibitemShut {NoStop}%
\bibitem [{\citenamefont {Chen}\ \emph {et~al.}(2020)\citenamefont {Chen},
  \citenamefont {Lin}, \citenamefont {Dinh}, \citenamefont {Zheng},
  \citenamefont {Shen}, \citenamefont {Ma}, \citenamefont {Chen}, \citenamefont
  {{Jarillo-Herrero}},\ and\ \citenamefont {Dai}}]{Chen2020}%
  \BibitemOpen
  \bibfield  {author} {\bibinfo {author} {\bibfnamefont {M.}~\bibnamefont
  {Chen}}, \bibinfo {author} {\bibfnamefont {X.}~\bibnamefont {Lin}}, \bibinfo
  {author} {\bibfnamefont {T.~H.}\ \bibnamefont {Dinh}}, \bibinfo {author}
  {\bibfnamefont {Z.}~\bibnamefont {Zheng}}, \bibinfo {author} {\bibfnamefont
  {J.}~\bibnamefont {Shen}}, \bibinfo {author} {\bibfnamefont {Q.}~\bibnamefont
  {Ma}}, \bibinfo {author} {\bibfnamefont {H.}~\bibnamefont {Chen}}, \bibinfo
  {author} {\bibfnamefont {P.}~\bibnamefont {{Jarillo-Herrero}}},\ and\
  \bibinfo {author} {\bibfnamefont {S.}~\bibnamefont {Dai}},\ }\bibfield
  {title} {\enquote {\bibinfo {title} {Configurable phonon polaritons in
  twisted {$\alpha$}-{{MoO3}}},}\ }\href
  {https://doi.org/10.1038/s41563-020-0732-6} {\bibfield  {journal} {\bibinfo
  {journal} {Nature Materials}\ }\textbf {\bibinfo {volume} {19}},\ \bibinfo
  {pages} {1307--1311} (\bibinfo {year} {2020})}\BibitemShut {NoStop}%
\bibitem [{\citenamefont {{Mart{\'i}n-S{\'a}nchez}}\ \emph
  {et~al.}(2021)\citenamefont {{Mart{\'i}n-S{\'a}nchez}}, \citenamefont {Duan},
  \citenamefont {{Taboada-Guti{\'e}rrez}}, \citenamefont
  {{\'A}lvarez-P{\'e}rez}, \citenamefont {Voronin}, \citenamefont {Prieto},
  \citenamefont {Ma}, \citenamefont {Bao}, \citenamefont {Volkov},
  \citenamefont {Hillenbrand}, \citenamefont {Nikitin},\ and\ \citenamefont
  {{Alonso-Gonz{\'a}lez}}}]{Martin-Sanchez2021}%
  \BibitemOpen
  \bibfield  {author} {\bibinfo {author} {\bibfnamefont {J.}~\bibnamefont
  {{Mart{\'i}n-S{\'a}nchez}}}, \bibinfo {author} {\bibfnamefont
  {J.}~\bibnamefont {Duan}}, \bibinfo {author} {\bibfnamefont {J.}~\bibnamefont
  {{Taboada-Guti{\'e}rrez}}}, \bibinfo {author} {\bibfnamefont
  {G.}~\bibnamefont {{\'A}lvarez-P{\'e}rez}}, \bibinfo {author} {\bibfnamefont
  {K.~V.}\ \bibnamefont {Voronin}}, \bibinfo {author} {\bibfnamefont
  {I.}~\bibnamefont {Prieto}}, \bibinfo {author} {\bibfnamefont
  {W.}~\bibnamefont {Ma}}, \bibinfo {author} {\bibfnamefont {Q.}~\bibnamefont
  {Bao}}, \bibinfo {author} {\bibfnamefont {V.~S.}\ \bibnamefont {Volkov}},
  \bibinfo {author} {\bibfnamefont {R.}~\bibnamefont {Hillenbrand}}, \bibinfo
  {author} {\bibfnamefont {A.~Y.}\ \bibnamefont {Nikitin}},\ and\ \bibinfo
  {author} {\bibfnamefont {P.}~\bibnamefont {{Alonso-Gonz{\'a}lez}}},\
  }\bibfield  {title} {\enquote {\bibinfo {title} {Focusing of in-plane
  hyperbolic polaritons in van der {{Waals}} crystals with tailored infrared
  nanoantennas},}\ }\href {https://doi.org/10.1126/sciadv.abj0127} {\bibfield
  {journal} {\bibinfo  {journal} {Science Advances}\ }\textbf {\bibinfo
  {volume} {7}},\ \bibinfo {pages} {eabj0127} (\bibinfo {year}
  {2021})}\BibitemShut {NoStop}%
\bibitem [{\citenamefont {Ma}\ \emph {et~al.}(2021)\citenamefont {Ma},
  \citenamefont {Hu}, \citenamefont {Hu}, \citenamefont {Chen}, \citenamefont
  {Sun}, \citenamefont {Zhang}, \citenamefont {Dai}, \citenamefont {Zeng},
  \citenamefont {Al{\`u}}, \citenamefont {Qiu},\ and\ \citenamefont
  {Li}}]{Ma2021}%
  \BibitemOpen
  \bibfield  {author} {\bibinfo {author} {\bibfnamefont {W.}~\bibnamefont
  {Ma}}, \bibinfo {author} {\bibfnamefont {G.}~\bibnamefont {Hu}}, \bibinfo
  {author} {\bibfnamefont {D.}~\bibnamefont {Hu}}, \bibinfo {author}
  {\bibfnamefont {R.}~\bibnamefont {Chen}}, \bibinfo {author} {\bibfnamefont
  {T.}~\bibnamefont {Sun}}, \bibinfo {author} {\bibfnamefont {X.}~\bibnamefont
  {Zhang}}, \bibinfo {author} {\bibfnamefont {Q.}~\bibnamefont {Dai}}, \bibinfo
  {author} {\bibfnamefont {Y.}~\bibnamefont {Zeng}}, \bibinfo {author}
  {\bibfnamefont {A.}~\bibnamefont {Al{\`u}}}, \bibinfo {author} {\bibfnamefont
  {C.-W.}\ \bibnamefont {Qiu}},\ and\ \bibinfo {author} {\bibfnamefont
  {P.}~\bibnamefont {Li}},\ }\bibfield  {title} {\enquote {\bibinfo {title}
  {Ghost hyperbolic surface polaritons in bulk anisotropic crystals},}\ }\href
  {https://doi.org/10.1038/s41586-021-03755-1} {\bibfield  {journal} {\bibinfo
  {journal} {Nature}\ }\textbf {\bibinfo {volume} {596}},\ \bibinfo {pages}
  {362--366} (\bibinfo {year} {2021})}\BibitemShut {NoStop}%
\bibitem [{\citenamefont {Hu}\ \emph {et~al.}(2023)\citenamefont {Hu},
  \citenamefont {Chen}, \citenamefont {Teng}, \citenamefont {Yu}, \citenamefont
  {Xue}, \citenamefont {Chen}, \citenamefont {Xiao}, \citenamefont {Qu},
  \citenamefont {Hu}, \citenamefont {Chen}, \citenamefont {Sun}, \citenamefont
  {Li}, \citenamefont {De~Abajo},\ and\ \citenamefont {Dai}}]{Hu2023}%
  \BibitemOpen
  \bibfield  {author} {\bibinfo {author} {\bibfnamefont {H.}~\bibnamefont
  {Hu}}, \bibinfo {author} {\bibfnamefont {N.}~\bibnamefont {Chen}}, \bibinfo
  {author} {\bibfnamefont {H.}~\bibnamefont {Teng}}, \bibinfo {author}
  {\bibfnamefont {R.}~\bibnamefont {Yu}}, \bibinfo {author} {\bibfnamefont
  {M.}~\bibnamefont {Xue}}, \bibinfo {author} {\bibfnamefont {K.}~\bibnamefont
  {Chen}}, \bibinfo {author} {\bibfnamefont {Y.}~\bibnamefont {Xiao}}, \bibinfo
  {author} {\bibfnamefont {Y.}~\bibnamefont {Qu}}, \bibinfo {author}
  {\bibfnamefont {D.}~\bibnamefont {Hu}}, \bibinfo {author} {\bibfnamefont
  {J.}~\bibnamefont {Chen}}, \bibinfo {author} {\bibfnamefont {Z.}~\bibnamefont
  {Sun}}, \bibinfo {author} {\bibfnamefont {P.}~\bibnamefont {Li}}, \bibinfo
  {author} {\bibfnamefont {F.~J.~G.}\ \bibnamefont {De~Abajo}},\ and\ \bibinfo
  {author} {\bibfnamefont {Q.}~\bibnamefont {Dai}},\ }\bibfield  {title}
  {\enquote {\bibinfo {title} {Gate-tunable negative refraction of mid-infrared
  polaritons},}\ }\href {https://doi.org/10.1126/science.adf1251} {\bibfield
  {journal} {\bibinfo  {journal} {Science}\ }\textbf {\bibinfo {volume}
  {379}},\ \bibinfo {pages} {558--561} (\bibinfo {year} {2023})}\BibitemShut
  {NoStop}%
\bibitem [{\citenamefont {Guo}\ \emph {et~al.}(2025)\citenamefont {Guo},
  \citenamefont {Esin}, \citenamefont {Li}, \citenamefont {Chen}, \citenamefont
  {Han}, \citenamefont {Liu}, \citenamefont {Edgar}, \citenamefont {Zhou},
  \citenamefont {Demler}, \citenamefont {Refael},\ and\ \citenamefont
  {Xia}}]{Guo2025}%
  \BibitemOpen
  \bibfield  {author} {\bibinfo {author} {\bibfnamefont {Q.}~\bibnamefont
  {Guo}}, \bibinfo {author} {\bibfnamefont {I.}~\bibnamefont {Esin}}, \bibinfo
  {author} {\bibfnamefont {C.}~\bibnamefont {Li}}, \bibinfo {author}
  {\bibfnamefont {C.}~\bibnamefont {Chen}}, \bibinfo {author} {\bibfnamefont
  {G.}~\bibnamefont {Han}}, \bibinfo {author} {\bibfnamefont {S.}~\bibnamefont
  {Liu}}, \bibinfo {author} {\bibfnamefont {J.~H.}\ \bibnamefont {Edgar}},
  \bibinfo {author} {\bibfnamefont {S.}~\bibnamefont {Zhou}}, \bibinfo {author}
  {\bibfnamefont {E.}~\bibnamefont {Demler}}, \bibinfo {author} {\bibfnamefont
  {G.}~\bibnamefont {Refael}},\ and\ \bibinfo {author} {\bibfnamefont
  {F.}~\bibnamefont {Xia}},\ }\bibfield  {title} {\enquote {\bibinfo {title}
  {Hyperbolic phonon-polariton electroluminescence in {{2D}}
  heterostructures},}\ }\href {https://doi.org/10.1038/s41586-025-08686-9}
  {\bibfield  {journal} {\bibinfo  {journal} {Nature}\ }\textbf {\bibinfo
  {volume} {639}},\ \bibinfo {pages} {915--921} (\bibinfo {year}
  {2025})}\BibitemShut {NoStop}%
\bibitem [{\citenamefont {Kamada}\ \emph {et~al.}(2026)\citenamefont {Kamada},
  \citenamefont {Shinokita}, \citenamefont {Zeng}, \citenamefont {Kitaura},
  \citenamefont {Watanabe}, \citenamefont {Taniguchi}, \citenamefont
  {Paarmann}, \citenamefont {Shibuta}, \citenamefont {Kumagai},\ and\
  \citenamefont {Nishida}}]{Kamada2026}%
  \BibitemOpen
  \bibfield  {author} {\bibinfo {author} {\bibfnamefont {K.}~\bibnamefont
  {Kamada}}, \bibinfo {author} {\bibfnamefont {K.}~\bibnamefont {Shinokita}},
  \bibinfo {author} {\bibfnamefont {F.}~\bibnamefont {Zeng}}, \bibinfo {author}
  {\bibfnamefont {R.}~\bibnamefont {Kitaura}}, \bibinfo {author} {\bibfnamefont
  {K.}~\bibnamefont {Watanabe}}, \bibinfo {author} {\bibfnamefont
  {T.}~\bibnamefont {Taniguchi}}, \bibinfo {author} {\bibfnamefont
  {A.}~\bibnamefont {Paarmann}}, \bibinfo {author} {\bibfnamefont
  {M.}~\bibnamefont {Shibuta}}, \bibinfo {author} {\bibfnamefont
  {T.}~\bibnamefont {Kumagai}},\ and\ \bibinfo {author} {\bibfnamefont
  {J.}~\bibnamefont {Nishida}},\ }\bibfield  {title} {\enquote {\bibinfo
  {title} {Ultrafast {{Nano-imaging}} and {{Optical Control}} of {{Hyperbolic
  Phonon Polaritons}} at {{hBN}}/{{WS2 Heterojunctions}}},}\ }\href
  {https://doi.org/10.1021/acs.nanolett.6c02448} {\bibfield  {journal}
  {\bibinfo  {journal} {Nano Letters}\ }\textbf {\bibinfo {volume} {26}},\
  \bibinfo {pages} {10446--10454} (\bibinfo {year} {2026})}\BibitemShut
  {NoStop}%
\bibitem [{\citenamefont {De~Abajo}\ \emph {et~al.}(2025)\citenamefont
  {De~Abajo}, \citenamefont {Basov}, \citenamefont {Koppens}, \citenamefont
  {Orsini}, \citenamefont {Ceccanti}, \citenamefont {Castilla}, \citenamefont
  {Cavicchi}, \citenamefont {Polini}, \citenamefont {Gon{\c c}alves},
  \citenamefont {Costa}, \citenamefont {Peres}, \citenamefont {Mortensen},
  \citenamefont {Bharadwaj}, \citenamefont {Jacob}, \citenamefont {Schuck},
  \citenamefont {Pasupathy}, \citenamefont {Delor}, \citenamefont {Liu},
  \citenamefont {Mugarza}, \citenamefont {Merino}, \citenamefont {Cuxart},
  \citenamefont {{Ch{\'a}vez-Angel}}, \citenamefont {{\v S}vec}, \citenamefont
  {Tizei}, \citenamefont {Dirnberger}, \citenamefont {Deng}, \citenamefont
  {Schneider}, \citenamefont {Menon}, \citenamefont {Deilmann}, \citenamefont
  {Chernikov}, \citenamefont {Thygesen}, \citenamefont {Abate}, \citenamefont
  {Terrones}, \citenamefont {Sangwan}, \citenamefont {Hersam}, \citenamefont
  {Yu}, \citenamefont {Chen}, \citenamefont {Heinz}, \citenamefont {Murthy},
  \citenamefont {Kroner}, \citenamefont {Smolenski}, \citenamefont {Thureja},
  \citenamefont {Chervy}, \citenamefont {Genco}, \citenamefont {Trovatello},
  \citenamefont {Cerullo}, \citenamefont {Dal~Conte}, \citenamefont {Timmer},
  \citenamefont {De~Sio}, \citenamefont {Lienau}, \citenamefont {Shang},
  \citenamefont {Hong}, \citenamefont {Liu}, \citenamefont {Sun}, \citenamefont
  {Rozema}, \citenamefont {Walther}, \citenamefont {Al{\`u}}, \citenamefont
  {Marini}, \citenamefont {Cotrufo}, \citenamefont {Queiroz}, \citenamefont
  {Zhu}, \citenamefont {Cox}, \citenamefont {Dias}, \citenamefont {Echarri},
  \citenamefont {Iyikanat}, \citenamefont {Herrmann}, \citenamefont {Tornow},
  \citenamefont {Klimmer}, \citenamefont {Wilhelm}, \citenamefont {Soavi},
  \citenamefont {Sun}, \citenamefont {Wu}, \citenamefont {Xiong}, \citenamefont
  {Matsyshyn}, \citenamefont {Krishna~Kumar}, \citenamefont {Song},
  \citenamefont {Bucher}, \citenamefont {Gorlach}, \citenamefont {Tsesses},
  \citenamefont {Kaminer}, \citenamefont {Schwab}, \citenamefont {Mangold},
  \citenamefont {Giessen}, \citenamefont {S{\'a}nchez}, \citenamefont {Efetov},
  \citenamefont {Low}, \citenamefont {{G{\'o}mez-Santos}}, \citenamefont
  {Stauber}, \citenamefont {{\'A}lvarez-P{\'e}rez}, \citenamefont {Duan},
  \citenamefont {{Mart{\'i}n-Moreno}}, \citenamefont {Paarmann}, \citenamefont
  {Caldwell}, \citenamefont {Nikitin}, \citenamefont {{Alonso-Gonz{\'a}lez}},
  \citenamefont {Mueller}, \citenamefont {Volkov}, \citenamefont {Jariwala},
  \citenamefont {Shegai}, \citenamefont {Van De~Groep}, \citenamefont
  {Boltasseva}, \citenamefont {Bondarev}, \citenamefont {Shalaev},
  \citenamefont {Simon}, \citenamefont {Fruhling}, \citenamefont {Shen},
  \citenamefont {Novko}, \citenamefont {Tan}, \citenamefont {Wang},
  \citenamefont {Petek}, \citenamefont {Mkhitaryan}, \citenamefont {Yu},
  \citenamefont {Manjavacas}, \citenamefont {Ortega}, \citenamefont {Cheng},
  \citenamefont {Tian}, \citenamefont {Mao}, \citenamefont {Van~Thourhout},
  \citenamefont {Gan}, \citenamefont {Dai}, \citenamefont {Sternbach},
  \citenamefont {Zhou}, \citenamefont {Hafezi}, \citenamefont {Litvinov},
  \citenamefont {Grzeszczyk}, \citenamefont {Novoselov}, \citenamefont
  {Koperski}, \citenamefont {Papadopoulos}, \citenamefont {Novotny},
  \citenamefont {Viti}, \citenamefont {Vitiello}, \citenamefont {Cottam},
  \citenamefont {Dewes}, \citenamefont {Makarovsky}, \citenamefont
  {Patan{\`e}}, \citenamefont {Song}, \citenamefont {Cai}, \citenamefont
  {Chen}, \citenamefont {Naveh}, \citenamefont {Jang}, \citenamefont {Park},
  \citenamefont {Xia}, \citenamefont {Jenke}, \citenamefont {Bajo},
  \citenamefont {Braun}, \citenamefont {Burch}, \citenamefont {Zhao},\ and\
  \citenamefont {Xu}}]{DeAbajo2025}%
  \BibitemOpen
  \bibfield  {author} {\bibinfo {author} {\bibfnamefont {F.~J.~G.}\
  \bibnamefont {De~Abajo}}, \bibinfo {author} {\bibfnamefont {D.~N.}\
  \bibnamefont {Basov}}, \bibinfo {author} {\bibfnamefont {F.~H.~L.}\
  \bibnamefont {Koppens}}, \bibinfo {author} {\bibfnamefont {L.}~\bibnamefont
  {Orsini}}, \bibinfo {author} {\bibfnamefont {M.}~\bibnamefont {Ceccanti}},
  \bibinfo {author} {\bibfnamefont {S.}~\bibnamefont {Castilla}}, \bibinfo
  {author} {\bibfnamefont {L.}~\bibnamefont {Cavicchi}}, \bibinfo {author}
  {\bibfnamefont {M.}~\bibnamefont {Polini}}, \bibinfo {author} {\bibfnamefont
  {P.~A.~D.}\ \bibnamefont {Gon{\c c}alves}}, \bibinfo {author} {\bibfnamefont
  {A.~T.}\ \bibnamefont {Costa}}, \bibinfo {author} {\bibfnamefont {N.~M.~R.}\
  \bibnamefont {Peres}}, \bibinfo {author} {\bibfnamefont {N.~A.}\ \bibnamefont
  {Mortensen}}, \bibinfo {author} {\bibfnamefont {S.}~\bibnamefont
  {Bharadwaj}}, \bibinfo {author} {\bibfnamefont {Z.}~\bibnamefont {Jacob}},
  \bibinfo {author} {\bibfnamefont {P.~J.}\ \bibnamefont {Schuck}}, \bibinfo
  {author} {\bibfnamefont {A.~N.}\ \bibnamefont {Pasupathy}}, \bibinfo {author}
  {\bibfnamefont {M.}~\bibnamefont {Delor}}, \bibinfo {author} {\bibfnamefont
  {M.~K.}\ \bibnamefont {Liu}}, \bibinfo {author} {\bibfnamefont
  {A.}~\bibnamefont {Mugarza}}, \bibinfo {author} {\bibfnamefont
  {P.}~\bibnamefont {Merino}}, \bibinfo {author} {\bibfnamefont {M.~G.}\
  \bibnamefont {Cuxart}}, \bibinfo {author} {\bibfnamefont {E.}~\bibnamefont
  {{Ch{\'a}vez-Angel}}}, \bibinfo {author} {\bibfnamefont {M.}~\bibnamefont
  {{\v S}vec}}, \bibinfo {author} {\bibfnamefont {L.~H.~G.}\ \bibnamefont
  {Tizei}}, \bibinfo {author} {\bibfnamefont {F.}~\bibnamefont {Dirnberger}},
  \bibinfo {author} {\bibfnamefont {H.}~\bibnamefont {Deng}}, \bibinfo {author}
  {\bibfnamefont {C.}~\bibnamefont {Schneider}}, \bibinfo {author}
  {\bibfnamefont {V.}~\bibnamefont {Menon}}, \bibinfo {author} {\bibfnamefont
  {T.}~\bibnamefont {Deilmann}}, \bibinfo {author} {\bibfnamefont
  {A.}~\bibnamefont {Chernikov}}, \bibinfo {author} {\bibfnamefont {K.~S.}\
  \bibnamefont {Thygesen}}, \bibinfo {author} {\bibfnamefont {Y.}~\bibnamefont
  {Abate}}, \bibinfo {author} {\bibfnamefont {M.}~\bibnamefont {Terrones}},
  \bibinfo {author} {\bibfnamefont {V.~K.}\ \bibnamefont {Sangwan}}, \bibinfo
  {author} {\bibfnamefont {M.~C.}\ \bibnamefont {Hersam}}, \bibinfo {author}
  {\bibfnamefont {L.}~\bibnamefont {Yu}}, \bibinfo {author} {\bibfnamefont
  {X.}~\bibnamefont {Chen}}, \bibinfo {author} {\bibfnamefont {T.~F.}\
  \bibnamefont {Heinz}}, \bibinfo {author} {\bibfnamefont {P.}~\bibnamefont
  {Murthy}}, \bibinfo {author} {\bibfnamefont {M.}~\bibnamefont {Kroner}},
  \bibinfo {author} {\bibfnamefont {T.}~\bibnamefont {Smolenski}}, \bibinfo
  {author} {\bibfnamefont {D.}~\bibnamefont {Thureja}}, \bibinfo {author}
  {\bibfnamefont {T.}~\bibnamefont {Chervy}}, \bibinfo {author} {\bibfnamefont
  {A.}~\bibnamefont {Genco}}, \bibinfo {author} {\bibfnamefont
  {C.}~\bibnamefont {Trovatello}}, \bibinfo {author} {\bibfnamefont
  {G.}~\bibnamefont {Cerullo}}, \bibinfo {author} {\bibfnamefont
  {S.}~\bibnamefont {Dal~Conte}}, \bibinfo {author} {\bibfnamefont
  {D.}~\bibnamefont {Timmer}}, \bibinfo {author} {\bibfnamefont
  {A.}~\bibnamefont {De~Sio}}, \bibinfo {author} {\bibfnamefont
  {C.}~\bibnamefont {Lienau}}, \bibinfo {author} {\bibfnamefont
  {N.}~\bibnamefont {Shang}}, \bibinfo {author} {\bibfnamefont
  {H.}~\bibnamefont {Hong}}, \bibinfo {author} {\bibfnamefont {K.}~\bibnamefont
  {Liu}}, \bibinfo {author} {\bibfnamefont {Z.}~\bibnamefont {Sun}}, \bibinfo
  {author} {\bibfnamefont {L.~A.}\ \bibnamefont {Rozema}}, \bibinfo {author}
  {\bibfnamefont {P.}~\bibnamefont {Walther}}, \bibinfo {author} {\bibfnamefont
  {A.}~\bibnamefont {Al{\`u}}}, \bibinfo {author} {\bibfnamefont
  {A.}~\bibnamefont {Marini}}, \bibinfo {author} {\bibfnamefont
  {M.}~\bibnamefont {Cotrufo}}, \bibinfo {author} {\bibfnamefont
  {R.}~\bibnamefont {Queiroz}}, \bibinfo {author} {\bibfnamefont {X.-Y.}\
  \bibnamefont {Zhu}}, \bibinfo {author} {\bibfnamefont {J.~D.}\ \bibnamefont
  {Cox}}, \bibinfo {author} {\bibfnamefont {E.~J.~C.}\ \bibnamefont {Dias}},
  \bibinfo {author} {\bibfnamefont {{\'A}.~R.}\ \bibnamefont {Echarri}},
  \bibinfo {author} {\bibfnamefont {F.}~\bibnamefont {Iyikanat}}, \bibinfo
  {author} {\bibfnamefont {P.}~\bibnamefont {Herrmann}}, \bibinfo {author}
  {\bibfnamefont {N.}~\bibnamefont {Tornow}}, \bibinfo {author} {\bibfnamefont
  {S.}~\bibnamefont {Klimmer}}, \bibinfo {author} {\bibfnamefont
  {J.}~\bibnamefont {Wilhelm}}, \bibinfo {author} {\bibfnamefont
  {G.}~\bibnamefont {Soavi}}, \bibinfo {author} {\bibfnamefont
  {Z.}~\bibnamefont {Sun}}, \bibinfo {author} {\bibfnamefont {S.}~\bibnamefont
  {Wu}}, \bibinfo {author} {\bibfnamefont {Y.}~\bibnamefont {Xiong}}, \bibinfo
  {author} {\bibfnamefont {O.}~\bibnamefont {Matsyshyn}}, \bibinfo {author}
  {\bibfnamefont {R.}~\bibnamefont {Krishna~Kumar}}, \bibinfo {author}
  {\bibfnamefont {J.~C.~W.}\ \bibnamefont {Song}}, \bibinfo {author}
  {\bibfnamefont {T.}~\bibnamefont {Bucher}}, \bibinfo {author} {\bibfnamefont
  {A.}~\bibnamefont {Gorlach}}, \bibinfo {author} {\bibfnamefont
  {S.}~\bibnamefont {Tsesses}}, \bibinfo {author} {\bibfnamefont
  {I.}~\bibnamefont {Kaminer}}, \bibinfo {author} {\bibfnamefont
  {J.}~\bibnamefont {Schwab}}, \bibinfo {author} {\bibfnamefont
  {F.}~\bibnamefont {Mangold}}, \bibinfo {author} {\bibfnamefont
  {H.}~\bibnamefont {Giessen}}, \bibinfo {author} {\bibfnamefont {M.~S.}\
  \bibnamefont {S{\'a}nchez}}, \bibinfo {author} {\bibfnamefont {D.~K.}\
  \bibnamefont {Efetov}}, \bibinfo {author} {\bibfnamefont {T.}~\bibnamefont
  {Low}}, \bibinfo {author} {\bibfnamefont {G.}~\bibnamefont
  {{G{\'o}mez-Santos}}}, \bibinfo {author} {\bibfnamefont {T.}~\bibnamefont
  {Stauber}}, \bibinfo {author} {\bibfnamefont {G.}~\bibnamefont
  {{\'A}lvarez-P{\'e}rez}}, \bibinfo {author} {\bibfnamefont {J.}~\bibnamefont
  {Duan}}, \bibinfo {author} {\bibfnamefont {L.}~\bibnamefont
  {{Mart{\'i}n-Moreno}}}, \bibinfo {author} {\bibfnamefont {A.}~\bibnamefont
  {Paarmann}}, \bibinfo {author} {\bibfnamefont {J.~D.}\ \bibnamefont
  {Caldwell}}, \bibinfo {author} {\bibfnamefont {A.~Y.}\ \bibnamefont
  {Nikitin}}, \bibinfo {author} {\bibfnamefont {P.}~\bibnamefont
  {{Alonso-Gonz{\'a}lez}}}, \bibinfo {author} {\bibfnamefont {N.~S.}\
  \bibnamefont {Mueller}}, \bibinfo {author} {\bibfnamefont {V.}~\bibnamefont
  {Volkov}}, \bibinfo {author} {\bibfnamefont {D.}~\bibnamefont {Jariwala}},
  \bibinfo {author} {\bibfnamefont {T.}~\bibnamefont {Shegai}}, \bibinfo
  {author} {\bibfnamefont {J.}~\bibnamefont {Van De~Groep}}, \bibinfo {author}
  {\bibfnamefont {A.}~\bibnamefont {Boltasseva}}, \bibinfo {author}
  {\bibfnamefont {I.~V.}\ \bibnamefont {Bondarev}}, \bibinfo {author}
  {\bibfnamefont {V.~M.}\ \bibnamefont {Shalaev}}, \bibinfo {author}
  {\bibfnamefont {J.}~\bibnamefont {Simon}}, \bibinfo {author} {\bibfnamefont
  {C.}~\bibnamefont {Fruhling}}, \bibinfo {author} {\bibfnamefont
  {G.}~\bibnamefont {Shen}}, \bibinfo {author} {\bibfnamefont {D.}~\bibnamefont
  {Novko}}, \bibinfo {author} {\bibfnamefont {S.}~\bibnamefont {Tan}}, \bibinfo
  {author} {\bibfnamefont {B.}~\bibnamefont {Wang}}, \bibinfo {author}
  {\bibfnamefont {H.}~\bibnamefont {Petek}}, \bibinfo {author} {\bibfnamefont
  {V.}~\bibnamefont {Mkhitaryan}}, \bibinfo {author} {\bibfnamefont
  {R.}~\bibnamefont {Yu}}, \bibinfo {author} {\bibfnamefont {A.}~\bibnamefont
  {Manjavacas}}, \bibinfo {author} {\bibfnamefont {J.~E.}\ \bibnamefont
  {Ortega}}, \bibinfo {author} {\bibfnamefont {X.}~\bibnamefont {Cheng}},
  \bibinfo {author} {\bibfnamefont {R.}~\bibnamefont {Tian}}, \bibinfo {author}
  {\bibfnamefont {D.}~\bibnamefont {Mao}}, \bibinfo {author} {\bibfnamefont
  {D.}~\bibnamefont {Van~Thourhout}}, \bibinfo {author} {\bibfnamefont
  {X.}~\bibnamefont {Gan}}, \bibinfo {author} {\bibfnamefont {Q.}~\bibnamefont
  {Dai}}, \bibinfo {author} {\bibfnamefont {A.}~\bibnamefont {Sternbach}},
  \bibinfo {author} {\bibfnamefont {Y.}~\bibnamefont {Zhou}}, \bibinfo {author}
  {\bibfnamefont {M.}~\bibnamefont {Hafezi}}, \bibinfo {author} {\bibfnamefont
  {D.}~\bibnamefont {Litvinov}}, \bibinfo {author} {\bibfnamefont
  {M.}~\bibnamefont {Grzeszczyk}}, \bibinfo {author} {\bibfnamefont {K.~S.}\
  \bibnamefont {Novoselov}}, \bibinfo {author} {\bibfnamefont {M.}~\bibnamefont
  {Koperski}}, \bibinfo {author} {\bibfnamefont {S.}~\bibnamefont
  {Papadopoulos}}, \bibinfo {author} {\bibfnamefont {L.}~\bibnamefont
  {Novotny}}, \bibinfo {author} {\bibfnamefont {L.}~\bibnamefont {Viti}},
  \bibinfo {author} {\bibfnamefont {M.~S.}\ \bibnamefont {Vitiello}}, \bibinfo
  {author} {\bibfnamefont {N.~D.}\ \bibnamefont {Cottam}}, \bibinfo {author}
  {\bibfnamefont {B.~T.}\ \bibnamefont {Dewes}}, \bibinfo {author}
  {\bibfnamefont {O.}~\bibnamefont {Makarovsky}}, \bibinfo {author}
  {\bibfnamefont {A.}~\bibnamefont {Patan{\`e}}}, \bibinfo {author}
  {\bibfnamefont {Y.}~\bibnamefont {Song}}, \bibinfo {author} {\bibfnamefont
  {M.}~\bibnamefont {Cai}}, \bibinfo {author} {\bibfnamefont {J.}~\bibnamefont
  {Chen}}, \bibinfo {author} {\bibfnamefont {D.}~\bibnamefont {Naveh}},
  \bibinfo {author} {\bibfnamefont {H.}~\bibnamefont {Jang}}, \bibinfo {author}
  {\bibfnamefont {S.}~\bibnamefont {Park}}, \bibinfo {author} {\bibfnamefont
  {F.}~\bibnamefont {Xia}}, \bibinfo {author} {\bibfnamefont {P.~K.}\
  \bibnamefont {Jenke}}, \bibinfo {author} {\bibfnamefont {J.}~\bibnamefont
  {Bajo}}, \bibinfo {author} {\bibfnamefont {B.}~\bibnamefont {Braun}},
  \bibinfo {author} {\bibfnamefont {K.~S.}\ \bibnamefont {Burch}}, \bibinfo
  {author} {\bibfnamefont {L.}~\bibnamefont {Zhao}},\ and\ \bibinfo {author}
  {\bibfnamefont {X.}~\bibnamefont {Xu}},\ }\bibfield  {title} {\enquote
  {\bibinfo {title} {Roadmap for {{Photonics}} with {{2D Materials}}},}\ }\href
  {https://doi.org/10.1021/acsphotonics.5c00353} {\bibfield  {journal}
  {\bibinfo  {journal} {ACS Photonics}\ }\textbf {\bibinfo {volume} {12}},\
  \bibinfo {pages} {3961--4095} (\bibinfo {year} {2025})}\BibitemShut {NoStop}%
\bibitem [{\citenamefont {Gershuni}\ and\ \citenamefont
  {Epstein}(2024)}]{Gershuni2024}%
  \BibitemOpen
  \bibfield  {author} {\bibinfo {author} {\bibfnamefont {Y.}~\bibnamefont
  {Gershuni}}\ and\ \bibinfo {author} {\bibfnamefont {I.}~\bibnamefont
  {Epstein}},\ }\bibfield  {title} {\enquote {\bibinfo {title} {In-plane
  exciton polaritons versus plasmon polaritons: {{Nonlocal}} corrections,
  confinement, and loss},}\ }\href
  {https://doi.org/10.1103/PhysRevB.109.L121408} {\bibfield  {journal}
  {\bibinfo  {journal} {Physical Review B}\ }\textbf {\bibinfo {volume}
  {109}},\ \bibinfo {pages} {L121408} (\bibinfo {year} {2024})}\BibitemShut
  {NoStop}%
\bibitem [{\citenamefont {Epstein}\ \emph {et~al.}(2020)\citenamefont
  {Epstein}, \citenamefont {Chaves}, \citenamefont {Rhodes}, \citenamefont
  {Frank}, \citenamefont {Watanabe}, \citenamefont {Taniguchi}, \citenamefont
  {Giessen}, \citenamefont {Hone}, \citenamefont {Peres},\ and\ \citenamefont
  {Koppens}}]{Epstein2020}%
  \BibitemOpen
  \bibfield  {author} {\bibinfo {author} {\bibfnamefont {I.}~\bibnamefont
  {Epstein}}, \bibinfo {author} {\bibfnamefont {A.~J.}\ \bibnamefont {Chaves}},
  \bibinfo {author} {\bibfnamefont {D.~A.}\ \bibnamefont {Rhodes}}, \bibinfo
  {author} {\bibfnamefont {B.}~\bibnamefont {Frank}}, \bibinfo {author}
  {\bibfnamefont {K.}~\bibnamefont {Watanabe}}, \bibinfo {author}
  {\bibfnamefont {T.}~\bibnamefont {Taniguchi}}, \bibinfo {author}
  {\bibfnamefont {H.}~\bibnamefont {Giessen}}, \bibinfo {author} {\bibfnamefont
  {J.~C.}\ \bibnamefont {Hone}}, \bibinfo {author} {\bibfnamefont {N.~M.~R.}\
  \bibnamefont {Peres}},\ and\ \bibinfo {author} {\bibfnamefont {F.~H.~L.}\
  \bibnamefont {Koppens}},\ }\bibfield  {title} {\enquote {\bibinfo {title}
  {Highly confined in-plane propagating exciton-polaritons on monolayer
  semiconductors},}\ }\href {https://doi.org/10.1088/2053-1583/ab8dd4}
  {\bibfield  {journal} {\bibinfo  {journal} {2D Materials}\ }\textbf {\bibinfo
  {volume} {7}},\ \bibinfo {pages} {035031} (\bibinfo {year}
  {2020})}\BibitemShut {NoStop}%
\bibitem [{\citenamefont {Meshulam}\ \emph {et~al.}(2026)\citenamefont
  {Meshulam}, \citenamefont {Kahlon}, \citenamefont {Gershuni}, \citenamefont
  {Poirier}, \citenamefont {Edgar}, \citenamefont {Tongay},\ and\ \citenamefont
  {Epstein}}]{Meshulam2026}%
  \BibitemOpen
  \bibfield  {author} {\bibinfo {author} {\bibfnamefont {M.}~\bibnamefont
  {Meshulam}}, \bibinfo {author} {\bibfnamefont {A.~A.}\ \bibnamefont
  {Kahlon}}, \bibinfo {author} {\bibfnamefont {Y.}~\bibnamefont {Gershuni}},
  \bibinfo {author} {\bibfnamefont {T.}~\bibnamefont {Poirier}}, \bibinfo
  {author} {\bibfnamefont {J.~H.}\ \bibnamefont {Edgar}}, \bibinfo {author}
  {\bibfnamefont {S.~A.}\ \bibnamefont {Tongay}},\ and\ \bibinfo {author}
  {\bibfnamefont {I.}~\bibnamefont {Epstein}},\ }\bibfield  {title} {\enquote
  {\bibinfo {title} {Temperature-{{Dependent Optical}} and {{Polaritonic
  Properties}} of {{Excitons}} in {{hBN}}-{{Encapsulated Monolayer TMDs}}},}\
  }\href {https://doi.org/10.1002/adom.202502535} {\bibfield  {journal}
  {\bibinfo  {journal} {Advanced Optical Materials}\ }\textbf {\bibinfo
  {volume} {14}},\ \bibinfo {pages} {e02535} (\bibinfo {year}
  {2026})}\BibitemShut {NoStop}%
\bibitem [{\citenamefont {Wang}\ \emph {et~al.}(2018)\citenamefont {Wang},
  \citenamefont {Chernikov}, \citenamefont {Glazov}, \citenamefont {Heinz},
  \citenamefont {Marie}, \citenamefont {Amand},\ and\ \citenamefont
  {Urbaszek}}]{Wang2018}%
  \BibitemOpen
  \bibfield  {author} {\bibinfo {author} {\bibfnamefont {G.}~\bibnamefont
  {Wang}}, \bibinfo {author} {\bibfnamefont {A.}~\bibnamefont {Chernikov}},
  \bibinfo {author} {\bibfnamefont {M.~M.}\ \bibnamefont {Glazov}}, \bibinfo
  {author} {\bibfnamefont {T.~F.}\ \bibnamefont {Heinz}}, \bibinfo {author}
  {\bibfnamefont {X.}~\bibnamefont {Marie}}, \bibinfo {author} {\bibfnamefont
  {T.}~\bibnamefont {Amand}},\ and\ \bibinfo {author} {\bibfnamefont
  {B.}~\bibnamefont {Urbaszek}},\ }\bibfield  {title} {\enquote {\bibinfo
  {title} {{\emph{Colloquium}} : {{Excitons}} in atomically thin transition
  metal dichalcogenides},}\ }\href
  {https://doi.org/10.1103/RevModPhys.90.021001} {\bibfield  {journal}
  {\bibinfo  {journal} {Reviews of Modern Physics}\ }\textbf {\bibinfo {volume}
  {90}},\ \bibinfo {pages} {021001} (\bibinfo {year} {2018})}\BibitemShut
  {NoStop}%
\bibitem [{\citenamefont {Eini}\ \emph {et~al.}(2022)\citenamefont {Eini},
  \citenamefont {Asherov}, \citenamefont {Mazor},\ and\ \citenamefont
  {Epstein}}]{Eini2022}%
  \BibitemOpen
  \bibfield  {author} {\bibinfo {author} {\bibfnamefont {T.}~\bibnamefont
  {Eini}}, \bibinfo {author} {\bibfnamefont {T.}~\bibnamefont {Asherov}},
  \bibinfo {author} {\bibfnamefont {Y.}~\bibnamefont {Mazor}},\ and\ \bibinfo
  {author} {\bibfnamefont {I.}~\bibnamefont {Epstein}},\ }\bibfield  {title}
  {\enquote {\bibinfo {title} {Valley-polarized hyperbolic exciton polaritons
  in few-layer two-dimensional semiconductors at visible frequencies},}\ }\href
  {https://doi.org/10.1103/PhysRevB.106.L201405} {\bibfield  {journal}
  {\bibinfo  {journal} {Physical Review B}\ }\textbf {\bibinfo {volume}
  {106}},\ \bibinfo {pages} {L201405} (\bibinfo {year} {2022})}\BibitemShut
  {NoStop}%
\bibitem [{\citenamefont {Chernikov}\ \emph {et~al.}(2014)\citenamefont
  {Chernikov}, \citenamefont {Berkelbach}, \citenamefont {Hill}, \citenamefont
  {Rigosi}, \citenamefont {Li}, \citenamefont {Aslan}, \citenamefont
  {Reichman}, \citenamefont {Hybertsen},\ and\ \citenamefont
  {Heinz}}]{Chernikov2014}%
  \BibitemOpen
  \bibfield  {author} {\bibinfo {author} {\bibfnamefont {A.}~\bibnamefont
  {Chernikov}}, \bibinfo {author} {\bibfnamefont {T.~C.}\ \bibnamefont
  {Berkelbach}}, \bibinfo {author} {\bibfnamefont {H.~M.}\ \bibnamefont
  {Hill}}, \bibinfo {author} {\bibfnamefont {A.}~\bibnamefont {Rigosi}},
  \bibinfo {author} {\bibfnamefont {Y.}~\bibnamefont {Li}}, \bibinfo {author}
  {\bibfnamefont {B.}~\bibnamefont {Aslan}}, \bibinfo {author} {\bibfnamefont
  {D.~R.}\ \bibnamefont {Reichman}}, \bibinfo {author} {\bibfnamefont {M.~S.}\
  \bibnamefont {Hybertsen}},\ and\ \bibinfo {author} {\bibfnamefont {T.~F.}\
  \bibnamefont {Heinz}},\ }\bibfield  {title} {\enquote {\bibinfo {title}
  {Exciton {{Binding Energy}} and {{Nonhydrogenic Rydberg Series}} in
  {{Monolayer WS}} 2},}\ }\href
  {https://doi.org/10.1103/PhysRevLett.113.076802} {\bibfield  {journal}
  {\bibinfo  {journal} {Physical Review Letters}\ }\textbf {\bibinfo {volume}
  {113}},\ \bibinfo {pages} {076802} (\bibinfo {year} {2014})}\BibitemShut
  {NoStop}%
\bibitem [{\citenamefont {Berkelbach}, \citenamefont {Hybertsen},\ and\
  \citenamefont {Reichman}(2015)}]{Berkelbach2015}%
  \BibitemOpen
  \bibfield  {author} {\bibinfo {author} {\bibfnamefont {T.~C.}\ \bibnamefont
  {Berkelbach}}, \bibinfo {author} {\bibfnamefont {M.~S.}\ \bibnamefont
  {Hybertsen}},\ and\ \bibinfo {author} {\bibfnamefont {D.~R.}\ \bibnamefont
  {Reichman}},\ }\bibfield  {title} {\enquote {\bibinfo {title} {Bright and
  dark singlet excitons via linear and two-photon spectroscopy in monolayer
  transition-metal dichalcogenides},}\ }\href
  {https://doi.org/10.1103/PhysRevB.92.085413} {\bibfield  {journal} {\bibinfo
  {journal} {Physical Review B}\ }\textbf {\bibinfo {volume} {92}},\ \bibinfo
  {pages} {085413} (\bibinfo {year} {2015})}\BibitemShut {NoStop}%
\bibitem [{\citenamefont {Liu}\ \emph {et~al.}(2019)\citenamefont {Liu},
  \citenamefont {Van~Baren}, \citenamefont {Taniguchi}, \citenamefont
  {Watanabe}, \citenamefont {Chang},\ and\ \citenamefont {Lui}}]{Liu2019}%
  \BibitemOpen
  \bibfield  {author} {\bibinfo {author} {\bibfnamefont {E.}~\bibnamefont
  {Liu}}, \bibinfo {author} {\bibfnamefont {J.}~\bibnamefont {Van~Baren}},
  \bibinfo {author} {\bibfnamefont {T.}~\bibnamefont {Taniguchi}}, \bibinfo
  {author} {\bibfnamefont {K.}~\bibnamefont {Watanabe}}, \bibinfo {author}
  {\bibfnamefont {Y.-C.}\ \bibnamefont {Chang}},\ and\ \bibinfo {author}
  {\bibfnamefont {C.~H.}\ \bibnamefont {Lui}},\ }\bibfield  {title} {\enquote
  {\bibinfo {title} {Magnetophotoluminescence of exciton rydberg states in
  monolayer {{WSe$_2$}}},}\ }\href {https://doi.org/10.1103/PhysRevB.99.205420}
  {\bibfield  {journal} {\bibinfo  {journal} {Physical Review B}\ }\textbf
  {\bibinfo {volume} {99}},\ \bibinfo {pages} {205420} (\bibinfo {year}
  {2019})}\BibitemShut {NoStop}%
\bibitem [{\citenamefont {Shi}\ \emph {et~al.}(2022)\citenamefont {Shi},
  \citenamefont {Lin}, \citenamefont {Zhu}, \citenamefont {Zhou}, \citenamefont
  {Xu},\ and\ \citenamefont {Xu}}]{Shi2022}%
  \BibitemOpen
  \bibfield  {author} {\bibinfo {author} {\bibfnamefont {J.}~\bibnamefont
  {Shi}}, \bibinfo {author} {\bibfnamefont {Z.}~\bibnamefont {Lin}}, \bibinfo
  {author} {\bibfnamefont {Z.}~\bibnamefont {Zhu}}, \bibinfo {author}
  {\bibfnamefont {J.}~\bibnamefont {Zhou}}, \bibinfo {author} {\bibfnamefont
  {G.~Q.}\ \bibnamefont {Xu}},\ and\ \bibinfo {author} {\bibfnamefont {Q.-H.}\
  \bibnamefont {Xu}},\ }\bibfield  {title} {\enquote {\bibinfo {title} {Probing
  {{Excitonic Rydberg States}} by {{Plasmon Enhanced Nonlinear Optical
  Spectroscopy}} in {{Monolayer WS$_2$}} at {{Room Temperature}}},}\ }\href
  {https://doi.org/10.1021/acsnano.2c02276} {\bibfield  {journal} {\bibinfo
  {journal} {ACS Nano}\ }\textbf {\bibinfo {volume} {16}},\ \bibinfo {pages}
  {15862--15872} (\bibinfo {year} {2022})}\BibitemShut {NoStop}%
\bibitem [{\citenamefont {Xiao}\ \emph {et~al.}(2012)\citenamefont {Xiao},
  \citenamefont {Liu}, \citenamefont {Feng}, \citenamefont {Xu},\ and\
  \citenamefont {Yao}}]{Xiao2012}%
  \BibitemOpen
  \bibfield  {author} {\bibinfo {author} {\bibfnamefont {D.}~\bibnamefont
  {Xiao}}, \bibinfo {author} {\bibfnamefont {G.-B.}\ \bibnamefont {Liu}},
  \bibinfo {author} {\bibfnamefont {W.}~\bibnamefont {Feng}}, \bibinfo {author}
  {\bibfnamefont {X.}~\bibnamefont {Xu}},\ and\ \bibinfo {author}
  {\bibfnamefont {W.}~\bibnamefont {Yao}},\ }\bibfield  {title} {\enquote
  {\bibinfo {title} {Coupled {{Spin}} and {{Valley Physics}} in {{Monolayers}}
  of {{MoS}} 2 and {{Other Group-VI Dichalcogenides}}},}\ }\href
  {https://doi.org/10.1103/PhysRevLett.108.196802} {\bibfield  {journal}
  {\bibinfo  {journal} {Physical Review Letters}\ }\textbf {\bibinfo {volume}
  {108}},\ \bibinfo {pages} {196802} (\bibinfo {year} {2012})}\BibitemShut
  {NoStop}%
\bibitem [{\citenamefont {Wannier}(1937)}]{Wannier1937}%
  \BibitemOpen
  \bibfield  {author} {\bibinfo {author} {\bibfnamefont {G.~H.}\ \bibnamefont
  {Wannier}},\ }\bibfield  {title} {\enquote {\bibinfo {title} {The
  {{Structure}} of {{Electronic Excitation Levels}} in {{Insulating
  Crystals}}},}\ }\href {https://doi.org/10.1103/PhysRev.52.191} {\bibfield
  {journal} {\bibinfo  {journal} {Physical Review}\ }\textbf {\bibinfo {volume}
  {52}},\ \bibinfo {pages} {191--197} (\bibinfo {year} {1937})}\BibitemShut
  {NoStop}%
\bibitem [{\citenamefont {Yu}\ and\ \citenamefont {Cardona}(2010)}]{Yu2010}%
  \BibitemOpen
  \bibfield  {author} {\bibinfo {author} {\bibfnamefont {P.~Y.}\ \bibnamefont
  {Yu}}\ and\ \bibinfo {author} {\bibfnamefont {M.}~\bibnamefont {Cardona}},\
  }\href {https://doi.org/10.1007/978-3-642-00710-1} {\emph {\bibinfo {title}
  {Fundamentals of {{Semiconductors}}: {{Physics}} and {{Materials
  Properties}}}}},\ Graduate {{Texts}} in {{Physics}}\ (\bibinfo  {publisher}
  {Springer Berlin Heidelberg},\ \bibinfo {address} {Berlin, Heidelberg},\
  \bibinfo {year} {2010})\BibitemShut {NoStop}%
\bibitem [{\citenamefont {Haug}\ and\ \citenamefont {Koch}(2009)}]{Haug2009}%
  \BibitemOpen
  \bibfield  {author} {\bibinfo {author} {\bibfnamefont {H.}~\bibnamefont
  {Haug}}\ and\ \bibinfo {author} {\bibfnamefont {S.~W.}\ \bibnamefont
  {Koch}},\ }\href {https://doi.org/10.1142/7184} {\emph {\bibinfo {title}
  {Quantum Theory of the Optical and Electronic Properties of
  Semiconductors}}},\ \bibinfo {edition} {5th}\ ed.\ (\bibinfo  {publisher}
  {WORLD SCIENTIFIC},\ \bibinfo {year} {2009})\BibitemShut {NoStop}%
\bibitem [{\citenamefont {Ye}\ \emph {et~al.}(2014)\citenamefont {Ye},
  \citenamefont {Cao}, \citenamefont {O'Brien}, \citenamefont {Zhu},
  \citenamefont {Yin}, \citenamefont {Wang}, \citenamefont {Louie},\ and\
  \citenamefont {Zhang}}]{Ye2014}%
  \BibitemOpen
  \bibfield  {author} {\bibinfo {author} {\bibfnamefont {Z.}~\bibnamefont
  {Ye}}, \bibinfo {author} {\bibfnamefont {T.}~\bibnamefont {Cao}}, \bibinfo
  {author} {\bibfnamefont {K.}~\bibnamefont {O'Brien}}, \bibinfo {author}
  {\bibfnamefont {H.}~\bibnamefont {Zhu}}, \bibinfo {author} {\bibfnamefont
  {X.}~\bibnamefont {Yin}}, \bibinfo {author} {\bibfnamefont {Y.}~\bibnamefont
  {Wang}}, \bibinfo {author} {\bibfnamefont {S.~G.}\ \bibnamefont {Louie}},\
  and\ \bibinfo {author} {\bibfnamefont {X.}~\bibnamefont {Zhang}},\ }\bibfield
   {title} {\enquote {\bibinfo {title} {Probing excitonic dark states in
  single-layer tungsten disulphide},}\ }\href
  {https://doi.org/10.1038/nature13734} {\bibfield  {journal} {\bibinfo
  {journal} {Nature}\ }\textbf {\bibinfo {volume} {513}},\ \bibinfo {pages}
  {214--218} (\bibinfo {year} {2014})}\BibitemShut {NoStop}%
\bibitem [{\citenamefont {Srivastava}\ and\ \citenamefont {Imamo{\u
  g}lu}(2015)}]{Srivastava2015}%
  \BibitemOpen
  \bibfield  {author} {\bibinfo {author} {\bibfnamefont {A.}~\bibnamefont
  {Srivastava}}\ and\ \bibinfo {author} {\bibfnamefont {A.}~\bibnamefont
  {Imamo{\u g}lu}},\ }\bibfield  {title} {\enquote {\bibinfo {title}
  {Signatures of {{Bloch-Band Geometry}} on {{Excitons}}: {{Nonhydrogenic
  Spectra}} in {{Transition-Metal Dichalcogenides}}},}\ }\href
  {https://doi.org/10.1103/PhysRevLett.115.166802} {\bibfield  {journal}
  {\bibinfo  {journal} {Physical Review Letters}\ }\textbf {\bibinfo {volume}
  {115}},\ \bibinfo {pages} {166802} (\bibinfo {year} {2015})}\BibitemShut
  {NoStop}%
\bibitem [{\citenamefont {Elliott}(1957)}]{Elliott1957}%
  \BibitemOpen
  \bibfield  {author} {\bibinfo {author} {\bibfnamefont {R.~J.}\ \bibnamefont
  {Elliott}},\ }\bibfield  {title} {\enquote {\bibinfo {title} {Intensity of
  {{Optical Absorption}} by {{Excitons}}},}\ }\href
  {https://doi.org/10.1103/PhysRev.108.1384} {\bibfield  {journal} {\bibinfo
  {journal} {Physical Review}\ }\textbf {\bibinfo {volume} {108}},\ \bibinfo
  {pages} {1384--1389} (\bibinfo {year} {1957})}\BibitemShut {NoStop}%
\bibitem [{\citenamefont {Hopfield}(1958)}]{Hopfield1958}%
  \BibitemOpen
  \bibfield  {author} {\bibinfo {author} {\bibfnamefont {J.~J.}\ \bibnamefont
  {Hopfield}},\ }\bibfield  {title} {\enquote {\bibinfo {title} {Theory of the
  {{Contribution}} of {{Excitons}} to the {{Complex Dielectric Constant}} of
  {{Crystals}}},}\ }\href {https://doi.org/10.1103/PhysRev.112.1555} {\bibfield
   {journal} {\bibinfo  {journal} {Physical Review}\ }\textbf {\bibinfo
  {volume} {112}},\ \bibinfo {pages} {1555--1567} (\bibinfo {year}
  {1958})}\BibitemShut {NoStop}%
\bibitem [{\citenamefont {Andersen}, \citenamefont {Latini},\ and\
  \citenamefont {Thygesen}(2015)}]{Andersen2015}%
  \BibitemOpen
  \bibfield  {author} {\bibinfo {author} {\bibfnamefont {K.}~\bibnamefont
  {Andersen}}, \bibinfo {author} {\bibfnamefont {S.}~\bibnamefont {Latini}},\
  and\ \bibinfo {author} {\bibfnamefont {K.~S.}\ \bibnamefont {Thygesen}},\
  }\bibfield  {title} {\enquote {\bibinfo {title} {Dielectric {{Genome}} of van
  der {{Waals Heterostructures}}},}\ }\href
  {https://doi.org/10.1021/acs.nanolett.5b01251} {\bibfield  {journal}
  {\bibinfo  {journal} {Nano Letters}\ }\textbf {\bibinfo {volume} {15}},\
  \bibinfo {pages} {4616--4621} (\bibinfo {year} {2015})}\BibitemShut {NoStop}%
\bibitem [{\citenamefont {Latini}, \citenamefont {Olsen},\ and\ \citenamefont
  {Thygesen}(2015)}]{Latini2015}%
  \BibitemOpen
  \bibfield  {author} {\bibinfo {author} {\bibfnamefont {S.}~\bibnamefont
  {Latini}}, \bibinfo {author} {\bibfnamefont {T.}~\bibnamefont {Olsen}},\ and\
  \bibinfo {author} {\bibfnamefont {K.~S.}\ \bibnamefont {Thygesen}},\
  }\bibfield  {title} {\enquote {\bibinfo {title} {Excitons in van der
  {{Waals}} heterostructures: {{The}} important role of dielectric
  screening},}\ }\href {https://doi.org/10.1103/PhysRevB.92.245123} {\bibfield
  {journal} {\bibinfo  {journal} {Physical Review B}\ }\textbf {\bibinfo
  {volume} {92}},\ \bibinfo {pages} {245123} (\bibinfo {year}
  {2015})}\BibitemShut {NoStop}%
\bibitem [{\citenamefont {Gjerding}\ \emph {et~al.}(2020)\citenamefont
  {Gjerding}, \citenamefont {Cavalcante}, \citenamefont {Chaves},\ and\
  \citenamefont {Thygesen}}]{Gjerding2020}%
  \BibitemOpen
  \bibfield  {author} {\bibinfo {author} {\bibfnamefont {M.~N.}\ \bibnamefont
  {Gjerding}}, \bibinfo {author} {\bibfnamefont {L.~S.~R.}\ \bibnamefont
  {Cavalcante}}, \bibinfo {author} {\bibfnamefont {A.}~\bibnamefont {Chaves}},\
  and\ \bibinfo {author} {\bibfnamefont {K.~S.}\ \bibnamefont {Thygesen}},\
  }\bibfield  {title} {\enquote {\bibinfo {title} {Efficient {{{\emph{Ab
  Initio}}}} {{Modeling}} of {{Dielectric Screening}} in {{2D}} van der {{Waals
  Materials}}: {{Including Phonons}}, {{Substrates}}, and {{Doping}}},}\ }\href
  {https://doi.org/10.1021/acs.jpcc.0c01635} {\bibfield  {journal} {\bibinfo
  {journal} {The Journal of Physical Chemistry C}\ }\textbf {\bibinfo {volume}
  {124}},\ \bibinfo {pages} {11609--11616} (\bibinfo {year}
  {2020})}\BibitemShut {NoStop}%
\bibitem [{\citenamefont {Katzer}\ \emph {et~al.}(2023)\citenamefont {Katzer},
  \citenamefont {Kovalchuk}, \citenamefont {Greben}, \citenamefont {Bolotin},
  \citenamefont {Selig},\ and\ \citenamefont {Knorr}}]{Katzer2023}%
  \BibitemOpen
  \bibfield  {author} {\bibinfo {author} {\bibfnamefont {M.}~\bibnamefont
  {Katzer}}, \bibinfo {author} {\bibfnamefont {S.}~\bibnamefont {Kovalchuk}},
  \bibinfo {author} {\bibfnamefont {K.}~\bibnamefont {Greben}}, \bibinfo
  {author} {\bibfnamefont {K.~I.}\ \bibnamefont {Bolotin}}, \bibinfo {author}
  {\bibfnamefont {M.}~\bibnamefont {Selig}},\ and\ \bibinfo {author}
  {\bibfnamefont {A.}~\bibnamefont {Knorr}},\ }\bibfield  {title} {\enquote
  {\bibinfo {title} {Impact of dark excitons on {{F\"orster-type}} resonant
  energy transfer between dye molecules and atomically thin semiconductors},}\
  }\href {https://doi.org/10.1103/PhysRevB.107.035304} {\bibfield  {journal}
  {\bibinfo  {journal} {Physical Review B}\ }\textbf {\bibinfo {volume}
  {107}},\ \bibinfo {pages} {035304} (\bibinfo {year} {2023})}\BibitemShut
  {NoStop}%
\bibitem [{\citenamefont {Thompson}\ \emph {et~al.}(2023)\citenamefont
  {Thompson}, \citenamefont {Gerhard}, \citenamefont {Witte},\ and\
  \citenamefont {Malic}}]{Thompson2023}%
  \BibitemOpen
  \bibfield  {author} {\bibinfo {author} {\bibfnamefont {J.~J.~P.}\
  \bibnamefont {Thompson}}, \bibinfo {author} {\bibfnamefont {M.}~\bibnamefont
  {Gerhard}}, \bibinfo {author} {\bibfnamefont {G.}~\bibnamefont {Witte}},\
  and\ \bibinfo {author} {\bibfnamefont {E.}~\bibnamefont {Malic}},\ }\bibfield
   {title} {\enquote {\bibinfo {title} {Optical signatures of
  {{F\"orster-induced}} energy transfer in organic/{{TMD}} heterostructures},}\
  }\href {https://doi.org/10.1038/s41699-023-00430-z} {\bibfield  {journal}
  {\bibinfo  {journal} {npj 2D Materials and Applications}\ }\textbf {\bibinfo
  {volume} {7}},\ \bibinfo {pages} {69} (\bibinfo {year} {2023})}\BibitemShut
  {NoStop}%
\bibitem [{\citenamefont {Lin}\ \emph {et~al.}(2023)\citenamefont {Lin},
  \citenamefont {Lo}, \citenamefont {Peng}, \citenamefont {Li}, \citenamefont
  {Huang}, \citenamefont {Chen},\ and\ \citenamefont {Cheng}}]{Lin2023}%
  \BibitemOpen
  \bibfield  {author} {\bibinfo {author} {\bibfnamefont {J.-D.}\ \bibnamefont
  {Lin}}, \bibinfo {author} {\bibfnamefont {P.-Y.}\ \bibnamefont {Lo}},
  \bibinfo {author} {\bibfnamefont {G.-H.}\ \bibnamefont {Peng}}, \bibinfo
  {author} {\bibfnamefont {W.-H.}\ \bibnamefont {Li}}, \bibinfo {author}
  {\bibfnamefont {S.-Y.}\ \bibnamefont {Huang}}, \bibinfo {author}
  {\bibfnamefont {G.-Y.}\ \bibnamefont {Chen}},\ and\ \bibinfo {author}
  {\bibfnamefont {S.-J.}\ \bibnamefont {Cheng}},\ }\bibfield  {title} {\enquote
  {\bibinfo {title} {Essential role of momentum-forbidden dark excitons in the
  energy transfer responses of monolayer transition-metal dichalcogenides},}\
  }\href {https://doi.org/10.1038/s41699-023-00414-z} {\bibfield  {journal}
  {\bibinfo  {journal} {npj 2D Materials and Applications}\ }\textbf {\bibinfo
  {volume} {7}},\ \bibinfo {pages} {51} (\bibinfo {year} {2023})}\BibitemShut
  {NoStop}%
\bibitem [{\citenamefont {Glazov}\ \emph {et~al.}(2014)\citenamefont {Glazov},
  \citenamefont {Amand}, \citenamefont {Marie}, \citenamefont {Lagarde},
  \citenamefont {Bouet},\ and\ \citenamefont {Urbaszek}}]{Glazov2014}%
  \BibitemOpen
  \bibfield  {author} {\bibinfo {author} {\bibfnamefont {M.~M.}\ \bibnamefont
  {Glazov}}, \bibinfo {author} {\bibfnamefont {T.}~\bibnamefont {Amand}},
  \bibinfo {author} {\bibfnamefont {X.}~\bibnamefont {Marie}}, \bibinfo
  {author} {\bibfnamefont {D.}~\bibnamefont {Lagarde}}, \bibinfo {author}
  {\bibfnamefont {L.}~\bibnamefont {Bouet}},\ and\ \bibinfo {author}
  {\bibfnamefont {B.}~\bibnamefont {Urbaszek}},\ }\bibfield  {title} {\enquote
  {\bibinfo {title} {Exciton fine structure and spin decoherence in monolayers
  of transition metal dichalcogenides},}\ }\href
  {https://doi.org/10.1103/PhysRevB.89.201302} {\bibfield  {journal} {\bibinfo
  {journal} {Physical Review B}\ }\textbf {\bibinfo {volume} {89}},\ \bibinfo
  {pages} {201302} (\bibinfo {year} {2014})}\BibitemShut {NoStop}%
\bibitem [{\citenamefont {Glazov}\ \emph {et~al.}(2015)\citenamefont {Glazov},
  \citenamefont {Ivchenko}, \citenamefont {Wang}, \citenamefont {Amand},
  \citenamefont {Marie}, \citenamefont {Urbaszek},\ and\ \citenamefont
  {Liu}}]{Glazov2015}%
  \BibitemOpen
  \bibfield  {author} {\bibinfo {author} {\bibfnamefont {M.~M.}\ \bibnamefont
  {Glazov}}, \bibinfo {author} {\bibfnamefont {E.~L.}\ \bibnamefont
  {Ivchenko}}, \bibinfo {author} {\bibfnamefont {G.}~\bibnamefont {Wang}},
  \bibinfo {author} {\bibfnamefont {T.}~\bibnamefont {Amand}}, \bibinfo
  {author} {\bibfnamefont {X.}~\bibnamefont {Marie}}, \bibinfo {author}
  {\bibfnamefont {B.}~\bibnamefont {Urbaszek}},\ and\ \bibinfo {author}
  {\bibfnamefont {B.~L.}\ \bibnamefont {Liu}},\ }\bibfield  {title} {\enquote
  {\bibinfo {title} {Spin and valley dynamics of excitons in transition metal
  dichalcogenide monolayers},}\ }\href {https://doi.org/10.1002/pssb.201552211}
  {\bibfield  {journal} {\bibinfo  {journal} {physica status solidi (b)}\
  }\textbf {\bibinfo {volume} {252}},\ \bibinfo {pages} {2349--2362} (\bibinfo
  {year} {2015})}\BibitemShut {NoStop}%
\bibitem [{\citenamefont {Cho}(1986)}]{Cho1986}%
  \BibitemOpen
  \bibfield  {author} {\bibinfo {author} {\bibfnamefont {K.}~\bibnamefont
  {Cho}},\ }\bibfield  {title} {\enquote {\bibinfo {title} {`{{ABC}}'-{{Free
  Theory}} of {{Polariton}}: {{From Semi-Infinite Medium}} to {{Quantum
  Well}}},}\ }\href {https://doi.org/10.1143/JPSJ.55.4113} {\bibfield
  {journal} {\bibinfo  {journal} {Journal of the Physical Society of Japan}\
  }\textbf {\bibinfo {volume} {55}},\ \bibinfo {pages} {4113--4121} (\bibinfo
  {year} {1986})}\BibitemShut {NoStop}%
\bibitem [{\citenamefont {Ishihara}\ and\ \citenamefont
  {Cho}(1990)}]{Ishihara1990}%
  \BibitemOpen
  \bibfield  {author} {\bibinfo {author} {\bibfnamefont {H.}~\bibnamefont
  {Ishihara}}\ and\ \bibinfo {author} {\bibfnamefont {K.}~\bibnamefont {Cho}},\
  }\bibfield  {title} {\enquote {\bibinfo {title}
  {Additional-boundary-condition-free theory of an exciton polariton in a
  slab},}\ }\href {https://doi.org/10.1103/PhysRevB.41.1424} {\bibfield
  {journal} {\bibinfo  {journal} {Physical Review B}\ }\textbf {\bibinfo
  {volume} {41}},\ \bibinfo {pages} {1424--1434} (\bibinfo {year}
  {1990})}\BibitemShut {NoStop}%
\bibitem [{\citenamefont {Inoue}\ \emph {et~al.}(2025)\citenamefont {Inoue},
  \citenamefont {Yamane}, \citenamefont {Tamura},\ and\ \citenamefont
  {Ishihara}}]{Inoue2025}%
  \BibitemOpen
  \bibfield  {author} {\bibinfo {author} {\bibfnamefont {S.}~\bibnamefont
  {Inoue}}, \bibinfo {author} {\bibfnamefont {H.}~\bibnamefont {Yamane}},
  \bibinfo {author} {\bibfnamefont {M.}~\bibnamefont {Tamura}},\ and\ \bibinfo
  {author} {\bibfnamefont {H.}~\bibnamefont {Ishihara}},\ }\bibfield  {title}
  {\enquote {\bibinfo {title} {Coherent coupling among plasmons, electron--hole
  pairs, and light: Energy transparency, imaging, and efficient hot-carrier
  generation},}\ }\href {https://doi.org/10.3788/PI.2025.R12} {\bibfield
  {journal} {\bibinfo  {journal} {Photonics Insights}\ }\textbf {\bibinfo
  {volume} {4}},\ \bibinfo {pages} {R12} (\bibinfo {year} {2025})}\BibitemShut
  {NoStop}%
\bibitem [{\citenamefont {Ashida}\ \emph {et~al.}(2026)\citenamefont {Ashida},
  \citenamefont {Okamoto}, \citenamefont {Matsuda},\ and\ \citenamefont
  {Ishihara}}]{Ashida2026}%
  \BibitemOpen
  \bibfield  {author} {\bibinfo {author} {\bibfnamefont {M.}~\bibnamefont
  {Ashida}}, \bibinfo {author} {\bibfnamefont {K.}~\bibnamefont {Okamoto}},
  \bibinfo {author} {\bibfnamefont {T.}~\bibnamefont {Matsuda}},\ and\ \bibinfo
  {author} {\bibfnamefont {H.}~\bibnamefont {Ishihara}},\ }\bibfield  {title}
  {\enquote {\bibinfo {title} {Accelerating {{Luminescence}} in
  {{Nanostructures}}: {{Exploring}} the {{Physical Limits}} and {{Impact}} of
  {{Ultrafast Emission}} in {{Nanoscale Materials}}},}\ }\href
  {https://doi.org/10.1002/adpr.202500127} {\bibfield  {journal} {\bibinfo
  {journal} {Advanced Photonics Research}\ }\textbf {\bibinfo {volume} {7}},\
  \bibinfo {pages} {e202500127} (\bibinfo {year} {2026})}\BibitemShut {NoStop}%
\bibitem [{\citenamefont {Haastrup}\ \emph {et~al.}(2018)\citenamefont
  {Haastrup}, \citenamefont {Strange}, \citenamefont {Pandey}, \citenamefont
  {Deilmann}, \citenamefont {Schmidt}, \citenamefont {Hinsche}, \citenamefont
  {Gjerding}, \citenamefont {Torelli}, \citenamefont {Larsen}, \citenamefont
  {{Riis-Jensen}}, \citenamefont {Gath}, \citenamefont {Jacobsen},
  \citenamefont {J{\o}rgen~Mortensen}, \citenamefont {Olsen},\ and\
  \citenamefont {Thygesen}}]{Haastrup2018}%
  \BibitemOpen
  \bibfield  {author} {\bibinfo {author} {\bibfnamefont {S.}~\bibnamefont
  {Haastrup}}, \bibinfo {author} {\bibfnamefont {M.}~\bibnamefont {Strange}},
  \bibinfo {author} {\bibfnamefont {M.}~\bibnamefont {Pandey}}, \bibinfo
  {author} {\bibfnamefont {T.}~\bibnamefont {Deilmann}}, \bibinfo {author}
  {\bibfnamefont {P.~S.}\ \bibnamefont {Schmidt}}, \bibinfo {author}
  {\bibfnamefont {N.~F.}\ \bibnamefont {Hinsche}}, \bibinfo {author}
  {\bibfnamefont {M.~N.}\ \bibnamefont {Gjerding}}, \bibinfo {author}
  {\bibfnamefont {D.}~\bibnamefont {Torelli}}, \bibinfo {author} {\bibfnamefont
  {P.~M.}\ \bibnamefont {Larsen}}, \bibinfo {author} {\bibfnamefont {A.~C.}\
  \bibnamefont {{Riis-Jensen}}}, \bibinfo {author} {\bibfnamefont
  {J.}~\bibnamefont {Gath}}, \bibinfo {author} {\bibfnamefont {K.~W.}\
  \bibnamefont {Jacobsen}}, \bibinfo {author} {\bibfnamefont {J.}~\bibnamefont
  {J{\o}rgen~Mortensen}}, \bibinfo {author} {\bibfnamefont {T.}~\bibnamefont
  {Olsen}},\ and\ \bibinfo {author} {\bibfnamefont {K.~S.}\ \bibnamefont
  {Thygesen}},\ }\bibfield  {title} {\enquote {\bibinfo {title} {The
  {{Computational 2D Materials Database}}: High-throughput modeling and
  discovery of atomically thin crystals},}\ }\href
  {https://doi.org/10.1088/2053-1583/aacfc1} {\bibfield  {journal} {\bibinfo
  {journal} {2D Materials}\ }\textbf {\bibinfo {volume} {5}},\ \bibinfo {pages}
  {042002} (\bibinfo {year} {2018})}\BibitemShut {NoStop}%
\bibitem [{\citenamefont {Momma}\ and\ \citenamefont
  {Izumi}(2011)}]{Momma2011}%
  \BibitemOpen
  \bibfield  {author} {\bibinfo {author} {\bibfnamefont {K.}~\bibnamefont
  {Momma}}\ and\ \bibinfo {author} {\bibfnamefont {F.}~\bibnamefont {Izumi}},\
  }\bibfield  {title} {\enquote {\bibinfo {title} {{{{\emph{VESTA}}}}{\emph{
  3}} for three-dimensional visualization of crystal, volumetric and morphology
  data},}\ }\href {https://doi.org/10.1107/S0021889811038970} {\bibfield
  {journal} {\bibinfo  {journal} {Journal of Applied Crystallography}\ }\textbf
  {\bibinfo {volume} {44}},\ \bibinfo {pages} {1272--1276} (\bibinfo {year}
  {2011})}\BibitemShut {NoStop}%
\end{thebibliography}
%

\end{document}